\documentclass[12pt]{article}
\usepackage[sort&compress,square,comma,numbers]{natbib}
\usepackage{amsmath,amssymb,graphicx,slashed,bm}
\usepackage{feynmp}
\usepackage{subfigure}

\usepackage[
	colorlinks=true,
	citecolor=black,
	linkcolor=black,
	urlcolor=blue,
	hypertexnames=false]{hyperref}

\makeatletter
\def\endfmffile{%
	\fmfcmd{\p@rcent\space the end.^^J%
		end.^^J%
		endinput;}%
	\if@fmfio
	\immediate\closeout\@outfmf
	\fi
	\ifnum\pdfshellescape>\z@
	\immediate\write18{mpost \thefmffile}%
	\fi}
\makeatother

\newcommand{\arXiv}[2]{\href{http://arxiv.org/pdf/#1}{{\tt #2/#1}}}
\newcommand{\arXivold}[1]{\href{http://arxiv.org/pdf/#1}{{\tt #1}}}

\newcommand{\beq}{\begin{eqnarray}}
\newcommand{\eeq}{\end{eqnarray}}

\begin{document}
\begin{center} 
{\huge \bf Schwinger vs Coleman:} \\
\vspace*{0.1cm}
{\huge \bf Magnetic Charge Renormalization
} 
\end{center}

\begin{center} 

{\bf Joshua Newey}$^1$, {\bf John Terning}$^2$ and {\bf Christopher B. Verhaaren}$^3$ \\
\end{center}
\vskip 8pt
\begin{center} 
$^1${\it Department of Mathematics, Brigham Young University, Provo, UT, 84602, USA}
$^2${\it Center for Quantum Mathematics and Physics (QMAP)\\Department of Physics, University of California, Davis, CA 95616}\\
$^3${\it Department of Physics and Astronomy, Brigham Young University, \\ Provo, UT 84602, USA}
\end{center}

\vspace*{0.1cm}
\begin{center} 
{\tt 
 \href{mailto:jnewey@mathematics.byu.edu}{jnewey@mathematics.byu.edu}\,
 \href{mailto:jterning@gmail.com}{jterning@gmail.com}\,
 \href{mailto:verhaaren@physics.byu.edu}{verhaaren@physics.byu.edu}}

\end{center}

\centerline{\large\bf Abstract}
\begin{quote}
The kinetic mixing of two $U(1)$ gauge theories can result in a massless photon that has perturbative couplings to both electric and magnetic charges. This framework can be used to  perturbatively calculate in a quantum field theory with both kinds of charge. Here we re-examine the running of the magnetic charge, where the calculations of Schwinger and Coleman sharply disagree.  We calculate  the running of both electric and magnetic couplings and show that the disagreement between Schwinger and Coleman is due to an incomplete summation of topological terms in the perturbation series. We present a momentum space prescription for calculating the loop corrections  in which the topological terms can be systematically separated for resummation.  Somewhat in the spirit of modern amplitude methods we avoid using a vector potential and use the field strength itself, thereby trading gauge redundancy for the geometric redundancy of Stokes surfaces. The resulting running of the couplings demonstrates that Dirac charge quantization is independent of renormalization scale, as Coleman predicted. As a simple application we also bound the parameter space of magnetically charged states through the experimental measurement of the running of electromagnetic coupling.
\end{quote}


\section{Introduction\label{s.Intro}}

In 1904 J.J. Thomson~\cite{Thomson:1904} recorded a significant fact regarding the interactions of electric and magnetic particles. He considered an electric charge that produces a Coulomb field
\beq
\vec{E}({\vec r})=\frac{eq}{4\pi }\frac{{\vec r}-{\vec r}_e}{|{\vec r}-{\vec r}_e|^3}~,
\eeq
 where $e$ is the electric coupling and $q$ is the actual (dimensionless) charge of the particle.  At some distance away he added a fixed magnetic charge which produces the monopole field
\beq
\vec{B}({\vec r})=\frac{bg}{4\pi }\frac{{\vec r}-{\vec r}_b}{|{\vec r}-{\vec r}_b|^3}~,
\eeq
where $b$ is the magnetic coupling and $g$ is the magnetic charge of this particle. Thomson then calculated the angular momentum carried by the electric and magnetic fields and found
\beq
\vec{L}=\frac{eqbg}{4\pi}\widehat{R}~,
\label{e.Thomson}
\eeq
where $\vec{R}=\vec{r}_b-\vec{r}_e$~.

This result is suggestive of two properties of electric-magnetic interactions. First, the quantization of angular momentum is directly tied to a quantized relationship between electric and magnetic couplings and charges discovered by Dirac~\cite{Dirac:1931kp}: $eqbg$ must be a multiple of $2 \pi$. Second, there is a topological aspect to the interaction of electric and magnetic charges. This follows from the angular momentum having no dependence on the distance between the charges. The direction of the angular momentum is given by the unit vector $\widehat{R}$ that points from the electric charge to the magnetic, but the magnitude of the angular momentum has no dependence on the geometry of the system.

It is now fairly widely understood that a the resolution of the Weinberg paradox \cite{Weinberg:1965rz} (the non-Lorentz invariance of tree-level electric-magnetic scattering amplitudes) relies on a careful accounting of topological effects \cite{Terning:2018udc}: the perturbative Lorentz breaking appears only in topological terms that re-sum to an overall phase, and the phase is trivial multiple of $2 \pi$ provided that Dirac charge quantization is satisfied. 
However when considering loop corrections that induce running of the electric coupling, another paradox arises.  The first calculation, due to Schwinger~\cite{Schwinger:1966zza,Schwinger:1966zzb}, indicated that the magnetic coupling runs just as the electric coupling does, but in that case the overall phase discussed above is trivial only at a particular renormalization scale and Lorentz violation would again rear its ugly head.

As emphasized by Coleman~\cite{Coleman:1982cx}, a field theorist confronted with Dirac's charge quantization should ask ``At what scale?" That is, given a product of couplings that run with energy scale,  why is  the product quantized at some particular renormalization scale? The answer is that Dirac charge quantization can only make sense if the product of electric and magnetic couplings is a renormalization group invariant. Thus Coleman concluded that the magnetic coupling runs inversely to the electric running.

Some might be tempted to dismiss the whole Schwinger-Coleman debate as unresolvable, since it must necessarily involve issues of strong coupling.  This is not really the case; we can make use of ``dark'' magnetic monopoles \cite{Terning:2018lsv}, a framework that allows us to work perturbatively with both electric and magnetic charges since the magnetic charge is suppressed by a dimensionless kinetic mixing between the ordinary photon and a massive ``dark" photon. The price we must pay is that the magnetic monopoles are confined by a ``dark'' magnetic flux tube (which can be described perturbatively in an effective theory below the mass of the ``dark" Higgs which provided the photon mass) and the fact that the re-summed  topological phases are no longer trivial (and depend on the flux tube orientation). However, the point is that the topological terms can still be identified and re-summed, independent of whether the resulting phase is a multiple of $2 \pi$ or not. 

We show in this paper that a proper accounting of topological terms (which should be re-summed into overall phases) allows one to show, using perturbative calculations in momentum space, that Coleman's argument is correct. However, a systematic analysis for separating the topological pieces from an amplitude is quite subtle. Here we are aided by recasting the calculation in terms of field strength propagators rather than gauge potential propagators. The former have the advantage that they are manifestly gauge invariant. They have been advocated in the past for position space calculations~\cite{Halpern:1978ik,Calucci:1981we,Calucci:1982fm,Calucci:1982wy,Blagojevic:1985yd}, but their couplings to currents are quite strange: nonlocal and seemingly dependent on the (non-unique) Stokes surface to which the field strengths couple (as opposed to the currents themselves).  Nevertheless, we find that the Stokes surface ambiguity does not appear in scattering amplitudes and we show how to write field strength interactions in a form suitable for momentum space calculations. Without this last step the formalism would just be formalism, unable to aid in practical calculations, for instance those related to the phenomenology of ``dark'' monopoles~\cite{GomezSanchez:2011orv,Hook:2017vyc,Terning:2019bhg,Terning:2020dzg,Graesser:2021vkr}. An essential ingredient is to have a simple momentum space description of a Stokes surface corresponding to particles with arbitrary momentum. 

Using these methods we are able to extract the running of both electric and magnetic couplings and verify that their product is renormalization scale independent. This quite general relation applies to any low-energy, effective theory in which the magnetic charge is perturbative. In particular, this running of the magnetic coupling is consistent with the results of Seiberg-Witten theory \cite{SeibergWitten} where light, dynamical monopoles can be under perturbative control in the low-energy effective theory and the one-loop running can be extracted from the exact solution, as shown in ref.~\cite{Colwell:2015wna}.  The essential point is that when the Compton wavelength of a composite state or soliton becomes much larger than the physical size of the object, then it can be treated as an ordinary particle in the low-energy, effective field theory. This is what happens in the Seiberg-Witten theory where the monopole mass goes to zero at a finite value of the vacuum expectation value of the field the breaks the gauge symmetry down to $U(1)$.

In the following section we review the past calculations of magnetic charge renormalization and the topological aspect of electric-magnetic interactions without invoking a specific Lagrangian formulation. In so doing we find that the topological part of the amplitude can be separated from the dynamical part when the Stokes surfaces bounded by electric or magnetic currents are used. These surfaces can be thought of as sources for the field strength or its dual. This naturally leads to a discussion of objects that map from two-forms, like surfaces and field strengths, to other two-forms. In Sec.~\ref{s.TwoForms} we consider these mappings generally and outline many of their properties. In Sec.~\ref{s.onePot} we show how these objects can be used to make the standard calculations of quantum electrodynamics (QED). 

We then generalize to theories with both electrically and magnetically charged particles, quantum electro-magnetodynamics\footnote{For a general review see ref. \cite{Blagojevic:1985yd}.} (QEMD), coupled to a single potential and show how the running described by Schwinger appears, along with extra contributions which signal that a subtle mistake has been made. The problem is that topological terms are embedded within the one-loop calculation such that they are difficult to recognize. The careful removal of the topological terms from the perturbation series is shown to produce exactly the renormalization condition obtained by Coleman. In Sec.~\ref{s.Zform}, we confirm these results using Zwanziger's two potential formalism~\cite{Zwanziger:1970hk} in two ways. In particular, we exhibit a particularly simple and concise language for these calculations in which electromagnetic dualities are manifest. These results are used, in Sec.~\ref{s.fineBound}, to find a simple phenomenological constraint on milli-magnetic charged particles from the measured running of the electric coupling. We offer conclusions and outlook in Sec.~\ref{s.con}. Technical details related to the derivation of field strength propagators, the surfaces that source field strengths, and the potential based renormalization in the Zwanziger formalism are provided in Appendices~\ref{a.Fprop},~\ref{a.Fsource}, and~\ref{a.PotRenorm}, respectively.

\section{Review of Past Claims and Counterclaims\label{s.Review}}
Schwinger's classic result on magnetic charge renormalization~\cite{Schwinger:1966zza,Schwinger:1966zzb}, subsequently supported by other calculations~\cite{Brandt:1977fa,Deans:1981qs,Panagiotakopoulos:1982fp}, is that the electric and magnetic couplings are renormalized by the same function
\beq
e(\mu)=\sqrt{Z_S(\mu)}e_0~, \ \ \ \ \ b(\mu)=\sqrt{Z_S(\mu)}b_0~,
\eeq
implying that the ratio of the couplings $e/b$ is a renormalization group invariant. In terms of charge quantization this seems to require that $Z_S(\mu)$ is a rational number, which is at odds with the usual understanding of a running coupling.

The alternative result, obtained by Coleman~\cite{Coleman:1982cx} and others~\cite{Calucci:1982wy,Jengo:1982wx,Goebel:1983we} is that the electric and magnetic couplings are renormalized exactly inversely
\beq
e(\mu)=\sqrt{Z(\mu)}e_0~, \ \ \ \ \ b(\mu)=\frac{b_0}{\sqrt{Z(\mu)}}~.
\eeq
In this case the product of the couplings $eb$ is a renormalization group invariant. The usual choice is to take
\beq
b(\mu)=\frac{4\pi}{e(\mu)}~,
\eeq 
leading to the Dirac quantization condition on the charges
\beq
qg=\frac{N}{2}~.
\eeq
This result agrees with our usual understanding of running couplings and also indicates why a theory with perturbative electric couplings to the photon implies a non-perturbative magnetic coupling to the photon. 

This non-perturbative aspect of theories with both types of charges leads to significant calculational difficulties. A simpler case to analyze is to consider a small kinetic mixing between two $U(1)$ gauge theories where one of the gauge bosons has acquired a mass through a Higgs mechanism. This is one of the standard scenarios for a class of dark matter models \cite {Holdom:1985ag}, and we refer to the massive photon as the ``dark'' photon. In this scenario the massless photon can have perturbative couplings to both types of charges~\cite{Brummer:2009cs,Bruemmer:2009ky,GomezSanchez:2011orv,DelZotto:2016fju,Hook:2017vyc,Terning:2018lsv,Hook:2022pcf}, though at least one of the types of particles must be confined by the interactions with the ``dark'' photon. In particular, if the magnetically charged particles are ``dark'' (i.e. they only couple to the ``dark'' photon in the absence of kinetic mixing), then their coupling to the ordinary photon can be arbitrarily small, suppressed by the kinetic mixing. The only ``dark'' electric charges required are those of the ``dark'' Higgs, which can be heavy, and integrated out of the theory. These results imply that there is a class of effective theories with perturbative interactions for both electric and magnetic particles with a photon. Still, non-perturbative information is a powerful guide, and the constraint that topological terms must become trivial when Dirac charge quantization is imposed is seen to be extremely helpful.

Theories with both electric and magnetic charges have equations of motion that are just the extended Maxwell equations:
\beq
\partial_\mu F^{\mu\nu}=eJ^\nu, \ \ \ \ \partial_\mu {}^\ast\! F^{\mu\nu}=bK^\nu~,\label{e.MaxEq}
\eeq
with $J^\mu$ the electric current and $K^\mu$ the magnetic current and where the dual field strength is defined by
\beq
{}^\ast\! F^{\mu\nu}=\frac12\varepsilon^{\mu\nu\alpha\beta}F_{\alpha\beta}~.
\eeq
One manifestation of the difficult nature of theories with both kinds of charges is the variety of Lagrangian formulations that have been employed to characterize them~\cite{Blagojevic:1985sh}. In general one has to sacrifice either manifest Lorentz invariance of manifest locality in order to even write a Lagrangian, so these theories have acquired the nickname ``mutually non-local'' theories.

Of course, Thomson's result~\eqref{e.Thomson} does not rely on any particular Lagrangian formulation of electrodynamics. Similarly, Weinberg determined a crucial aspect of the quantum field theory of electric-magnetic interactions without assuming any Lagrangian~\cite{Weinberg:1965rz}. After defining two vector potentials, one $A_\mu$ with local couplings to electric currents and one $B_\mu$ with local couplings to magnetic currents, he found the propagator between them to be\footnote{We use slightly different notation than Weinberg, but an equivalent result.}
\beq
\int d^4x e^{-ik\cdot(x-y)}\langle \bm{0}| T\left\{ A_\mu(x) B_\mu(y) \right\}|\bm{0}\rangle\equiv\Delta^{AB}_{\mu\nu}=-i\frac{\varepsilon_{\mu\nu\alpha\beta}n^\alpha k^\beta}{k^2(n\cdot k)}~,
\eeq
where $k^\mu$ is the propagating momentum and $n^\mu$ is a constant vector. It is, of course, the dependence on $n^\mu$ (and the associated spurious pole) that is most unsettling.

However, this perplexing dependence on $n^\mu$ can be tied to the topological nature of Thomson's calculation. Consider the leading order correlation of an external electric current $J_e$ with an external magnetic current $K_e$
\beq
i\mathcal{M}^{JK}_0\equiv -ieJ_e^\mu(-i)\frac{\varepsilon_{\mu\nu\alpha\beta}n^\alpha k^\beta}{k^2(n\cdot k)}(-ib)K_e^\nu=ieb\frac{\varepsilon_{\mu\nu\alpha\beta}J_e^\mu K_e^\nu n^\alpha k^\beta}{k^2(n\cdot k)}~.\label{e.WeinAB}
\eeq
Where we have split the currents into dynamical and external background pieces 
\beq 
J^\mu=J_d^\mu+J_e^\mu ~,\quad\quad K^\nu=K_d^\mu+K_e^\mu~.
\eeq
Using the worldline formulation~\cite{Calucci:1981we,Calucci:1982fm,Calucci:1982wy,Strassler:1992zr,Schubert:1996jj} we can write the external currents as
\beq
J_e^\mu(x)=q\int_{C_J}\,dz_J^\mu \delta^{(4)}(x-z_J)~, \ \ \ \ K_e^\mu(x)=g\int_{C_K}\,dz_K^\mu \delta^{(4)}(x-z_K)~, 
\eeq
where $C_J$ and $C_K$ are the spacetime paths (worldlines) of the electric and magnetic charges, respectively. The mixed propagator in position space is
\beq
\Delta^{AB}_{\mu\nu}(x-y)=-\Delta^{BA}_{\mu\nu}(x-y)=\frac{1}{4\pi^2}\frac{\varepsilon_{\mu\nu\alpha\beta}n^\alpha\partial^\beta}{n\cdot\partial}\frac{1}{(x-y)^2+i\epsilon}~.
\eeq
We can then write the lowest order tree-level interaction of an electric and a magnetic current as
\begin{align}
\mathcal{M}^{JK}_0=&\int d^4x\,d^4y\,J_e^\mu(x)\Delta_{\mu\nu}^{AB}(x-y)K_e^\nu(y)\nonumber\\
=&\frac{ebqg}{4\pi^2}\int dz_J^\mu dz_K^\nu\frac{\varepsilon_{\mu\nu\sigma\rho}n^\sigma\partial^\rho}{n\cdot\partial}\frac{1}{(z_J-z_K)^2}~.
\end{align}
Using Stokes' theorem,
\beq
\int_{\partial S}dz^\mu A_\mu =\int_S d^2\sigma_{\alpha\beta}\varepsilon^{\alpha\beta\mu\nu}\partial_\mu A_\nu ~,
\eeq
 on both particle trajectories we find
\begin{align}
\mathcal{M}^{JK}_0=&\frac{ebqg}{4\pi^2}\int d\sigma_{\alpha\beta}^Jd\sigma_{\gamma\delta}^K\varepsilon^{\alpha\beta\mu\lambda}\partial_\mu\varepsilon^{\gamma\delta\nu\xi}\partial_\nu\frac{\varepsilon_{\lambda\xi\sigma\rho}n^\sigma\partial^\rho}{n\cdot\partial}\frac{1}{(z_J-z_K)^2}\nonumber\\
=&\frac{2ebqg}{4\pi^2}\int d\sigma_{\alpha\beta}^Jd\sigma_{\gamma\delta}^K\varepsilon^{\alpha\beta\mu\lambda}\partial_\mu\left(\delta^\gamma_\lambda\partial^\delta n\cdot \partial-\delta^\gamma_\lambda n^\delta\partial^2 \right)\frac{1}{n\cdot\partial}\frac{1}{(z_J-z_K)^2}~.\label{e.JKcor}
\end{align}
Significantly, the first term is independent of $n^\mu$ and has the usual massless pole of photon exchange. The second term does depend on $n^\mu$ but does not have the usual massless pole. This latter term has been shown to be a pure number given by $4\pi^2$ times the number of intersections of the Stokes surface bounded by the worldline of the electric particle and the surface swept out by the magnetic worldline and the ray along $n^\mu$, see refs.~\cite{Brandt:1977be,Brandt:1978wc,Terning:2018udc}.  These papers show that with $n^\mu$ chosen to be spacelike it simply represents the direction of the Dirac string,  and the surface is just the world-sheet of the Dirac string. The number of intersections is equivalent to the 4D topological linking number of a worldline and a Stokes surface \cite{Terning:2018udc}. 

That such a topological term should appear may be surprising, but it can be understood as a manifestation of Thomson's discovery about the interactions of electric and magnetic charges. In worldline formulations~\cite{Brandt:1977be,Brandt:1978wc,Terning:2018udc} the topological term appears as a constant term in the action and hence a global phase multiplying the path integral. When recognized as a global phase it can be straightforward to separate the dynamics of a theory from the topological phase. When we perform perturbative calculations we are interested in the series
\beq
\left(1+ie^2 a_1+ie^4a_2+\ldots \right)e^{i\varphi}~,
\eeq
which captures the dynamics of the currents and fields and keeps the global phase separately multiplying each term.
When theories with electric and magnetic particles are written in terms of surfaces, or two-form interactions, this topological phase $\varphi$ can be separated from the other field dynamics in just this way, keeping it from appearing in the perturbative calculations. When the currents are directly coupled to the potentials, however, the topological terms cannot be fully separated in the action. This ends up confusing the perturbation series captured by Feynman diagrams, producing the series
\beq
1+i\varphi+ie^2 a_1-e^2a_1\varphi+ie^4a_2-\frac12\varphi^2 +\ldots~.
\eeq
In this work we show how to recognize all manifestations of terms like $a_1\varphi$, which we want to carefully separate from the field dynamics so that they can be re-summed as an overall phase. We show that standard potential calculations very nearly accomplish this, but run afoul of this mixing between the phase and dynamics in loop effects. By organizing the perturbation series in a new way we show how to completely separate the effects of topological terms. This novel organization is completely equal, algebraically, to the usual potential based calculations. However, the seemingly complicated way of writing the amplitudes, and in separating out the topological parts, are most easily understood in the language of mappings from two-forms to two-forms.\footnote{In this work we use the term two-form to apply to antisymmetric tensors like $F_{\mu\nu}$ even when the basis of differentials $F_{\mu\nu} dx^\mu\wedge dx^\nu$ are not explicitly included.}

The mappings we are interested can be found in the way the photon field connects the Stokes surfaces bounded by currents. 
In momentum space the mapping from two-forms to two forms related to photon propagation is
\beq
{}^{\alpha\beta}_{(k)}P^{\gamma\delta}_{(k)}=\frac{1}{2 k^2}\left(\eta^{\alpha\gamma}k^\beta k^\delta-\eta^{\alpha\delta}k^\beta k^\gamma-\eta^{\beta\gamma}k^\alpha k^\delta+\eta^{\beta\delta}k^\alpha k^\gamma \right)~,
\eeq
where $\eta_{\mu\nu}$ is the Minkowski metric, $\eta_{\mu\nu}=\text{Diag}(1,-1,-1,-1)$.
While this object is not familiar to most, we see in the following sections that it can be interpreted as being simply proportional to the field strength propagator in QED.

We also see that this mapping ${}^{\alpha\beta}_{(k)}P^{\gamma\delta}_{(k)}$ appears when we write the first term, in QEMD, of the leading electric-magnetic interaction given in Eq.~\eqref{e.JKcor} in momentum space: 
\begin{align}
&\frac{eqbg}{2\pi^2}\int d\sigma_{\alpha\beta}^Jd\sigma_{\gamma\delta}^K\varepsilon^{\alpha\beta\mu\gamma}\partial_\mu\partial^\delta\frac{1}{(z_J-z_K)^2}\nonumber\\
=&\frac{qg}{2\pi^2}\int d\sigma_{\alpha\beta}^Jd\sigma_{\gamma\delta}^K \frac14\varepsilon^{\alpha\beta\mu\nu}\left(\delta_\nu^\gamma\partial_\mu\partial^\delta+\delta_\mu^\delta\partial_\nu\partial^\gamma-\delta_\nu^\delta\partial_\mu\partial^\gamma-\delta_\mu^\gamma\partial_\nu\partial^\delta\right)\frac{1}{(z_J-z_K)^2}~,\nonumber\\
=&eqbg\int d\sigma_{\alpha\beta}^J\frac12\varepsilon^{\alpha\beta\mu\nu}d\sigma_{\gamma\delta}^K\frac12\varepsilon^{\gamma\delta\sigma\rho} d^4ke^{-ik\cdot(z_J-z_K)}(-2i){}_{\mu\nu}^{(k)}P_{\ast\sigma\rho}^{(k)}~,\label{e.JKpole}
\end{align}
where we have used a standard result for the double summation of Levi-Civita tensors:
\beq
-\frac12\varepsilon^{\alpha\beta\mu\nu}\frac12\varepsilon_{\mu\nu\gamma\delta}=\frac12\left(\eta^{\alpha}_{\gamma}\eta^{\beta}_{\delta}-\eta^{\alpha}_{\delta}\eta^{\beta}_{\gamma} \right)~.
\label{e.doubleLC}
\eeq

The second term in Eq.~\eqref{e.JKcor}, what we somewhat loosely call the topological part of electric-magnetic interactions\footnote{To be strictly topological the surface must be closed, which can be achieved by splitting the Dirac string into two strings that go off in opposite directions, but the extra effort hardly seems worth it.}, can be written as
\begin{align}
&-\frac{eqbg}{2\pi^2}\int d\sigma_{\alpha\beta}^J\frac12\varepsilon^{\alpha\beta\mu\nu}d\sigma_{\gamma\delta}^K\frac{1}{2n\cdot\partial}\left(\delta_\nu^\gamma \partial_\mu n^\delta-\delta_\mu^\gamma \partial_\nu n^\delta-\delta_\nu^\delta \partial_\mu n^\gamma+\delta_\mu^\delta \partial_\nu n^\gamma\right)\partial^2\frac{1}{(z_J-z_K)^2}\nonumber\\
=&eqbg\int d\sigma_{\alpha\beta}^J\frac12\varepsilon^{\alpha\beta\mu\nu}d\sigma_{\gamma\delta}^K\frac12\varepsilon^{\gamma\delta\sigma\rho}d^4ke^{-ik\cdot(z_J-z_K)}(2i){}_{\mu\nu}^{(k)}P_{\ast\sigma\rho}^{(n)}~,
\end{align}
where we have used the fact that
\beq
\partial^2\frac{1}{(z_J-z_K)^2}=-i4\pi^2\delta^{(4)}(z_J-z_K)~,
\eeq
and the notation
\beq
{}^{\alpha\beta}_{(n)}P^{\gamma\delta}_{(k)}=\frac{1}{2n\cdot k}\left(\eta^{\alpha\gamma}n^\beta k^\delta-\eta^{\alpha\delta}n^\beta k^\gamma-\eta^{\beta\gamma}n^\alpha k^\delta+\eta^{\beta\delta}n^\alpha k^\gamma \right)~.
\eeq
This shows that, remarkably, the topological term, which does not contribute to the propagation of physical degrees of freedom, can also be written as a mapping from two-forms to two-forms. Therefore, using such objects allows one to concisely navigate quantum field theories with electric and magnetic interactions. For instance the full mixed electric-magnetic charge interaction can be written as
\begin{align}
&-2ieqbg\int d\sigma_{\alpha\beta}^J\frac12\varepsilon^{\alpha\beta\mu\nu}d\sigma_{\gamma\delta}^K\frac12\varepsilon^{\gamma\delta\sigma\rho}d^4ke^{-ik\cdot(z_J-z_K)}\left({}_{\mu\nu}^{(k)}P_{\ast\sigma\rho}^{(k)}-{}_{\mu\nu}^{(k)}P_{\ast\sigma\rho}^{(n)} \right)\nonumber\\
=&-2ieqbg\int d\sigma_{\alpha\beta}^J\frac12\varepsilon^{\alpha\beta\mu\nu}d\sigma_{\gamma\delta}^K\frac12\varepsilon^{\gamma\delta\sigma\rho}d^4ke^{-ik\cdot(z_J-z_K)}{}_{\mu\nu}^{(k)}P_{\lambda\kappa}^{(k)}{}^{\ast\lambda\kappa}_{(n)}P_{\sigma\rho}^{(k)}~.\label{e.mixedProp}
\end{align}
This last version of the interaction does not naively appear to easily separate into topological and dynamical terms, which can make separating out the topological contribution difficult. However, contributions exactly like this term appear repeatedly in the potential formulation of loop effects in the quantum field theory of electric-magnetic interactions. It is by identifying and dealing with these terms in one-loop processes that we are able to determine the correct running of electric and magnetic couplings: the running predicted by Coleman.

\section{Mappings on the Space of Two-Forms\label{s.TwoForms}}
In what follows we typically consider interactions within the six dimensional space of spacetime two-forms. This is essential for readers that want to work through the algebra in the subsequent section. However, those that are satisfied to accept those calculations as correct may skip this section without losing the logical progression of the paper.

The space of two-forms has an identity element and metric given by
\beq
{}_{\;\alpha\beta}I_{\gamma\delta}=\frac12\left(\eta_{\alpha\gamma}\eta_{\beta\delta}-\eta_{\alpha\delta}\eta_{\beta\gamma} \right)~,
\eeq
which can be used to raise or lower the pairs of indices belonging to a two-form. There is also the duality operator on any two-form $F_{\alpha\beta}$
\beq
\frac12\varepsilon^{\alpha\beta\gamma\delta}F_{\alpha\beta}\equiv {}^\ast\! F^{\gamma\delta}~.
\eeq
When the free indices are contracted which two-form is associated with the duality operator is a matter of choice, that is
\beq
{}^\ast\!A_{\mu\nu}B^{\mu\nu}=A_{\mu\nu}{}^\ast\!B^{\mu\nu}~.
\eeq

We note that the duality operator is related to the identity by
\beq
\frac12\varepsilon^{\alpha\beta\mu\nu}\frac12\varepsilon_{\mu\nu\gamma\delta}=-{}^{\;\alpha\beta}I_{\gamma\delta}~,\label{e.eeIdent}
\eeq
as we saw in Eq.~(\ref{e.doubleLC}).
This means that when any dualized indices are contracted with other dualized indices we find
\beq
{}^\ast\!A_{\mu\nu}{}^\ast\!B^{\mu\nu}=-A_{\mu\nu}B^{\mu\nu}~.
\eeq
Any product of two dualizing operators can be written in terms of the identity operator according to
\begin{align}
&-\frac14\varepsilon_{\alpha\beta\gamma\delta}\varepsilon_{\mu\nu\sigma\rho}=\\
&{}_{\;\alpha\beta}I_{\mu\nu}\,{}_{\;\gamma\delta}I_{\sigma\rho} +{}_{\;\gamma\delta}I_{\mu\nu}\,{}_{\;\alpha\beta}I_{\sigma\rho}-{}_{\;\alpha\gamma}I_{\mu\nu}\,{}_{\;\beta\delta}I_{\sigma\rho} -{}_{\;\beta\delta}I_{\mu\nu}\,{}_{\;\alpha\gamma}I_{\sigma\rho}+{}_{\;\alpha\delta}I_{\mu\nu}\,{}_{\;\beta\gamma}I_{\sigma\rho} +{}_{\;\beta\gamma}I_{\mu\nu}\,{}_{\;\alpha\delta}I_{\sigma\rho} ~,\nonumber
\end{align}
which also leads to all the usual contracted identities, such as 
\beq
\varepsilon^{\gamma\delta\nu\rho}\varepsilon_{\alpha\beta\mu\rho}=-\delta^\gamma_\alpha\left(\delta^\delta_\beta\delta^\nu_\mu-\delta^\delta_\mu\delta^\nu_\beta \right)-\delta^\gamma_\beta\left(\delta^\delta_\mu\delta^\nu_\alpha-\delta^\delta_\alpha\delta^\nu_\mu \right)-\delta^\gamma_\mu\left(\delta^\delta_\alpha\delta^\nu_\beta-\delta^\delta_\beta\delta^\nu_\alpha \right)~.\label{e.epsilonIdentity}
\eeq

As discussed in Sec.~\ref{s.Review}, we are particularly interested in a class of operators that map two-forms to two-forms, based on two one-forms $q_\mu$ and $r_\mu$. This class is defined by
\beq
{}^{\;\alpha\beta}_{(q)}P_{(r)}^{\gamma\delta}\equiv\frac{1}{2q\cdot r}\left(\eta^{\alpha\gamma}q^\beta r^\delta-\eta^{\alpha\delta}q^\beta r^\gamma-\eta^{\beta\gamma}q^\alpha r^\delta+\eta^{\beta\delta}q^\alpha r^\gamma \right)={}^{\;\gamma\delta}_{(r)}P_{(q)}^{\alpha\beta}~.
\eeq
Notice that this operator is antisymmetric in both the first two and last two indices as required for a mapping from two-forms to two-forms.

Consider the action of this operator on the two-form $s\wedge t=\frac12(s_\mu t_\nu-s_\nu t_\mu)$
\begin{align}
{}^{\;\alpha\beta}_{(q)}P_{(r)}^{\gamma\delta}\frac12\left(s_\gamma t_\delta-s_\delta t_\gamma\right)=&\frac{r\cdot t}{2r\cdot q}\left(s^\alpha q^\beta-s^\beta q^\alpha \right)+\frac{r\cdot s}{2r\cdot q}\left(q^\alpha t^\beta-q^\beta t^\alpha \right)~,\\
\frac12\left(s_\alpha t_\beta-s_\beta t_\alpha\right){}^{\;\alpha\beta}_{(q)}P_{(r)}^{\gamma\delta}=&\frac{q\cdot t}{2r\cdot q}\left(s^\gamma r^\delta-s^\delta r^\gamma \right)+\frac{q\cdot s}{2r\cdot q}\left(r^\gamma t^\delta-r^\delta t^\gamma \right)~.
\end{align}
The first line projects into the space of two-forms created by taking the wedge product of a vector with $q^\mu$ while the second projects into wedge products with $r^\mu$. The coefficients of each term are related to the scalar products of $r^\mu$ and $q^\mu$ for the first and second lines, respectively. 

The composition of two of these operators is
\beq
{}^{\;\alpha\beta}_{(q)}P_{(r)}^{\mu\nu}{}_{\;\mu\nu}^{(s)}P^{(t)}_{\gamma\delta}=\frac{(q\cdot t)(r\cdot s)}{(q\cdot r)(s\cdot t)}{}^{\;\alpha\beta}_{(q)}P^{(t)}_{\gamma\delta}+\frac{1}{2(q\cdot r)(s\cdot t)}\left(q^\alpha s^\beta-q^\beta s^\alpha \right)\left(r_\gamma t_\delta-r_\delta t_\gamma \right)~.
\eeq
A few useful special cases are
\beq
{}^{\;\alpha\beta}_{(q)}P_{(r)}^{\mu\nu}{}_{\;\mu\nu}^{(r)}P^{(q)}_{\gamma\delta}=\frac{q^2 r^2}{(q\cdot r)^2}{}^{\;\alpha\beta}_{(q)}P^{(q)}_{\gamma\delta}-\frac{1}{2(q\cdot r)^2}\left(q^\alpha r^\beta-q^\beta r^\alpha \right)\left(q_\gamma r_\delta-q_\delta r_\gamma \right)~,
\eeq
as well as
\beq
{}^{\;\alpha\beta}_{(q)}P_{(q)}^{\mu\nu}{}_{\;\mu\nu}^{(r)}P^{(q)}_{\gamma\delta}={}^{\;\alpha\beta}_{(q)}P_{(r)}^{\mu\nu}{}_{\;\mu\nu}^{(q)}P^{(q)}_{\gamma\delta}={}^{\;\alpha\beta}_{(q)}P^{(q)}_{\gamma\delta}~,
\eeq
and
\beq
{}^{\;\alpha\beta}_{(q)}P_{(r)}^{\mu\nu}{}_{\;\mu\nu}^{(q)}P^{(r)}_{\gamma\delta}={}^{\;\alpha\beta}_{(q)}P^{(r)}_{\gamma\delta}~.
\eeq

These operators can also be composed with the duality transformation
\beq
\frac12\varepsilon^{\alpha\beta\mu\nu}{}_{\;\mu\nu}^{(q)}P^{(r)}_{\gamma\delta}\equiv{}^{\;\ast\alpha\beta}_{(q)}P^{(r)}_{\gamma\delta}~, \ \ \ \ {}_{\;\alpha\beta}^{(q)}P^{(r)}_{\mu\nu}\frac12\varepsilon^{\mu\nu\gamma\delta}\equiv{}_{\;\alpha\beta}^{(q)}P_{(r)}^{\ast\gamma\delta}~.
\eeq
Notice that the location of the $\ast$ denotes which two indices are contracted with the epsilon tensor. This transformation can also be applied to both sides at once, leading to the identity
\beq
\frac{1}{2q\cdot r}\varepsilon^{\alpha\beta\mu}_{\phantom{\alpha\beta\mu}\rho}\varepsilon^{\gamma\delta\nu\rho}q_\mu r_\nu={}_{(q)}^{\ast\alpha\beta}P_{(r)}^{\ast\gamma\delta}~.
\eeq
One can then explicitly calculate that
\begin{align}
{}_{\;\alpha\beta}^{(q)}P^{(r)}_{\gamma\delta}-{}_{\;\ast\alpha\beta}^{(r)}P^{(q)}_{\ast\gamma\delta}=&{}_{\;\alpha\beta}I_{\gamma\delta}~,\label{e.complete}\\
{}_{\;\ast\alpha\beta}^{(q)}P^{(r)}_{\gamma\delta}+{}_{\;\alpha\beta}^{(r)}P^{(q)}_{\ast\gamma\delta}=&\frac12\varepsilon_{\alpha\beta\gamma\delta}~.\label{e.intDef}
\end{align}

This last identity is related to intersection numbers and linking numbers. If we integrate the Fourier transform of Eq.~\eqref{e.intDef} over the surfaces bounded by the $J_e$ and $K_e$ current worldlines we find
\beq
\int d\sigma^J_{\alpha\beta} d\sigma^K_{\gamma\delta}\,d^4ke^{-ik\cdot(z_J-z_K)}\left({}^{\;\ast\alpha\beta}_{(n)}P_{(k)}^{\gamma\delta}+{}^{\;\alpha\beta}_{(k)}P_{(n)}^{\ast\gamma\delta}\right)=\int d\sigma^J_{\alpha\beta} d\sigma^K_{\gamma\delta}\,d^4ke^{-ik\cdot(z_J-z_K)}\frac12\varepsilon^{\alpha\beta\gamma\delta}~.\label{e.LinkIdent}
\eeq
The quantity on the right of the equals sign is the intersection number, it counts the (signed) intersections of the surfaces bounded by $J_e$ and $K_e$, which depends on the choice of surfaces. 

The two terms on the left-hand side of Eq.~\eqref{e.LinkIdent} also correspond to intersection numbers. The first is the intersections of the $K_e$ surface with the surface produced by extending the $J_e$ trajectory along the direction $n^\mu$, which is related to the linking number between the $K_e$ trajectory and the surface of $J_e$ extended along $n^\mu$. If the surface were to close by extending along $n^\mu$ and $-n^\mu$ it would be the topological linking number of the two surfaces. What appears in the integral is not quite topological because the surface does not extend in both directions. Just as a closed path and an open path can always be unlinked by continuous deformations in three dimensions, so can a closed path and an open surface in four dimensions. This means the intersection number is geometrical, its value depends on the integration surfaces. 

The second term on the left side of Eq.~\eqref{e.LinkIdent} is the intersection number between the $J_e$ surface and the extension of $K_e$ along $n^\mu$, which is similarly related to the linking number of $J_e$ trajectory and the surface of $K_e$ extended along $n^\mu$ and $-n^\mu$. When $K_e$ is only extended along positive $n^\mu$ the intersection only has a geometrical interpretation. The simple algebraic result in Eq.~\eqref{e.intDef} shows how these three intersection numbers are related.

The above identities can also be used to determine that
\beq
{}^{\;\alpha\beta}_{(q)}P_{(r)}^{\mu\nu}{}_{\;\ast\mu\nu}^{(s)}P^{(t)}_{\gamma\delta}={}^{\;\alpha\beta}_{(q)}P_{(r)}^{\ast\gamma\delta}-\frac{(q\cdot s)(r\cdot t)}{(q\cdot r)(s\cdot t)}{}^{\;\alpha\beta}_{(q)}P_{(s)}^{\ast\gamma\delta}+\frac{\varepsilon_{\mu\nu\gamma\delta}r^\mu s^\nu}{2(q\cdot r)(s\cdot t)}\left(q^\alpha t^\beta-q^\beta t^\alpha\right)~.
\eeq
An important special case that we make great use of when considering the interactions of electric and magnetic currents is
\begin{align}
{}^{\;\alpha\beta}_{(q)}P_{(q)}^{\mu\nu}{}_{\;\ast\mu\nu}^{(r)}P_{(q)}^{\gamma\delta}=&{}^{\;\alpha\beta}_{(q)}P_{(q)}^{\ast\gamma\delta}-{}^{\;\alpha\beta}_{(q)}P_{(r)}^{\ast\gamma\delta}~\nonumber\\
=&-{}^{\ast\alpha\beta}_{(q)}P_{(q)}^{\gamma\delta}+{}^{\ast\alpha\beta}_{(r)}P_{(q)}^{\gamma\delta}~.\label{e.MixedIdent}
\end{align}
It is these two equal forms that encode the topological contributions to electric-magnetic interactions when we take $q^\mu$ to be the photon momentum, $q^\mu=k^\mu$, and $r^\mu$  to be the reference vector $r^\mu=n^\mu$. 

Our understanding of these operators is enhanced by considering ${}_{\;\alpha\beta}^{(k)}P^{(k)}_{\gamma\delta}$. The results above make clear that
\beq
{}_{\;\alpha\beta}^{(k)}P_{(k)}^{\mu\nu}{}_{\;\mu\nu}^{(k)}P^{(k)}_{\gamma\delta}={}_{\;\alpha\beta}^{(k)}P^{(k)}_{\gamma\delta}~, \ \ \ \  {}_{\;\alpha\beta}^{(k)}P_{(k)}^{\mu\nu}{}_{\;\ast\mu\nu}^{(k)}P^{(k)}_{\gamma\delta}=0~ \ \ \ \ {}_{\;\mu\nu}^{(k)}P_{(k)}^{\mu\nu}=3~.
\eeq
This indicates that ${}_{\;\alpha\beta}^{(k)}P^{(k)}_{\gamma\delta}$ acts as a projector into a three dimensional subspace of two-forms. What subspace is this? Notice that 
\beq
{}_{\;\alpha\beta}^{(k)}P_{(k)}^{\mu\nu}\frac12\left(k_\mu q_\nu-k_\nu q_\mu \right)=\frac12\left(k_\alpha q_\beta-k_\beta q_\alpha \right)~.
\eeq
This shows that the projector acts as the identity on $k_\mu$ wedged with any one-form. If we imagine spacetime spanned by a basis composed of $k_\mu$ and three other vectors then the three dimensional two-form space spanned by the wedge product of $k_\mu$ and each of the other vectors is exactly the space ${}_{\;\alpha\beta}^{(k)}P_{(k)}^{\mu\nu}$ projects into. The orthogonal space is projected into by\footnote{Technically the orthogonal complement projector is $-{}_{\;\ast\alpha\beta}^{(k)}P^{(k)}_{\ast\gamma\delta}$ as is clear from the completeness relation in Eq.~\eqref{e.complete}.} ${}_{\;\ast\alpha\beta}^{(k)}P^{(k)}_{\ast\gamma\delta}$.

\section{Field Strength Based Perturbation Series using a Single Potential \label{s.onePot}}
In this section we show how the the usual QED perturbation series, using a single gauge potential $A_\mu$, can be rewritten in terms of the field strength $F_{\mu\nu}=\partial_\mu A_\nu-\partial_\nu A_\mu$. This is shown for both tree-level and one loop calculations. The higher loop generalization is straightforward. We find that a projector of the type outlined in the previous section naturally occurs in defining the field-strength propagator. We then show how the theory must be modified to consistently incorporate magnetic currents.

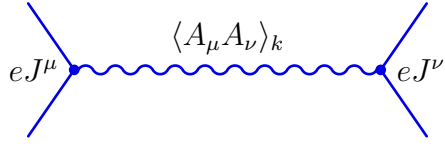
\begin{figure}
\center
\begin{fmffile}{treePotJJ}
\begin{fmfgraph*}(150,50)
\fmfpen{1.0}
\fmfstraight
\fmfset{arrow_len}{3mm}
\fmfleft{i1,i2} \fmfright{o1,o2}
\fmf{plain,tension=1.0,fore=(0,,0,,0.9)}{i1,v1,i2}
\fmf{photon,tension=0.3,label=$\langle A_\mu A_\nu\rangle_k$,label.side=left,fore=(0,,0,,0.9)}{v1,v2}
\fmf{plain,tension=1.0,fore=(0,,0,,0.9)}{o1,v2,o2}
\fmfv{decor.shape=circle,decor.filled=full,decor.size=1.5thick,l=$eJ^\mu$,fore=(0,,0,,0.9)}{v1} 
\fmfv{decor.shape=circle,decor.filled=full,decor.size=1.5thick,l=$eJ^\nu$,fore=(0,,0,,0.9)}{v2} 
\end{fmfgraph*}
\end{fmffile}
\caption{\label{f.treeJJpot}Standard one-photon exchange between electric currents. The blue current denotes that it is electric. The blue photon line denotes a coupling to electric charges without using the Levi-Civita tensor.}
\end{figure}

We begin with the usual QED Lagrangian for electric currents
\beq
\mathcal{L}=-\frac14 F_{\mu\nu}F^{\mu\nu}-eA_\mu J^\mu~.\label{e.MaxLag}
\eeq
This, along with a gauge fixing term, leads to the standard propagator
\beq
\int d^4x\, e^{-ik\cdot(x-y)}\langle\bm{0}|T\left\{ A_\mu(x)A_\nu(y)\right\}|\bm{0}\rangle=-i\frac{\eta_{\mu\nu}-(1-\xi)\frac{k_\mu k_\nu}{k^2}}{k^2-i\varepsilon}\equiv\langle A_\mu A_\nu\rangle_k~.
\eeq
The last definition is simply a way to quickly describe the momentum space propagator. The leading order correlation of two currents, see Fig.~\ref{f.treeJJpot}, is the gauge invariant amplitude
\beq
i\mathcal{M}^{JJ}_0=-ieJ_e^\mu(-i)\frac{\eta_{\mu\nu}-(1-\xi)\frac{k_\mu k_\nu}{k^2}}{k^2-i\varepsilon}(-ie)J_e^\nu=ie^2\frac{J_e^\mu J_{e\mu}}{k^2-i\varepsilon}~,
\eeq
where we have used the fact that $k_\mu J_e^\mu=0$. In what follows we typically do not write the $i\varepsilon$ terms for simplicity.

The leading correction to this correlation function due to a virtual current loop from a dynamical, electrically charged particle is
\begin{align}
i\mathcal{M}^{JJ}_{1_J}=&-ieJ_e^\mu(-i)\frac{\eta_{\mu\alpha}-(1-\xi)\frac{k_\mu k_\alpha}{k^2}}{k^2}i\Pi_J^{\alpha\beta}(-i)\frac{\eta_{\beta\nu}-(1-\xi)\frac{k_\beta k_\nu}{k^2}}{k^2}(-ie)J_e^\nu\nonumber\\
=&ie^2J_{e\alpha}\frac{k^2\eta^{\alpha\beta}-k^\alpha k^\beta }{k^4}\Pi_J J_{e\beta}\nonumber\\
=&ie^2\Pi_J\frac{J_e^\mu J_{e\mu}}{k^2}~.
\end{align}
Here we have used the Ward identity to write
\beq
\Pi_J^{\mu\nu}=\int d^4x\, e^{-ik\cdot(x-y)}\langle\bm{0}|T\left\{ J_d^\mu(x)J_d^\nu(y)\right\}|\bm{0}\rangle=\Pi_J(k^2)\left(k^2\eta^{\mu\nu}-k^\mu k^\nu \right)~.\label{e.WardForm}
\eeq
Combining the leading term and first correction we have
\beq
\mathcal{M}^{JJ}_0+\mathcal{M}^{JJ}_{1_J}=e^2\frac{J_e^\mu J_{e\mu}}{k^2}\left[1+\Pi_J \right]\approx e^2\frac{J_e^\mu J_{e\mu}}{k^2[1-\Pi_J]}~.
\eeq
As expected, we see that the loop correction preserves the form of the interaction and only modifies the coupling. That is, we find that
\beq
Z_e(\mu)=\frac{1}{1-\Pi_J(\mu^2)}~.
\eeq

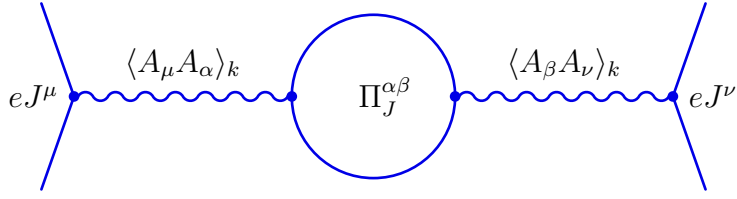
\begin{figure}
\center
\begin{fmffile}{loopPotJJ}
\begin{fmfgraph*}(250,70)
\fmfpen{1.0}
\fmfstraight
\fmfset{arrow_len}{3mm}
\fmfleft{i1,i2} \fmfright{o1,o2}
\fmf{plain,tension=1.0,fore=(0,,0,,0.9)}{i1,v1,i2}
\fmf{photon,tension=0.3,label=$\langle A_\mu A_\alpha\rangle_k$,label.side=left,fore=(0,,0,,0.9)}{v1,v2}
\fmf{plain,right=1,tension=0.2,fore=(0,,0,,0.9)}{v2,v3,v2}
\fmf{photon,tension=0.3,label=$\langle A_\beta A_\nu\rangle_k$,label.side=left,fore=(0,,0,,0.9)}{v3,v4}
\fmf{plain,tension=1.0,fore=(0,,0,,0.9)}{o1,v4,o2}
\fmfv{decor.shape=circle,decor.filled=full,decor.size=1.5thick,l=$eJ^\mu$,fore=(0,,0,,0.9)}{v1} 
\fmfv{decor.shape=circle,decor.filled=full,decor.size=1.5thick,l=$\Pi_J^{\alpha\beta}$,l.d=25,l.a=0,fore=(0,,0,,0.9)}{v2} 
\fmfv{decor.shape=circle,decor.filled=full,decor.size=1.5thick,fore=(0,,0,,0.9)}{v3} 
\fmfv{decor.shape=circle,decor.filled=full,decor.size=1.5thick,l=$eJ^\nu$,fore=(0,,0,,0.9)}{v4} 
\end{fmfgraph*}
\end{fmffile}
\caption{\label{f.loopJJpot}One electric loop correction to photon exchange between electric currents.}
\end{figure}

Using this photon field formalism one can also determine the propagator of two field strengths. A heuristic derivation is
\begin{align}
\langle F^{\alpha\beta}F^{\gamma\delta}\rangle_k=&k^\alpha k^\gamma\langle A^\beta A^\delta\rangle_k-k^\alpha k^\delta\langle A^\beta A^\gamma\rangle_k-k^\beta k^\gamma\langle A^\alpha A^\delta\rangle_k+k^\beta k^\delta\langle A^\alpha A^\gamma\rangle_k \nonumber\\
=&-2i\,{}^{\;\alpha\beta}_{(k)}P^{\gamma\delta}_{(k)}~,
\end{align}
while two different, careful derivations of the same result are given in Appendix~\ref{a.Fprop}. Note that this propagator is a projector, a mapping of two-forms to two-forms. The space it projects into is the space of two-forms of the form $V\wedge k$ for any one-form $V_\mu$ and $k_\mu$ the propagating momentum. We can understand why the field strength propagator might have this form because the field strength, in momentum space, is simply
\beq
F_{\mu\nu}=i\left(k_\mu A_\nu-k_\nu A_\mu \right)=2i k\wedge A~.
\eeq

As $k_\mu$ does not belonging to the ${}^{(k)}_{\ast\mu\nu}P^{(k)}_{\ast\sigma\rho}$ subspace, we see that
\begin{equation}
F_{\mu\nu}={}_{\;\mu\nu}I_{\alpha\beta}F^{\alpha\beta}=\left({}_{\;\mu\nu}^{(k)}P^{(k)}_{\alpha\beta}-{}_{\;\ast\mu\nu}^{(k)}P^{(k)}_{\ast\alpha\beta}\right)F^{\alpha\beta}={}_{\;\mu\nu}^{(k)}P^{(k)}_{\alpha\beta}F^{\alpha\beta}~.
\end{equation}
In other words, the field strength is completely contained within the ${}_{\;\mu\nu}^{(k)}P^{(k)}_{\alpha\beta}$ subspace. This also means that the sourceless Lagrangian can be written as
\beq
\mathcal{L}=-\frac14F_{\mu\nu}F^{\mu\nu}=-\frac14F_{\alpha\beta}{}^{\;\alpha\beta}I^{\gamma\delta}F_{\gamma\delta}=-\frac14F_{\alpha\beta}{}_{(k)}^{\;\alpha\beta}P_{(k)}^{\gamma\delta}F_{\gamma\delta}~.
\eeq
Within the subspace ${}_{(k)}^{\;\alpha\beta}P_{(k)}^{\gamma\delta}$ projects onto it acts as the metric and is, therefore, its own inverse. Therefore, we have some similarity to the usual result that a propagator is given by the inverse of the operator appearing in the quadratic part of the Lagrangian. 

Of course, to go through the usual path integral calculation of the two-point function we need a source for $F_{\mu\nu}$. Seeing as $J_\mu$ is a one-form and $\frac12\varepsilon_{\mu\nu\alpha\beta}J^\alpha$ is a three-form neither can couple directly to $F_{\mu\nu}$. However, the usual Maxwell equation in momentum space is
\beq
ik_\nu F^{\mu\nu}=eJ^\mu~.
\eeq
One, particular, solution of this is
\beq
F^{\mu\nu}_J=-ie\frac{J^\mu n^\nu-J^\nu n^\mu}{n\cdot k}~,
\eeq
We can then motivate that within the Lagrangian
\beq
-A_\mu J^\mu\to-A_\mu\partial_\nu F_J^{\mu\nu}\to \partial_\nu (A_\mu)F_J^{\mu\nu}=\frac{e}{2}F_{\nu\mu}(-i)\frac{J^\mu n^\nu-J^\nu n^\mu}{n\cdot k}=ieF_{\mu\nu}\frac{J^\mu n^\nu-J^\nu n^\mu}{2n\cdot k}~,\label{e.twoFormCoupling}
\eeq
where the second arrow denotes integration by parts. 

With this source term we can define a Feynman rule for how the field strength is generated by electric currents, though it comes at the cost of introducing a reference vector $n^\mu$. As shown in Appendix~\ref{a.Fsource}, we interpret these vectors as defining the surface bounded by the worldlines associated with the currents $J^\mu$. Though, in principle, one can imagine different surfaces defined for each current we can also choose to use the same vector to define them all. 

\begin{figure}
\center
\includegraphics[width=0.6\textwidth]{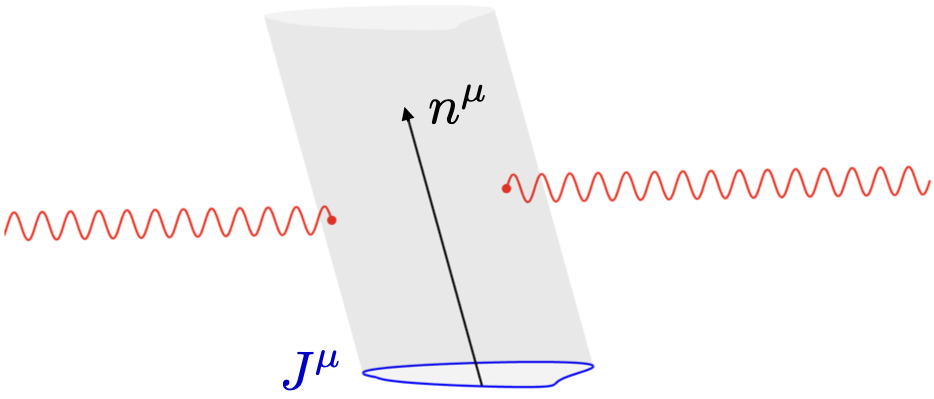}
\caption{Schematic illustration of the surfaces the field strength couples to. The infinite surface is bounded by the worldline associated with the current $J^\mu$ and lies along the direction $n^\mu$. The surface formally closes at the point at infinity.\label{f.surface}}
\end{figure}

These surfaces are of a special type, as shown in Fig.~\ref{f.surface}. The surface erupts from the particle trajectory, which is its boundary, in the direction $n^\mu$ and extends out to infinity. Thus, the surface only closes at infinity. Also, as we take \emph{the} point at infinity to truly be a completion of Minkowski space, every surface of this type, even those associated to different $n^\mu$ vectors, intersect at infinity.

That the field strength is sourced by an area is expected. But Stokes' theorem holds for any surface that has the same boundary. It is true, as shown in Appendix~\ref{a.Fsource}, that the theory is independent of the surface chosen. This turns out to simply follow from the field strength being a closed two-form, or in other words it satisfies the Bianchi identity $\partial_\mu{}^\ast\!F^{\mu\nu}=0$.\footnote{The Appendix also shows that QEMD theories are surface independent even through the field strength does not satisfy the Bianchi identity in this case.} However, finite surfaces in position space do not lead to simple descriptions in momentum space. The family of surfaces we use are somewhat singular in position space, but are simple to express in momentum space.

We can also obtain these results from a different starting point. Consider the Lagrangian
\beq
\mathcal{L}=-\frac14F_{\mu\nu}F^{\mu\nu}-F_{\mu\nu}\Sigma_J^{\mu\nu}~,
\eeq
where $\Sigma_J^{\mu\nu}$ is the as yet unspecified source of $F_{\mu\nu}$. As any symmetric part of this source disappears from the Lagrangian we simply assume that it is antisymmetric, like $F_{\mu\nu}$. The equations of motion for this theory are
\beq
-\frac12 F^\mu-\Sigma_J^{\mu\nu}=0~\Rightarrow~F^{\mu\nu}=-2\Sigma_J^{\mu\nu}~.
\eeq
This is equivalent to the Maxwell equations if
\beq
-2\partial_\mu\Sigma_J^{\mu\nu}=J^\nu~,
\eeq
which leads directly to the same form obtained above. 

\begin{figure}
\center
\begin{fmffile}{treeFJJ}
\begin{fmfgraph*}(150,50)
\fmfpen{1.0}
\fmfstraight
\fmfset{arrow_len}{3mm}
\fmfleft{p1,p2,p3,p4,p5} \fmfright{p6,p7,p8,p9,p10}
\fmf{phantom,tension=1.5}{p2,v1}
\fmfv{decor.shape=circle,decor.filled=0,decor.size=75,fore=(0,,0,,0.9),background=(0.675,,0.675,,0.675)}{p3}
\fmfv{decor.shape=circle,decor.filled=0,decor.size=75,fore=(0,,0,,0.9),background=(0.675,,0.675,,0.675)}{p8}
\fmf{photon,rubout=0.5,tension=0.2,label=$\langle F_{\mu\nu} F_{\sigma\rho}\rangle_k$,label.side=left,fore=(0,,0,,0.9)}{v1,v2}
\fmf{phantom,tension=1}{v2,p9}
\fmfv{decor.shape=circle,decor.filled=full,decor.size=1.5thick,l=$e\Sigma_J^{\mu\nu}$,fore=(0,,0,,0.9)}{v1} 
\fmfv{decor.shape=circle,decor.filled=full,decor.size=1.5thick,l=$e\Sigma_J^{\sigma\rho}$,fore=(0,,0,,0.9)}{v2} 
\end{fmfgraph*}
\end{fmffile}
\caption{\label{f.treeJJF}One-photon exchange between electric currents (blue) using field strengths. Note that the photon lines couple to the surfaces bounded by the worldlines associated with the currents.}
\end{figure}
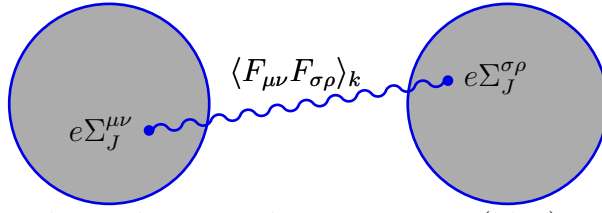

With this Lagrangian
\beq
\mathcal{L}_\text{QED}=-\frac14 F_{\mu\nu}F^{\mu\nu}-eF_{\mu\nu}\frac{1}{2n\cdot\partial}\left( J^\mu n^\nu-J^\nu n^\mu\right)~,
\eeq
we can use the standard Feynman rules to calculate amplitudes. This still uses the photon degrees of freedom associated with $A_\mu$; we are simply organizing the calculation in a slightly different way. First, consider the scattering of two electric currents by the exchange of a single photon, as in Fig.~\ref{f.treeJJF}. The amplitude is calculated to be
\begin{equation}
i\mathcal{M}^{JJ}_0=ie\frac{J_e^\alpha n_1^\beta-J_e^\beta n_1^\alpha}{2n_1\cdot k}(-2i){}^{(k)}_{\;\alpha\beta}P^{(k)}_{\gamma\delta}~ie\frac{J_e^\gamma n_2^\delta-J_e^\delta n_2^\gamma}{2n_2\cdot k}=ie^2\frac{J_e^2}{k^2}~,
\end{equation}
which agrees exactly with the usual calculations made with the potential coupling and propagator. This result is also completely independent of the (in principle) distinct reference vectors $n_{1,2}^\mu$ used to define the two Stokes surfaces. In what follows we typically take all surfaces to lie along the same $n^\mu$ for simplicity.

Turning to renormalization, we include a closed loop due to the electrically charged particle within the exchanged photon line. To calculate this quantity we must determine the form of the four index loop function. However, the closed loop still has the same functional form no matter how we describe the photon coupling, it is simply determined by the current. Therefore, we can use this known form, given in Eq.~\eqref{e.WardForm}, to find
\begin{align}
&{}^{\mu\nu}\Pi^{\sigma\rho}\equiv\frac{n^\mu\eta^{\nu\alpha}-n^\nu\eta^{\mu\alpha}}{2n\cdot k}\frac{n^\sigma\eta^{\rho\beta}-n^\rho\eta^{\sigma\beta}}{2n\cdot k}\Pi(k^2)\left(k^2\eta_{\alpha\beta}-k_\alpha k_\beta \right)\nonumber\\
&=\frac{\Pi(k^2)}{4(n\cdot k)^2}\left[k^2\left(n^\mu n^\sigma\eta^{\nu\rho}-n^\mu n^\rho\eta^{\nu\sigma}-n^\nu n^\sigma\eta^{\mu\rho}+n^\nu n^\rho\eta^{\mu\sigma} \right)-\left(n^\mu k^\nu-n^\nu k^\mu \right) \left(n^\sigma k^\rho-n^\rho k^\sigma \right) \right]\nonumber\\
&=\frac12\frac{\Pi(k^2)}{(n\cdot k)^2}\left[k^2n^2{}^{\;\mu\nu}_{(n)}P_{(n)}^{\sigma\rho}-\frac12\left(n^\mu k^\nu-n^\nu k^\mu \right) \left(n^\sigma k^\rho-n^\rho k^\sigma \right)\right]\nonumber\\
&=\frac12\Pi(k^2)~{}^{\;\mu\nu}_{(n)}P_{(k)}^{\alpha\beta}\,{}_{\alpha\beta}I_{\gamma\delta}\, {}^{\;\gamma\delta}_{(k)}P_{(n)}^{\sigma\rho}~.\label{e.LoopFunc}
\end{align}

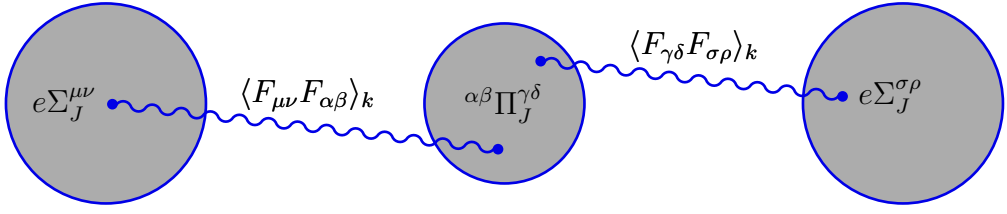
\begin{figure}
\center
\begin{fmffile}{loopFJJ}
\begin{fmfgraph*}(300,70)
\fmfpen{1.0}
\fmfstraight
\fmfset{arrow_len}{3mm}
\fmfleft{p1,p2,p3,p4,p5} \fmfright{p6,p7,p8,p9,p10}
\fmf{phantom,tension=1}{p2,p11,p7}
\fmf{phantom,tension=1}{p4,p12,p9}
\fmf{phantom,tension=1}{p3,v5,p8}
\fmffreeze
\fmf{phantom,tension=3.0}{p3,v3}
\fmf{phantom,tension=1.5}{v4,p8}
\fmf{phantom,tension=3.0}{p11,v1}
\fmf{phantom,tension=2.5}{p12,v2}
\fmfv{decor.shape=circle,decor.filled=0,decor.size=75,fore=(0,,0,,0.9),background=(0.675,,0.675,,0.675)}{p3}
\fmfv{decor.shape=circle,decor.filled=0,decor.size=75,fore=(0,,0,,0.9),background=(0.675,,0.675,,0.675)}{p8}
\fmfv{decor.shape=circle,decor.filled=0,decor.size=60,fore=(0,,0,,0.9),background=(0.675,,0.675,,0.675),l=${}^{\alpha\beta}\Pi_J^{\gamma\delta}$,l.a=0,l.d=-15}{v5}
\fmf{photon,rubout=0.5,tension=0.05,label=$\langle F_{\mu\nu} F_{\alpha\beta}\rangle_k$,label.side=left,fore=(0,,0,,0.9)}{v3,v1}
\fmf{photon,rubout=0.5,tension=0.3,label=$\langle F_{\gamma\delta} F_{\sigma\rho}\rangle_k$,label.side=left,fore=(0,,0,,0.9)}{v2,v4}
\fmfv{decor.shape=circle,decor.filled=full,decor.size=1.5thick,fore=(0,,0,,0.9)}{v1} 
\fmfv{decor.shape=circle,decor.filled=full,decor.size=1.5thick,fore=(0,,0,,0.9)}{v2} 
\fmfv{decor.shape=circle,decor.filled=full,decor.size=1.5thick,l=$e\Sigma_J^{\mu\nu}$,fore=(0,,0,,0.9)}{v3} 
\fmfv{decor.shape=circle,decor.filled=full,decor.size=1.5thick,l=$e\Sigma_J^{\sigma\rho}$,fore=(0,,0,,0.9)}{v4} 
\end{fmfgraph*}
\end{fmffile}
\caption{\label{f.loopJJF}One electric loop correction to photon exchange between electric currents using field strengths.}
\end{figure}

From the last equality, and the results of Sec.~\ref{s.TwoForms}, it is simple to calculate that
\beq
{}^{(k)}_{\;\alpha\beta}P^{(k)}_{\mu\nu}{}^{\mu\nu}\Pi^{\sigma\rho}{}^{(k)}_{\;\sigma\rho}P^{(k)}_{\gamma\delta}=\frac12\Pi(k^2){}^{(k)}_{\;\alpha\beta}P^{(k)}_{\gamma\delta}~,\label{e.PpiP}
\eeq
which is independent of $n^\mu$! The full one-loop correction amplitude, shown schematically in Fig.~\ref{f.loopJJF}, is
\beq
\mathcal{M}_{1_J}^{JJ}=e^2\Pi_J\frac{J_e^2}{k^2}~,
\eeq
in agreement with the standard potential propagator calculations. We emphasize again that this result is $n^\mu$ independent, as expected. It is also unaffected by taking the $n^\mu$s of the external and virtual currents to be different from each other. 

Let us here take stock of what has been accomplished. We have outlined a formalism for calculating in QED that is gauge independent. The definition of the Feynman vertices does depend on specific Stokes surfaces, but the final results are completely independent of what surfaces are chosen. While this may be somewhat diverting when calculating with only electric charges we see in what follows that the true utility comes when both electric and magnetic charges are present.

\subsection{Including Magnetic Charges}
Turning to QEMD, how should we include magnetic currents $K^\mu$? First, let us discuss what types of currents we consider. The simplest addition (in some sense) is that of a so-called Dirac monopole, a point-like magnetic charge. As emphasized in~\cite{Reece:2023czb} the fact that the field configuration for such an object has an infinite classical self-energy is not qualitatively different from electric point particles, this is merely a mass renorrmalization. Often, however, this construct is a proxy for a more complicated object with a finite physical size. In such a scenario there can still be an effective theory in which a massive solitonic monopole (for instance) appears point-like at low-energies as long as the Compton wavelength of the monopole is much larger than the physical size. In what follows we remain agnostic to the structure\textemdash if any\textemdash of the monopole but confine our analysis to those theories in which it looks point-like to the low-energy photon.

One simple, and naive, guess as to how to include the magnetic current is to only include the coupling term
 \beq
 ib{}^\ast\!F_{\mu\nu}\frac{K^\mu n^\nu-K^\nu n^\mu}{2n\cdot k}~,
 \eeq
 in the usual one potential Lagrangian~\eqref{e.MaxLag}. One might guess that $K^\mu$ is sourced by ${}^\ast\!F_{\mu\nu}$ in the same way that $F_{\mu\nu}$ is sourced by $J^\mu$. We then find the propagator for ${}^\ast\!F_{\mu\nu}$ by applying factors of $\frac12\varepsilon_{\alpha\beta\gamma\delta}$ to the known $F_{\mu\nu}$ propagator. (This assumes, mistakenly, that the $F_{\mu\nu}$ propagator is unchanged when magnetic currents are included in the theory.) In short, one would naively expect
\begin{align}
\langle F_{\alpha\beta}F_{\gamma\delta}\rangle_k=&-2i{}^{(k)}_{\;\alpha\beta}P^{(k)}_{\gamma\delta}~,&
\langle F_{\alpha\beta}{}^\ast\!F_{\gamma\delta}\rangle_k=&-2i{}^{(k)}_{\;\alpha\beta}P^{(k)}_{\ast\gamma\delta}~,\\
\langle {}^\ast\!F_{\alpha\beta}F_{\gamma\delta}\rangle_k=&-2i{}^{(k)}_{\ast\alpha\beta}P^{(k)}_{\gamma\delta}~,&
\langle {}^\ast\!F_{\alpha\beta}{}^\ast\!F_{\gamma\delta}\rangle_k=&-2i{}^{(k)}_{\ast\alpha\beta}P^{(k)}_{\ast\gamma\delta}~,
\end{align}
though we see below that these are \emph{not correct}.

To see the problem with these assumptions, consider the one-photon exchange between two magnetic currents. The naive amplitude is
\begin{align}
i\mathcal{M}_0^{KK}=&ib\frac{K_e^\alpha n^\beta-K_e^\beta n^\alpha}{2n\cdot k}(-2i){}^{(k)}_{\ast\alpha\beta}P^{(k)}_{\ast\gamma\delta}~ib\frac{K_e^\gamma n^\delta-K_e^\delta n^\gamma}{2n\cdot k}\nonumber\\
=&2ib^2\frac{K_e^\alpha n^\beta-K_e^\beta n^\alpha}{2n\cdot k}\left({}^{(k)}_{\;\alpha\beta}P^{(k)}_{\gamma\delta}-{}_{\;\alpha\beta}I_{\gamma\delta}\right)~\frac{K_e^\gamma n^\delta-K_e^\delta n^\gamma}{2n\cdot k}\nonumber\\
=&ib^2\frac{K_e^2}{k^2}-ib^2\frac{K_e^2n^2-(K_e\cdot n)^2}{2(n\cdot k)^2}~,
\label{e.extracontact}
\end{align}
where we have used the completeness relation~\eqref{e.complete} in the second line. The additional ``contact'' term, with no photon pole, in the final result contradicts our expectation for magnetic-magnetic scattering from electric-magnetic duality. That is, we expect that magnetic-magnetic scattering should have the same form as electric-electric scattering.

 Remember, however, that we assumed that the field strength propagator was unchanged when we introduce magnetic currents. This assumption must be reexamined. To more carefully include magnetic currents we follow Dirac's~\cite{Dirac:1948um} one potential description as extended in~\cite{Blagojevic:1978zv,Blagojevic:1979bm} and usefully summarized in the review~\cite{Blagojevic:1985sh}. Dirac's formulation of a quantum theory of electric and magnetic particles leads to a Lagrangian ``contact'' term between the magnetic currents:
\beq
ib^2\frac{K^2n^2-(K\cdot n)^2}{2(n\cdot k)^2}~.
\eeq
This additional ``contact'' term must be included in our calculations, cancelling the unexpected term in (\ref{e.extracontact}) and ensuring that the final results agree with electric-magnetic duality.

The essential point is that if we want to include magnetic charges then we cannot assume that the field strength has the form $F_{\mu\nu}=\partial_\mu A_\nu-\partial_\nu A_\mu$ as this leads to $\partial_\mu{}^\ast\!F^{\mu\nu}=0$. Following in the spirit of~\cite{Blagojevic:1978zv,Blagojevic:1979bm} we consider
\beq
F_{\mu\nu}=\partial_\mu A_\nu-\partial_\nu A_\mu-\frac{b}{2}\varepsilon_{\mu\nu\alpha\beta}\frac{n^\alpha K^\beta-n^\beta K^\alpha}{n\cdot \partial}\equiv F^A_{\mu\nu}-\frac{b}{2}\varepsilon_{\mu\nu\alpha\beta}\frac{n^\alpha K^\beta-n^\beta K^\alpha}{n\cdot \partial}~,
\label{e.DiracF}
\eeq
where $n^\mu$ is any constant vector and 
\beq
F^X_{\mu\nu}=\partial_\mu X_\nu-\partial_\nu X_\mu~,
\eeq
for any vector field $X^\mu$. In this definition the operator $(n\cdot \partial)^{-1}$ is understood to act on the magnetic current. This form of the field strength implies that 
\beq
{}^\ast\!F^{\mu\nu}=b\frac{n^\mu K^\nu-n^\nu K^\mu}{n\cdot \partial}+\varepsilon^{\mu\nu\alpha\beta}\partial_\alpha A_\beta~,
\eeq
and hence produces the correct Maxwell equation
\beq
\partial_\mu{}^\ast\!F^{\mu\nu}=bK^\nu~,
\eeq
for the conserved magnetic current, $\partial_\mu K^\mu=0$.

Substituting the form of $F_{\mu\nu}$ given in Eq.~(\ref{e.DiracF}) into the usual Lagrangian we find
\begin{align}
\mathcal{L}=&-\frac14F_{\mu\nu}F^{\mu\nu}-eA_\mu J^\mu~,\\
=&-\frac14 F^A_{\mu\nu}F^{A\mu\nu}-eA_\mu J^\mu+b\partial_\beta(A_\mu)\frac{1}{n\cdot\partial}\varepsilon^{\mu\nu\alpha\beta}K_\nu n_\alpha+\frac{b^2}{2(n\cdot\partial)^2}\left[ n^2K^2-(n\cdot K)^2\right]~,\nonumber
\end{align}
where each of the $(n\cdot\partial)^{-1}$ operators are acting one magnetic current.
In this case the photon field has the usual coupling to electric currents and a nonlocal coupling to magnetic currents that involves the Levi-Civita tensor. In addition, there is a nonlocal ``contact'' interaction between magnetic currents. Despite these new magnetic-magnetic interactions the quadratic part of the $A_\mu$ Lagrangian is unchanged. Consequently, we find the same photon propagator and can reproduce all the standard (correct) results for electric-electric interactions. 

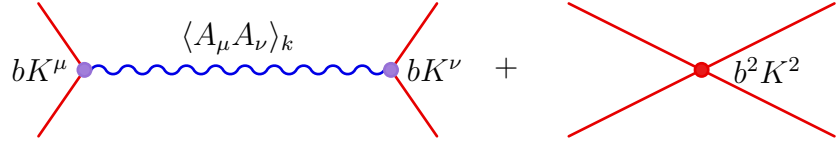
\begin{figure}
\center
\begin{fmffile}{treePotKK}
\begin{fmfgraph*}(150,50)
\fmfpen{1.0}
\fmfstraight
\fmfset{arrow_len}{3mm}
\fmfleft{i1,i2} \fmfright{o1,o2}
\fmf{plain,tension=1.0,fore=(0.9,,0,,0)}{i1,v1,i2}
\fmf{photon,tension=0.3,label=$\langle A_\mu A_\nu\rangle_k$,label.side=left,fore=(0,,0,,0.9)}{v1,v2}
\fmf{plain,tension=1.0,fore=(0.9,,0,,0)}{o1,v2,o2}
\fmfv{decor.shape=circle,decor.filled=90,decor.size=2.5thick,l=$bK^\mu$,fore=(0.59,,0.44,,0.84)}{v1} 
\fmfv{decor.shape=circle,decor.filled=90,decor.size=2.5thick,l=$bK^\nu$,fore=(0.59,,0.44,,0.84)}{v2} 
\end{fmfgraph*}
\end{fmffile}
\raisebox{0.8cm}{\;\;\; $+$ \;}
\begin{fmffile}{contactKK}
\begin{fmfgraph*}(100,50)
\fmfpen{1.0}
\fmfstraight
\fmfset{arrow_len}{3mm}
\fmfleft{i1,i2} \fmfright{o1,o2}
\fmf{plain,tension=1.0,fore=(0.9,,0,,0)}{i1,v1,i2}
\fmf{plain,tension=1.0,fore=(0.9,,0,,0)}{o1,v1,o2}
\fmfv{decor.shape=circle,decor.filled=90,decor.size=2.5thick,l=$b^2K^2$,l.d=12,fore=(0.9,,0,,0)}{v1} 
\end{fmfgraph*}
\end{fmffile}
\caption{\label{f.treeKKpot}One-photon exchange between magnetic currents and the current-current ``contact'' term. The photon line is colored blue to indicate that it has the standard, local coupling to electric currents. The photon coupling to (red) magnetic currents includes a factor of the Levi-Civita tensor, which we denote with a purple coloring.}
\end{figure}

In addition, we can now evaluate the leading contributions to electric-magnetic scattering
\begin{align}
i\mathcal{M}^{JK}_0&=-ieJ_e^\gamma(-i)\frac{\eta_{\gamma\mu}-(1-\xi)\frac{k_\gamma k_\mu}{k^2}}{k^2}(-ib)\varepsilon^{\mu\nu\alpha\beta}\frac{K_{e\nu} n_\alpha k_\beta}{n\cdot k} \nonumber\\
&=ieb\frac{\varepsilon^{\mu\nu\alpha\beta}J_{e\mu} K_{e\nu} n_\alpha k_\beta}{k^2n\cdot k}~.
\end{align}
This agrees exactly with Weinberg's Lagrangian independent result~\eqref{e.WeinAB}. We also find the magnetic-magnetic scattering to be
\begin{align}
i\mathcal{M}^{KK}_0=&-ib\varepsilon^{\mu\rho\sigma\lambda}\frac{K_{e\rho} n_\sigma k_\lambda}{n\cdot k}(-i)\frac{\eta_{\mu\nu}-(1-\xi)\frac{k_\mu k_\nu}{k^2}}{k^2}(-ib)\varepsilon^{\nu\alpha\beta\gamma}\frac{K_{e\alpha} n_\beta k_\gamma}{n\cdot k} +ib^2\frac{n^2K_e^2-\left( n\cdot K_e\right)^2}{(n\cdot k)^2}  \nonumber\\
=&-\frac{ib^2}{k^2(n\cdot k)^2}\left[K_e^2\left(n^2k^2-(n\cdot k)^2 \right)-k^2\left(n\cdot K_e\right)^2 \right]+ib^2\frac{n^2K_e^2-\left( n\cdot K_e\right)^2}{(n\cdot k)^2}  \nonumber\\
=&ib^2\frac{K_e^\mu K_{e\mu}}{k^2}~.
\end{align}
Notice that the magnetic-magnetic ``contact'' term, see Fig.~\ref{f.treeKKpot}, plays a crucial role in this calculation. It cancels contributions from the contracted Levi-Civita tensors in order to produce an amplitude of the same form as in electric-electric scattering, as required by electric-magnetic duality. 

One might wonder why it is so much more involved to obtain this simple result for magnetic currents as compared to electric currents. This follows from our choice to give the electric currents priority in the sense of simple local couplings to a conserved current. This imbalance in how we treat potential-current couplings leads to the difference in calculations, even though the end results are essentially identical. Using a more balanced formalism, such as Zwanziger's~\cite{Zwanziger:1970hk}, the calculations for each type of charge are much more similar. We illustrate this in the following section.

Turning to the field strength formalism, we note that the propagator for $F^A_{\mu\nu}$, not $F_{\mu\nu}$, is given by
\beq
\langle F^A_{\alpha\beta}F^A_{\gamma\delta}\rangle_k=-2i\,{}_{\;\alpha\beta}^{(k)}P_{\gamma\delta}^{(k)}~,
\eeq
because the $A_\mu$ propagator has not changed. The form of the $F_{\mu\nu}$ propagator must be different as it does not belong to the subspace of objects that are wedge products with $k_\mu$. If it were it would satisfy the Bianchi identity, implying no magnetic charges. Therefore, we see that when magnetic currents are included in the theory the form of the $F_{\mu\nu}$ propagator must differ from the standard QED result. 

However, we do not need to know the $F_{\mu\nu}$ propagator to calculate one-photon exchange between two currents. The $F^A_{\mu\nu}$ field strength itself has couplings to electric and magnetic currents
\begin{align}
\mathcal{L}=&-\frac14 F^A_{\mu\nu}F^{A\mu\nu}+eF^A_{\mu\nu} \frac{J^\mu n^\nu-J^\nu n^\mu}{2n\cdot\partial}+bF^A_{\mu\nu}\frac{\varepsilon^{\mu\nu\alpha\beta} K_\alpha n_\beta}{2n\cdot\partial}+b^2\frac{n^2K^2-(n\cdot K)^2}{2(n\cdot\partial)^2}\nonumber\\
=&-\frac14 F^A_{\mu\nu}F^{A\mu\nu}+eF^A_{\mu\nu} \frac{J^\mu n^\nu-J^\nu n^\mu}{2n\cdot\partial}+b{}^\ast F^A_{\mu\nu}\frac{K^\mu n^\nu-K^\nu n^\mu}{2n\cdot\partial}+b^2\frac{n^2K^2-(n\cdot K)^2}{2(n\cdot\partial)^2}~.
\end{align} 
This way of expressing the Lagrangian is almost what we had guessed for including magnetic currents in QED, except for the ``contact'' term between magnetic currents. Again, there is no corresponding ``contact'' term for electric currents because we have chosen a formalism in which electric currents, and only electric currents, have a local coupling to the potential $A_\mu$.

The calculation of electric-electric scattering $\mathcal{M}_0^{JJ}$ is identical to what we found in QED. The tree-level amplitude for magnetic-magnetic scattering is 
\begin{align}
i\mathcal{M}_0^{KK}=&ib\frac{K_e^\alpha n^\beta-K_e^\beta n^\alpha}{2n\cdot k}(-2i){}^{(k)}_{\ast\alpha\beta}P^{(k)}_{\ast\gamma\delta}(ib)\frac{K_e^\gamma n^\delta-K_e^\delta n^\gamma}{2n\cdot k}+ib^2\frac{K_e^2n^2-(K_e\cdot n)^2}{(n\cdot k)^2}\nonumber\\
=&2ib^2\frac{K_e^\alpha n^\beta-K_e^\beta n^\alpha}{2n\cdot k}\left({}^{(k)}_{\;\alpha\beta}P^{(k)}_{\gamma\delta}-{}_{\;\alpha\beta}I_{\gamma\delta}\right)~\frac{K_e^\gamma n^\delta-K_e^\delta n^\gamma}{2n\cdot k}+ib^2\frac{K_e^2n^2-(K_e\cdot n)^2}{(n\cdot k)^2}\nonumber\\
=&ib^2\frac{K_e^2}{k^2}-ib^2\frac{K_e^2n^2-(K_e\cdot n)^2}{(n\cdot k)^2}+ib^2\frac{K_e^2n^2-(K_e\cdot n)^2}{(n\cdot k)^2}=ib^2\frac{K_e^2}{k^2}~.
\end{align}
This calculation is algebraically identical to what we evaluated before. However, it is arguably a simpler way of organizing the calculation. In particular, the need for the ``contact'' term between magnetic currents is immediately obvious. 

We can also understand the scattering of electric and magnetic charges. Because $F^A_{\mu\nu}$ is sourced by $J^\mu$ and ${}^\ast\!F^A_{\mu\nu}$ is sourced by $K^\mu$, the propagators connecting electric and magnetic currents are
\beq
\langle F^A_{\alpha\beta}{}^\ast\!F^A_{\gamma\delta}\rangle_k=&-2i{}^{(k)}_{\;\alpha\beta}P^{(k)}_{\ast\gamma\delta}~, \ \ \ \ \langle {}^\ast\!F^A_{\alpha\beta}F^A_{\gamma\delta}\rangle_k=&-2i{}^{(k)}_{\ast\alpha\beta}P^{(k)}_{\gamma\delta}~.
\eeq
There is, however, something of an ambiguity because 
\beq
{}_{(k)}^{\;\alpha\beta}P_{(k)}^{\ast\gamma\delta}=\frac12\varepsilon^{\alpha\beta\gamma\delta}-{}_{(k)}^{\ast\alpha\beta}P_{(k)}^{\gamma\delta}~.\label{e.EpsIdent}
\eeq
This implies that the electric-magnetic scattering amplitude can be expressed in multiple, equivalent, ways. Allowing the surfaces bounded by the two currents to correspond to distinct vectors $n_J^\mu$ and $n_K^\mu$ we find
\begin{align}
i\mathcal{M}_0^{JK}=&ie\frac{J_e^\alpha n_J^\beta-J_e^\beta n_J^\alpha}{2n_J\cdot k}(-2i){}^{(k)}_{\;\alpha\beta}P^{(k)}_{\ast\gamma\delta}(ib)\frac{K_e^\gamma n_K^\delta-K_e^\delta n_K^\gamma}{2n_K\cdot k}\nonumber\\
=&ieb\frac{\varepsilon_{\alpha\beta\gamma\delta}J_e^\alpha K_e^\beta n_K^\gamma k^\delta}{k^2n_K\cdot k}\\
=&ie\frac{J_e^\alpha n_J^\beta-J_e^\beta n_J^\alpha}{2n_J\cdot k}\left(2i{}^{(k)}_{\ast\alpha\beta}P^{(k)}_{\gamma\delta}-i\varepsilon_{\alpha\beta\gamma\delta}\right)ib\frac{K_e^\gamma n_K^\delta-K_e^\delta n_K^\gamma}{2n_K\cdot k}\nonumber\\
=&ieb\frac{\varepsilon_{\alpha\beta\gamma\delta}J_e^\alpha K_e^\beta n_J^\gamma k^\delta}{k^2n_J\cdot k}-ieb\frac{\varepsilon_{\alpha\beta\gamma\delta}J_e^\alpha K_e^\beta n_J^\gamma n_K^\delta}{(n_J\cdot k)(n_K\cdot k)}~.
\end{align}

Notice that unlike the interactions of like charges, these scattering amplitudes depend on the $n_i^\mu$, or equivalently on the Stokes surface bounded by the current. The term with the photon pole ($1/k^{2}$) only depends on one of the surface reference vectors. In the second line it depends on the magnetic Stokes surface. We can, however, exchange the magnetic surface vector for the electric one by introducing a term related to the intersection of the two surfaces, as discussed in Sec.~\ref{s.TwoForms}. If we choose $n_J^\mu=n_K^\mu$ any difference between the two results vanishes. 

\subsection{Renormalization Done Wrong\label{ss.RenWr}}
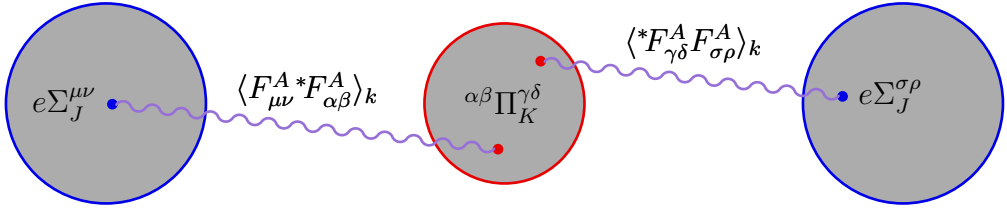
\begin{figure}
\center
\begin{fmffile}{KloopFJJ}
\begin{fmfgraph*}(300,70)
\fmfpen{1.0}
\fmfstraight
\fmfset{arrow_len}{3mm}
\fmfleft{p1,p2,p3,p4,p5} \fmfright{p6,p7,p8,p9,p10}
\fmf{phantom,tension=1}{p2,p11,p7}
\fmf{phantom,tension=1}{p4,p12,p9}
\fmf{phantom,tension=1}{p3,v5,p8}
\fmffreeze
\fmf{phantom,tension=3.0}{p3,v3}
\fmf{phantom,tension=1.5}{v4,p8}
\fmf{phantom,tension=3.0}{p11,v1}
\fmf{phantom,tension=2.5}{p12,v2}
\fmfv{decor.shape=circle,decor.filled=0,decor.size=75,fore=(0,,0,,0.9),background=(0.675,,0.675,,0.675)}{p3}
\fmfv{decor.shape=circle,decor.filled=0,decor.size=75,fore=(0,,0,,0.9),background=(0.675,,0.675,,0.675)}{p8}
\fmfv{decor.shape=circle,decor.filled=0,decor.size=60,l=${}^{\alpha\beta}\Pi_K^{\gamma\delta}$,l.a=0,l.d=-15,fore=(0.9,,0,,0),background=(0.675,,0.675,,0.675)}{v5}
\fmf{photon,rubout=0.5,tension=0.05,label=$\langle F_{\mu\nu}^A {}^\ast\!F_{\alpha\beta}^A\rangle_k$,label.side=left,fore=(0.59,,0.44,,0.84)}{v3,v1}
\fmf{photon,rubout=0.5,tension=0.3,label=$\langle {}^\ast\!F_{\gamma\delta}^A F_{\sigma\rho}^A\rangle_k$,label.side=left,fore=(0.59,,0.44,,0.84)}{v2,v4}
\fmfv{decor.shape=circle,decor.filled=full,decor.size=1.5thick,fore=(0.9,,0,,0)}{v1} 
\fmfv{decor.shape=circle,decor.filled=full,decor.size=1.5thick,fore=(0.9,,0,,0)}{v2} 
\fmfv{decor.shape=circle,decor.filled=full,decor.size=1.5thick,l=$e\Sigma_J^{\mu\nu}$,fore=(0,,0,,0.9)}{v3} 
\fmfv{decor.shape=circle,decor.filled=full,decor.size=1.5thick,l=$e\Sigma_J^{\sigma\rho}$,fore=(0,,0,,0.9)}{v4} 
\end{fmfgraph*}
\end{fmffile}
\caption{\label{f.KloopJJF}One magnetic (red) loop correction to photon exchange between electric (blue) currents using field strengths. The field strength $F^A_{\mu\nu}$ has direct coupling to electric currents while ${}^\ast\!F^A_{\mu\nu}$ couples directly to magnetic currents. The mixed propagator includes the Levi-Civita tensor and is colored purple.}
\end{figure}

Having determined how to account for mutually nonlocal charge interactions we can now consider the contribution of a dynamic magnetic loop to the interaction of external electric currents. We use the field strength formalism to organize the calculation because it simplifies the algebra. Importantly for this calculation, the magnetic current ``contact'' term plays no role because two distinct magnetic currents never exchange a photon. The heart of this calculation, see Fig.~\ref{f.KloopJJF}, is captured by
\begin{align}
{}^{(k)}_{\;\alpha\beta}P^{(k)}_{\ast\mu\nu}{}^{\mu\nu}\Pi^{\sigma\rho}{}^{(k)}_{\ast\sigma\rho}P^{(k)}_{\gamma\delta}=&\frac12\Pi(k^2){}^{(k)}_{\;\alpha\beta}P^{(k)}_{\ast\mu\nu}\,{}_{(n)}^{\;\mu\nu}P_{(k)}^{\zeta\theta}{}_{\zeta\theta}I_{\kappa\lambda}\,{}_{(k)}^{\;\kappa\lambda}P_{(n)}^{\sigma\rho} {}^{(k)}_{\ast\sigma\rho}P^{(k)}_{\gamma\delta}\label{e.hidTopTerms}\\
=&\frac12\Pi(k^2){}^{(k)}_{\;\alpha\beta}P^{(k)}_{\gamma\delta}-\frac12\Pi(k^2){}_{\;\alpha\beta}^{(k)}P_{(n)}^{\mu\nu}{}_{\mu\nu}I_{\sigma\rho} {}^{\;\sigma\rho}_{(n)}P^{(k)}_{\gamma\delta}~.
\end{align}

For a dynamic magnetic loop we have the magnetic vacuum polarization
\beq
\Pi_K^{\mu\nu}=\int d^4x\, e^{-ik\cdot(x-y)}\langle\bm{0}|T\left\{ K_d^\mu(x)K_d^\nu(y)\right\}|\bm{0}\rangle=\Pi_K(k^2)\left(k^2\eta^{\mu\nu}-k^\mu k^\nu \right)~.\label{e.MagWardForm}
\eeq
In contrast to what we found in Eq.~\eqref{e.PpiP} for a virtual loop of electric charge, we immediately obtain troubling results. In particular, there is an additional term that does have the same form as the tree-level scattering. One might hope that when contracted with external currents this extra term vanishes, but the one-loop amplitude is actually found to be
\begin{align}
i\mathcal{M}_{1_K}^{JJ}=ie^2\Pi_K\frac{J_e^2}{k^2}-ie^2\Pi_K\frac{J_e^2n^2-(J_e\cdot n)^2}{(k\cdot n)^2}~,
\end{align}
where we have suppressed the argument of $\Pi_K$, the vacuum polarization function.
If one focuses only on the first term then one is tempted to claim that electric and magnetic loop lead to the same kind of renormalization, as Schwinger found. That is, we have
\beq
\mathcal{M}^{JJ}_0+\mathcal{M}^{JJ}_{1_J}+\mathcal{M}^{JJ}_{1_K}=e^2\frac{J_e^2}{k^2}\left[1+\Pi_J+\Pi_K \right]-e^2\Pi_K\frac{J_e^2n^2-(J_e\cdot n)^2}{(k\cdot n)^2}~,\label{e.wrongJJ}
\eeq
with both electric and magnetic loops contributing in the same way to the running. However, the fact that this second term (which looks like a ``contact'' term between electric currents) is generated signals that further thought is required.

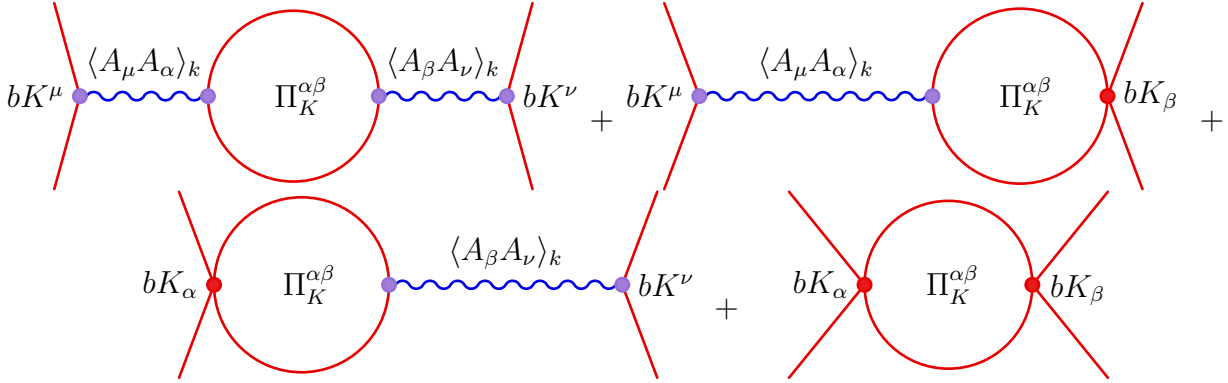
\begin{figure}
\center
\begin{fmffile}{loopPotKK}
\begin{fmfgraph*}(180,70)
\fmfpen{1.0}
\fmfstraight
\fmfset{arrow_len}{3mm}
\fmfleft{i1,i2} \fmfright{o1,o2}
\fmf{plain,tension=1.0,fore=(0.9,,0,,0)}{i1,v1,i2}
\fmf{photon,tension=0.4,label=$\langle A_\mu A_\alpha\rangle_k$,label.side=left,fore=(0,,0,,0.9)}{v1,v2}
\fmf{plain,right=1,tension=0.15,fore=(0.9,,0,,0)}{v2,v3,v2}
\fmf{photon,tension=0.4,label=$\langle A_\beta A_\nu\rangle_k$,label.side=left,fore=(0,,0,,0.9)}{v3,v4}
\fmf{plain,tension=1.0,fore=(0.9,,0,,0)}{o1,v4,o2}
\fmfv{decor.shape=circle,decor.filled=90,decor.size=2.5thick,l=$bK^\mu$,fore=(0.59,,0.44,,0.84)}{v1} 
\fmfv{decor.shape=circle,decor.filled=90,decor.size=2.5thick,l=$\Pi_K^{\alpha\beta}$,l.d=25,l.a=0,fore=(0.59,,0.44,,0.84)}{v2} 
\fmfv{decor.shape=circle,decor.filled=90,decor.size=2.5thick,fore=(0.59,,0.44,,0.84)}{v3} 
\fmfv{decor.shape=circle,decor.filled=90,decor.size=2.5thick,l=$bK^\nu$,fore=(0.59,,0.44,,0.84)}{v4} 
\end{fmfgraph*}
\end{fmffile}
\raisebox{0.8cm}{\;\;\; $+$ \;}
\begin{fmffile}{loopPotKKcont}
\begin{fmfgraph*}(180,70)
\fmfpen{1.0}
\fmfstraight
\fmfset{arrow_len}{3mm}
\fmfleft{i1,i2} \fmfright{o1,o2}
\fmf{plain,tension=1.0,fore=(0.9,,0,,0)}{i1,v1,i2}
\fmf{photon,tension=0.3,label=$\langle A_\mu A_\alpha\rangle_k$,label.side=left,fore=(0,,0,,0.9)}{v1,v2}
\fmf{plain,right=1,tension=0.2,fore=(0.9,,0,,0)}{v2,v3,v2}
\fmf{plain,tension=1.0,fore=(0.9,,0,,0)}{o1,v3,o2}
\fmfv{decor.shape=circle,decor.filled=90,decor.size=2.5thick,l=$bK^\mu$,fore=(0.59,,0.44,,0.84)}{v1} 
\fmfv{decor.shape=circle,decor.filled=90,decor.size=2.5thick,l=$\Pi_K^{\alpha\beta}$,l.d=25,l.a=0,fore=(0.59,,0.44,,0.84)}{v2} 
\fmfv{decor.shape=circle,decor.filled=90,decor.size=2.5thick,l=$bK_\beta$,fore=(0.9,,0,,0)}{v3} 
\end{fmfgraph*}
\end{fmffile}
\raisebox{0.8cm}{\;\;\; $+$ \;}\\
\begin{fmffile}{contLoopPotKK}
\begin{fmfgraph*}(180,70)
\fmfpen{1.0}
\fmfstraight
\fmfset{arrow_len}{3mm}
\fmfleft{i1,i2} \fmfright{o1,o2}
\fmf{plain,tension=1.0,fore=(0.9,,0,,0)}{i1,v1,i2}
\fmf{plain,right=1,tension=0.2,fore=(0.9,,0,,0)}{v1,v2,v1}
\fmf{photon,tension=0.3,label=$\langle A_\beta A_\nu\rangle_k$,label.side=left,fore=(0,,0,,0.9)}{v2,v3}
\fmf{plain,tension=1.0,fore=(0.9,,0,,0)}{o1,v3,o2}
\fmfv{decor.shape=circle,decor.filled=90,decor.size=2.5thick,l=$bK_\alpha$,fore=(0.9,,0,,0)}{v1} 
\fmfv{decor.shape=circle,decor.filled=90,decor.size=2.5thick,l=$\Pi_K^{\alpha\beta}$,l.d=20,l.a=180,fore=(0.59,,0.44,,0.84)}{v2} 
\fmfv{decor.shape=circle,decor.filled=90,decor.size=2.5thick,l=$bK^\nu$,fore=(0.59,,0.44,,0.84)}{v3} 
\end{fmfgraph*}
\end{fmffile}
\raisebox{0.8cm}{\;\;\; $+$ \;}
\begin{fmffile}{contLoopPotKKcont}
\begin{fmfgraph*}(120,70)
\fmfpen{1.0}
\fmfstraight
\fmfset{arrow_len}{3mm}
\fmfleft{i1,i2} \fmfright{o1,o2}
\fmf{plain,tension=1.0,fore=(0.9,,0,,0)}{i1,v1,i2}
\fmf{plain,right=1,tension=0.45,fore=(0.9,,0,,0)}{v1,v2,v1}
\fmf{plain,tension=1.0,fore=(0.9,,0,,0)}{o1,v2,o2}
\fmfv{decor.shape=circle,decor.filled=90,decor.size=2.5thick,l=$bK_\alpha$,fore=(0.9,,0,,0)}{v1} 
\fmfv{decor.shape=circle,decor.filled=90,decor.size=2.5thick,l=$\Pi_K^{\alpha\beta}$,l.d=20,l.a=180,fore=(0.9,,0,,0)}{v2} 
\fmfv{l=$bK_\beta$,l.a=-110,l.d=33}{o2} 
\end{fmfgraph*}
\end{fmffile}
\caption{\label{f.loopKKpot}One magnetic loop corrections to photon exchange between magnetic currents using the one potential formalism.}
\end{figure}

One loop corrections to magnetic-magnetic interactions follow a similar form. In this case, however, one must include all four diagrams, see Fig.~\ref{f.loopKKpot}, when considering the virtual magnetic loop contribution because of the ``contact'' term. That is, one must include the ``contact'' terms between the each external magnetic current and the virtual current. The final result is
\beq
\mathcal{M}^{KK}_0+\mathcal{M}^{KK}_{1_J}+\mathcal{M}^{KK}_{1_K}=b^2\frac{K_e^2}{k^2}\left[1+\Pi_J+\Pi_K \right]-b^2\Pi_J\frac{K_e^2n^2-(K_e\cdot n)^2}{(k\cdot n)^2}~.\label{e.wrongKK}
\eeq
The form matches the electric-electric case, supporting the notion of electric-magnetic duality within this formalism.
In addition, if one does not consider the final term, it appears that both electric and magnetic loops renormalize the magnetic coupling in the same way. Finally, by comparing Eqs.~\eqref{e.wrongJJ} and~\eqref{e.wrongKK} it seems (if we only consider the terms with the usual photon pole) that both the electric and magnetic couplings are renormalized in the same way, again as Schwinger concluded.

What about renormalization of the mixed propagator? The calculation corresponding to electric-magnetic scattering for a virtual electric loop is
\begin{align}
{}^{(k)}_{\;\alpha\beta}P^{(k)}_{\mu\nu}{}^{\mu\nu}\Pi^{\sigma\rho}{}^{(k)}_{\sigma\rho}P^{(k)}_{\ast\gamma\delta}=&\frac12\Pi(k^2){}^{(k)}_{\;\alpha\beta}P^{(k)}_{\mu\nu}\,{}_{(n)}^{\;\mu\nu}P_{(k)}^{\zeta\theta}{}_{\zeta\theta}I_{\kappa\lambda}\,{}_{(k)}^{\;\kappa\lambda}P_{(n)}^{\sigma\rho} {}^{(k)}_{\sigma\rho}P^{(k)}_{\ast\gamma\delta}\label{e.hidTopTermsMix}~\\
=&\frac{\Pi(k^2)}{2}{}^{(k)}_{\;\alpha\beta}P^{(k)}_{\ast\gamma\delta}~.
\end{align}
This implies that
\beq
i\mathcal{M}^{JK}_{1_J}=ieb\Pi_J\frac{\varepsilon_{\alpha\beta\gamma\delta}J_e^\alpha K_e^\beta n_K^\gamma k^\delta}{k^2n_K\cdot k}~.
\eeq
Note that this is independent of the virtual loop reference vector $n_V^\mu$. 

When considering a virtual magnetic loop we must include an additional diagram to include the effects of the magnetic ``contact'' term. Once this has been included we find
\beq
i\mathcal{M}^{JK}_{1_K}=ieb\Pi_K\frac{\varepsilon_{\alpha\beta\gamma\delta}J_e^\alpha K_e^\beta n_K^\gamma k^\delta}{k^2n_K\cdot k}~.
\eeq
In total we have
\beq
\mathcal{M}^{JK}_0+\mathcal{M}^{JK}_{1_J}+\mathcal{M}^{JK}_{1_K}=eb\frac{\varepsilon_{\alpha\beta\gamma\delta}J_e^\alpha K_e^\beta n_K^\gamma k^\delta}{k^2n_K\cdot k}\left[1+\Pi_J+\Pi_K \right]~,
\eeq
which again seems to support Schwinger's result, this time with no additional worrying terms. However, as shown in the following subsection, all of these results are \emph{incorrect}. They include a subtle contribution from topological terms. When these terms are removed the results are significantly modified and agree with Coleman's characterization of the running couplings. 

\subsection{Renormalization Done Right\label{ss.RenR}}
The loop generated ``contact'' terms in Eqs.~\eqref{e.wrongJJ} and~\eqref{e.wrongKK} are the loud signal that something is amiss. Let us return to Eq.~\eqref{e.hidTopTerms} when calculating the magnetic loop correction to electric-electric interactions. 
\begin{align}
\mathcal{M}^{JJ}_{1_K}=&ie\frac{J_e^\alpha n^\beta-J_e^\beta n^\alpha}{2n\cdot k}(-2i){}^{(k)}_{\;\alpha\beta}P^{(k)}_{\ast\mu\nu}\,i\,{}^{\mu\nu}\Pi^{\sigma\rho}(-2i){}^{(k)}_{\ast\sigma\rho}P^{(k)}_{\gamma\delta}ie\frac{J_e^\gamma n^\delta-J_e^\delta n^\gamma}{2n\cdot k}\nonumber\\
=&2ie^2\Pi_K\frac{J_e^\alpha n^\beta-J_e^\beta n^\alpha}{2n\cdot k}{}^{(k)}_{\;\alpha\beta}P^{(k)}_{\ast\mu\nu}\,{}_{(n)}^{\;\mu\nu}P_{(k)}^{\zeta\theta}{}_{\zeta\theta}I_{\kappa\lambda}\,{}_{(k)}^{\;\kappa\lambda}P_{(n)}^{\sigma\rho} {}^{(k)}_{\ast\sigma\rho}P^{(k)}_{\gamma\delta}\frac{J_e^\gamma n^\delta-J_e^\delta n^\gamma}{2n\cdot k}~.
\end{align}
Using the identity in Eq.~\eqref{e.MixedIdent}, we note that ${}^{(k)}_{\;\alpha\beta}P^{(k)}_{\ast\mu\nu}\,{}_{(n)}^{\;\mu\nu}P_{(k)}^{\zeta\theta}$ can be written in two equal ways
\begin{align}
{}^{(k)}_{\;\alpha\beta}P^{(k)}_{\ast\mu\nu}\,{}_{(n)}^{\;\mu\nu}P_{(k)}^{\zeta\theta}&={}^{(k)}_{\;\alpha\beta}P_{(k)}^{\ast\zeta\theta}-{}^{(k)}_{\;\alpha\beta}P_{(n)}^{\ast\zeta\theta}~,\label{e.MixedStarRight1}\\
&=-{}^{(k)}_{\ast\alpha\beta}P_{(k)}^{\zeta\theta}+{}^{(n)}_{\ast\alpha\beta}P_{(k)}^{\zeta\theta}~.\label{e.MixedStarLeft1}
\end{align}
The second term on the right in each of these lines is the topological part of electric-magnetic scattering. It is this term that we need to separate out from the perturbation series. But which of these two ways of expressing the topological term should we use? If we use Eq.~\eqref{e.MixedStarLeft1} then the topological term is projected out because 
\beq
\frac{J_e^\alpha n^\beta-J_e^\beta n^\alpha}{2n\cdot k}{}^{(n)}_{\ast\alpha\beta}P_{(k)}^{\zeta\theta}=0~.
\eeq
This is simply due to the ${}^{(n)}_{\ast\alpha\beta}P_{(k)}^{\zeta\theta}$ mapping being from the space orthogonal to objects that are a wedge product with $n^\mu$. 

This projecting out might seem at first to be a benefit if it is understood as removing the topological term from the calculation. However, because this term projects out when coupled to the current, the remaining term is still equivalent to the Eq.~\eqref{e.MixedStarRight1} with the topological term included. Therefore, we can only truly separate the topological term from this calculation by using Eq.~\eqref{e.MixedStarRight1} and then removing the topological term. 

Similarly, we can write the combination of terms of the right-side of the identity operator in two ways:
\begin{align}
{}_{(k)}^{\;\kappa\lambda}P_{(n)}^{\sigma\rho} {}^{(k)}_{\ast\sigma\rho}P^{(k)}_{\gamma\delta}&={}_{(k)}^{\;\kappa\lambda}P^{(k)}_{\ast\gamma\delta}-{}_{(k)}^{\;\kappa\lambda}P^{(n)}_{\ast\gamma\delta}~,\label{e.MixedStarRight2}\\
&=-{}_{(k)}^{\ast\kappa\lambda}P^{(k)}_{\gamma\delta}+{}_{(n)}^{\ast\kappa\lambda}P^{(k)}_{\gamma\delta}~.\label{e.MixedStarLeft2}
\end{align}
If we use the top equality then the topological term is projected out and, counterintuitively, it is equivalent to keeping the topological effects. Therefore, we must use Eq.~\eqref{e.MixedStarLeft2} and remove the topological term by hand. If we do not eliminate these topological terms we are, in effect, including four unintended, topological contributions to the scattering we are interested in.

When the topological terms are removed we find the magnetic one-loop contribution to electric-electric interaction is
\beq
\mathcal{M}^{JJ}_{1_K}=-2ie^2\Pi_K\frac{J_e^\alpha n^\beta-J_e^\beta n^\alpha}{2n\cdot k}{}^{(k)}_{\;\alpha\beta}P^{(k)}_{\ast\mu\nu}\, {}_{(k)}^{\ast\mu\nu}P^{(k)}_{\gamma\delta}\frac{J_e^\gamma n^\delta-J_e^\delta n^\gamma}{2n\cdot k}=-ie^2\Pi_K\frac{J_e^2}{k^2}~.
\eeq 
Significantly, no ``contact'' term is included, there is only a contribution with the expected photon pole, $1/k^2$. We also see that the sign of the contribution from a magnetic loop is opposite to that of an electric loop. 

The full one-loop amplitude, with all topological terms removed, is
\beq
\mathcal{M}^{JJ}_0+\mathcal{M}^{JJ}_{1_J}+\mathcal{M}^{JJ}_{1_K}=e^2\frac{J_e^2}{k^2}\left[1+\Pi_J-\Pi_K \right]\approx e^2\frac{J_e^2}{k^2\left[1-\Pi_J+\Pi_K \right]}~.\label{e.rightJJ}
\eeq
Similarly, the one-loop corrections to magnetic-magnetic scattering with all topological terms removed is
\beq
\mathcal{M}^{KK}_0+\mathcal{M}^{KK}_{1_J}+\mathcal{M}^{KK}_{1_K}=b^2\frac{K_e^2}{k^2}\left[1-\Pi_J+\Pi_K \right]~.\label{e.rightKK}
\eeq
Considering the electric and magnetic results, we see that the two couplings run exactly inversely, in agreement with Coleman's analysis. 

This behavior is confirmed by the one-loop corrections to electric-magnetic scattering. The contribution of a virtual electric loop is
\begin{align}
\mathcal{M}^{JK}_{1_J}=&ie\frac{J_e^\alpha n^\beta-J_e^\beta n^\alpha}{2n\cdot k}(-2i){}^{(k)}_{\;\alpha\beta}P^{(k)}_{\mu\nu}\,i\,{}^{\mu\nu}\Pi^{\sigma\rho}(-2i){}^{(k)}_{\;\sigma\rho}P^{(k)}_{\ast\gamma\delta}ig\frac{K_e^\gamma n^\delta-K_e^\delta n^\gamma}{2n\cdot k}\nonumber\\
=&2ieg\Pi_J(k^2)\frac{J_e^\alpha n^\beta-J_e^\beta n^\alpha}{2n\cdot k}{}^{(k)}_{\;\alpha\beta}P^{(k)}_{\mu\nu}{}^{\;\mu\nu}_{(n)}P^{\zeta\theta}_{(k)}{}_{\;\zeta\theta}I_{\kappa\lambda}{}^{\;\kappa\lambda}_{(k)}P^{\sigma\rho}_{(n)} {}^{(k)}_{\;\sigma\rho}P^{(k)}_{\ast\gamma\delta}\frac{K_e^\gamma n^\delta-K_e^\delta n^\gamma}{2n\cdot k}\nonumber\\
=&2ieg\Pi_J(k^2)\frac{J_e^\alpha n^\beta-J_e^\beta n^\alpha}{2n\cdot k}{}^{(k)}_{\;\alpha\beta}P^{(k)}_{\mu\nu}{}^{\;\mu\nu}_{(k)}P^{\sigma\rho}_{(n)} \left(\frac12\varepsilon_{\sigma\rho\gamma\delta}- {}^{(k)}_{\ast\sigma\rho}P^{(k)}_{\gamma\delta}\right)\frac{K_e^\gamma n^\delta-K_e^\delta n^\gamma}{2n\cdot k}~,
\end{align}
where in the last line we have used the identity in Eq.~\eqref{e.EpsIdent}. Notice that the term with a lone epsilon tensor vanishes because it produces 
\beq
{}^{\;\mu\nu}_{(k)}P_{\ast\gamma\delta}^{(n)} \frac{K_e^\gamma n^\delta-K_e^\delta n^\gamma}{2n\cdot k}=0~.
\eeq
The remaining term can be written as
\beq
\mathcal{M}^{JK}_{1_J}=-2ieg\Pi_J(k^2)\frac{J_e^\alpha n^\beta-J_e^\beta n^\alpha}{2n\cdot k}{}^{(k)}_{\;\alpha\beta}P^{(k)}_{\mu\nu}\left(-{}_{(k)}^{\ast\mu\nu}P_{\gamma\delta}^{(k)}+{}_{(n)}^{\ast\mu\nu}P_{\gamma\delta}^{(k)} \right)\frac{K_e^\gamma n^\delta-K_e^\delta n^\gamma}{2n\cdot k}~,
\eeq
so that the topological term does not automatically project out. We note, however, that the other term vanishes because
\beq
{}^{(k)}_{\;\alpha\beta}P^{(k)}_{\mu\nu}{}_{(k)}^{\ast\mu\nu}P_{\gamma\delta}^{(k)}=0~.
\eeq
This means that only the topological term contributes. When this term is removed we simply find that the one-loop contribution to renormalization vanishes. A similar calculation (which includes an additional diagram to include the magnetic ``contact'' term) also shows that $\mathcal{M}^{JK}_{1_K}=0$. In total, the one-loop corrections to the mixed propagator vanish, so
\beq
\mathcal{M}^{JK}_0+\mathcal{M}^{JK}_{1_J}+\mathcal{M}^{JK}_{1_K}=eb\frac{\varepsilon_{\alpha\beta\gamma\delta}J_e^\alpha K_e^\beta n^\gamma k^\delta}{k^2 n\cdot k}~,
\eeq
or in other words, the mixed charge interaction is not renormalized! This, again, is consistent with $e$ and $b$ renormalizing inversely, and the renormalization group invariance of Dirac charge quantization, as argued by Coleman.

In these two subsections we have seen how to understand the apparent contradiction between Schwinger's and Coleman's results and the method to reconcile them. The two results are equal up to terms that arise when topological phase terms are included in loop effects. The signal of this problem is the apparent loss of multiplicative renormalization in the one-loop amplitudes. When these topological pieces are removed, Coleman's result is recovered. We note also that the naive calculation using the potential $A_\mu$ includes the topological phase in loop effects. This is the origin of the extra terms that appeared. The field strength formalism, though not absolutely essential, makes it much simpler to identify and remove the topological terms and to understand them as arising when a worldline and a Stokes surface associated with different types of charges are topologically linked, or when an electric Stokes surface and a magnetic Stokes surface intersect.

For all this success, the single potential formalism is somewhat unwieldy due to the tree-level ``contact'' term between magnetic currents that must be included. We show in the next sections how the two potential formalism due to Zwanziger~\cite{Zwanziger:1970hk}, can be expressed in terms of field strengths without introducing current-current ``contact'' terms. Indeed, we find the language of two-form maps is a concise and intuitive way to describe Zwanziger's Lagrangian and the resulting dynamics. The isolation and removal of the topological terms is also somewhat simplified compared to the one potential formulation.

\section{Zwanziger's Two-Potential Formalism\label{s.Zform}}
In the previous section, the single potential Lagrangian gave electric charges a preference in that the gauge potential only had a local coupling to the electric current. Zwanziger has developed a formalism~\cite{Zwanziger:1970hk} that keeps electric and magnetic currents on a more equal footing, each with local couplings to one gauge potential. However, both of these potentials must be constrained to describe the same photon degrees of freedom. 

Using the notation of~\cite{Terning:2018lsv} we can write the Lagrangian describing local electric and magnetic couplings to the photon by
\begin{align}
\mathcal{L}=&-\frac{n^\alpha n^\mu}{2 n^2}\eta^{\beta\nu}\left(F^A_{\alpha\beta}F^A_{\mu\nu}+F^B_{\alpha\beta}F^B_{\mu\nu} \right)+\frac{n^\alpha n_\mu}{4n^2}\varepsilon^{\mu\nu\gamma\delta} \left( F^B_{\alpha\nu}F^A_{\gamma\delta}-F^A_{\alpha\nu}F^B_{\gamma\delta}\right)\nonumber\\
&-eA_\mu J^\mu-bB_\mu K^\mu,
\end{align}
where $A_\mu$ and $B_\mu$ are the potentials with local couplings to the electric and magnetic currents, respectively. The Euler-Lagrange equations lead to the usual Maxwell equations of Eq.~\eqref{e.MaxEq} with
\begin{align}
F_{\mu\nu}=&\frac{n^\alpha}{n^2}\left(n_\mu F^A_{\alpha\nu}-n_\nu F^A_{\alpha\mu}-\varepsilon_{\mu\nu\alpha\beta}n_\gamma F^{B\gamma\beta} \right),\label{e.Fdef}\\
{}^\ast\! F_{\mu\nu}=&\frac{n^\alpha}{n^2}\left(n_\mu F^B_{\alpha\nu}-n_\nu F^B_{\alpha\mu}+\varepsilon_{\mu\nu\alpha\beta}n_\gamma F^{A\gamma\beta} \right).\label{e.Fstardef}
\end{align}

However, all these results can be written somewhat more concisely using the two-form mappings defined in Sec.~\ref{s.TwoForms}, in particular ${}^{(n)}_{\;\alpha\beta}P^{(n)}_{\gamma\delta}$. Recall that such an operator projects into the space of two-forms that have the form of a wedge product with $n^\mu$. The mapping $-{}^{(n)}_{\ast\alpha\beta}P^{(n)}_{\ast\gamma\delta}$ maps from and into the orthogonal complement space with
\beq
{}^{(n)}_{\;\alpha\beta}P^{(n)}_{\gamma\delta}-{}^{(n)}_{\ast\alpha\beta}P^{(n)}_{\ast\gamma\delta}={}_{\;\alpha\beta}I_{\gamma\delta}~.
\eeq
 It is a straightforward exercise to show that
\begin{align}
F_{\mu\nu}={}^{(n)}_{\;\mu\nu}P^{(n)}_{\alpha\beta}F^{A\alpha\beta}+{}^{(n)}_{\ast\mu\nu}P^{(n)}_{\ast\alpha\beta}{}^\ast\!F^{B\alpha\beta}~, \ \ \ \ {}^\ast\!F_{\mu\nu}={}^{(n)}_{\;\mu\nu}P^{(n)}_{\alpha\beta}F^{B\alpha\beta}-{}^{(n)}_{\ast\mu\nu}P^{(n)}_{\ast\alpha\beta}{}^\ast\!F^{A\alpha\beta}~.\label{e.FdefN}
\end{align}
This clarifies the role of the reference vector $n^\mu$ in this formalism. The six dimensional space of two-forms is divided into two three dimensional subspaces with $\displaystyle{}^{(n)}_{\;\mu\nu}P^{(n)}_{\alpha\beta}$ and $\displaystyle-{}^{(n)}_{\ast\mu\nu}P^{(n)}_{\ast\alpha\beta}$ playing the role of projectors onto each of them. The field strength $F_{\mu\nu}$ generically has components in each of the subspaces. The part of the field strength that belongs to the $\displaystyle{}^{(n)}_{\,\mu\nu}P^{(n)}_{\alpha\beta}$ subspace is $F^A_{\mu\nu}$ while the part in the orthogonal complement is $-{}^\ast\!F^B_{\mu\nu}$.

An essential aspect of this formalism is that changes in $n^\mu$ have no physical consequence. Suppose we change $n_\mu\to n'_\mu$. We may as well keep the length fixed because it divides out of every quantity. This implies that ${}_{(n)}^{\alpha\beta}P_{(n)}^{\gamma\delta}\to {}_{(n')}^{\alpha\beta}P_{(n')}^{\gamma\delta}$. We then require 
\begin{align}
{}^{(n')}_{\;\alpha\beta}P^{(n')}_{\mu\nu}F^{\mu\nu}\equiv& {}^{(n')}_{\;\alpha\beta}P^{(n')}_{\mu\nu}F^{A'\mu\nu}={}^{(n')}_{\;\alpha\beta}P^{(n')}_{\mu\nu}{}_{(n)}^{\;\mu\nu}P_{(n)}^{\gamma\delta}F^A_{\gamma\delta}+{}^{(n')}_{\alpha\beta}P^{(n')}_{\mu\nu}{}_{(n)}^{\ast\mu\nu}P_{(n)}^{\ast\gamma\delta}{}^\ast\!F^B_{\gamma\delta}\\
{}^{(n')}_{\;\alpha\beta}P^{(n')}_{\mu\nu}{}^\ast\!F^{\mu\nu}\equiv& {}^{(n')}_{\;\alpha\beta}P^{(n')}_{\mu\nu}F^{B'\mu\nu}={}^{(n')}_{\;\alpha\beta}P^{(n')}_{\mu\nu}{}_{(n)}^{\;\mu\nu}P_{(n)}^{\gamma\delta}F^B_{\gamma\delta}-{}^{(n')}_{\;\alpha\beta}P^{(n')}_{\mu\nu}{}_{(n)}^{\ast\mu\nu}P_{(n)}^{\ast\gamma\delta}{}^\ast\!F^A_{\gamma\delta}
\end{align}
This shows that if $n^\mu$ is changed that a suitable definition of $A'_\mu$ and $B'_\mu$, those that fulfill the equations above, keeps $F_{\mu\nu}$ unchanged. So, while the vector potentials change with $n^\mu$ the physical fields do not. In this sense a change in $n^\mu$ is something like a gauge transformation of the potentials. This is separate, however, from the invariance of the Zwanziger Lagrangian under the usual, and distinct, gauge transformations of  $A_\mu$ and $B_\mu$
\beq
A_\mu\to A_{\mu}+\partial_\mu \Lambda_A~, \ \ \ \ B_\mu\to B_{\mu}+\partial_\mu \Lambda_B~.
\eeq

Because the definition of the field strengths in Eq.~\eqref{e.FdefN} is in terms of projectors it is not easily invertible. However, we do note that 
\beq
F^{\mu\nu}{}^{(n)}_{\mu\nu}P^{(n)}_{\alpha\beta}=F^{A\mu\nu}{}^{(n)}_{\mu\nu}P^{(n)}_{\alpha\beta}~, \ \ \ \ {}^\ast\!F^{\mu\nu}{}^{(n)}_{\mu\nu}P^{(n)}_{\alpha\beta}=F^{B\mu\nu}{}^{(n)}_{\mu\nu}P^{(n)}_{\alpha\beta}~.\label{e.FtoFaOrb}
\eeq 
We can also write the kinetic part of the Lagrangian as
\begin{align}
\mathcal{L}=&-\frac{1}{4}{}_{(n)}^{\;\alpha\beta}P_{(n)}^{\mu\nu}\left( F^A_{\alpha\beta}F^A_{\mu\nu}+F^B_{\alpha\beta}F^B_{\mu\nu} \right)-\frac{1}{4}{}_{(n)}^{\ast\alpha\beta}P_{(n)}^{\mu\nu}\left(F^{B}_{\alpha\beta}F^{A}_{\mu\nu}-F^A_{\alpha\beta}F^B_{\mu\nu} \right)\\
=&-\frac{1}{4}\left(F^A_{\alpha\beta},\,F^B_{\alpha\beta} \right)\left(\begin{array}{cc}
{}_{(n)}^{\;\alpha\beta}P_{(n)}^{\mu\nu} & -{}_{(n)}^{\ast\alpha\beta}P_{(n)}^{\mu\nu}\\
{}_{(n)}^{\ast\alpha\beta}P_{(n)}^{\mu\nu}& {}_{(n)}^{\;\alpha\beta}P_{(n)}^{\mu\nu}
\end{array} \right)\left(\begin{array}{c}
F^A_{\mu\nu} \\
F^B_{\mu\nu}
\end{array} \right)~\\
=&-\frac{1}{4}F^A_{\mu\nu}F^{\mu\nu}-\frac{1}{4}F^B_{\mu\nu}{}^\ast\!F^{\mu\nu}~.
\end{align}
This last way of expressing the Lagrangian is perhaps the most reminiscent of the usual QED Lagrangian. 

We can imagine motivating the couplings of two-form fields to the currents similar to Eq.~\eqref{e.twoFormCoupling}. The electric current is a source for $F^A_{\mu\nu}$ 
\begin{align}
-eA_\mu J^\mu\to i eF^A_{\mu\nu}\frac{J^\mu n^\nu-J^\nu n^\mu}{2n\cdot k}~.
\end{align}
Similarly, the magnetic current is a source for $F^B_{\mu\nu}$
\beq
-bB_\mu K^\mu\to i bF^B_{\mu\nu}\frac{K^\mu n^\nu-K^\nu n^\mu}{2n\cdot k}~.
\eeq
The form of loop functions, for both electric or magnetic particles, is also identical to what was found in the previous section~\eqref{e.LoopFunc}.

One first finds that the propagators (see Appendix~\ref{a.Fprop} for the derivation) for both types of field strength are identical
\beq
\langle F^A_{\mu\nu}F^A_{\sigma\rho}\rangle_k=\langle F^B_{\mu\nu}F^B_{\sigma\rho}\rangle_k=-2i{}^{(k)}_{\mu\nu}P^{(k)}_{\sigma\rho}~.
\eeq
This, in contrast to the single potential formalism, illustrates how both types of charges are treated in the same way, or that the Zwanziger formalism is duality covariant~\cite{Terning:2018lsv}. The propagators quickly lead to the tree level results 
\beq
\mathcal{M}^{JJ}_0=e^2\frac{J_e^\mu J_{e\mu}}{k^2}~, \ \ \ \ \mathcal{M}^{KK}_0=b^2\frac{K_e^\mu K_{e\mu}}{k^2}~,
\eeq
without the need for any current-current ``contact'' terms. 

The mixed propagator, however, has a very significant form which differs from the single potential propagator. Instead, it agrees exactly with the position space version of Weinberg's mixed potential propagator~\eqref{e.mixedProp}
\begin{align}
\langle F^A_{\mu\nu}F^B_{\sigma\rho}\rangle_k=&-2i{}^{(k)}_{\mu\nu}P^{(k)}_{\ast\sigma\rho}+2i{}^{(k)}_{\mu\nu}P^{(n)}_{\ast\sigma\rho}~\\
=&2i{}^{(k)}_{\ast\mu\nu}P^{(k)}_{\sigma\rho}-2i{}^{(n)}_{\ast\mu\nu}P^{(k)}_{\sigma\rho}~\\
=&-2i{}^{(k)}_{\mu\nu}P_{(k)}^{\alpha\beta}{}_{\ast\alpha\beta}^{(n)}P^{(k)}_{\sigma\rho}~.
\end{align}
The first two lines show that this propagator includes a term that looks like the single potential mixed propagator, which is $n^\mu$ independent. They also contain the topological $n^\mu$ dependent term without a photon pole. The first two lines are, of course, equal which, as seen in the previous section, can introduce ambiguity regarding which form to use in calculations. The final line shows an unambiguous way of encoding this information, but the topological term is not explicitly separated out. As we saw in Sec.~\ref{ss.RenR}, when removing the topological pieces from loop amplitudes all ambiguity is removed from mixed charge interactions.

While we have emphasized the topological character of the $n^\mu$ dependent terms we also note that for tree-level calculations the topological terms project out when connected to the external currents defined above. This might lead one to expect that any subtleties related to these terms has been eliminated from the formalism. However, as we saw with the one potential calculations, this is not quite true of one-loop calculations. 

In Appendix~\ref{a.PotRenorm} we calculate the naive renormalization of the Zwanziger propagators, employing the usual formalism that one might use in standard QED. We find 
\begin{align}
\mathcal{M}^{JJ}_0+\mathcal{M}^{JJ}_{1_J}+\mathcal{M}^{JJ}_{1_K}=&e^2\frac{J_e^2}{k^2}\left[1+\Pi_J(k^2)+\Pi_K(k^2) \right]-e^2\Pi_K(k^2)\frac{J_e^2n^2-(J_e\cdot n)^2}{(n\cdot k)^2}~,\nonumber\\
\mathcal{M}^{KK}_0+\mathcal{M}^{KK}_{1_J}+\mathcal{M}^{KK}_{1_K}=&b^2\frac{K_e^2}{k^2}\left[1+\Pi_J(k^2)+\Pi_K(k^2) \right]-b^2\Pi_J(k^2)\frac{K_e^2n^2-(K_e\cdot n)^2}{(n\cdot k)^2}~,\nonumber\\
\mathcal{M}^{JK}_0+\mathcal{M}^{JK}_{1_J}+\mathcal{M}^{JK}_{1_K}=&eb\frac{\varepsilon_{\alpha\beta\gamma\delta}J_e^\alpha K_e^\beta n^\gamma k^\delta}{k^2(n\cdot k)}\left[1+\Pi_J(k^2)+\Pi_K(k^2) \right]~,
\end{align}
which is exactly the \emph{incorrect} result we found in Sec.~\ref{ss.RenWr} by using the one potential formalism and leaving in the topological terms. This emphasizes that the standard methods of renormalization, when using potentials, include the effects of the topological part of electric-magnetic interactions. The value of the field strength methods is that they enable the identification and separation of the topological terms. 

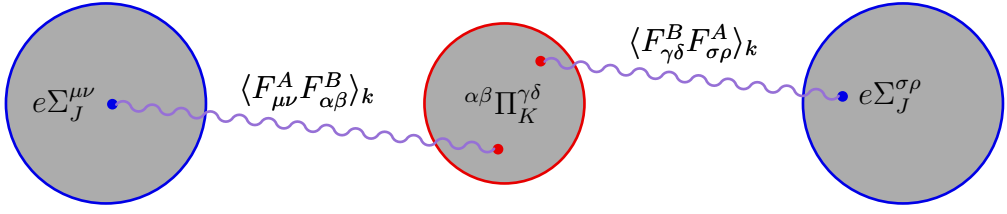
\begin{figure}
\center
\begin{fmffile}{KloopFAFBJJ}
\begin{fmfgraph*}(300,70)
\fmfpen{1.0}
\fmfstraight
\fmfset{arrow_len}{3mm}
\fmfleft{p1,p2,p3,p4,p5} \fmfright{p6,p7,p8,p9,p10}
\fmf{phantom,tension=1}{p2,p11,p7}
\fmf{phantom,tension=1}{p4,p12,p9}
\fmf{phantom,tension=1}{p3,v5,p8}
\fmffreeze
\fmf{phantom,tension=3.0}{p3,v3}
\fmf{phantom,tension=1.5}{v4,p8}
\fmf{phantom,tension=3.0}{p11,v1}
\fmf{phantom,tension=2.5}{p12,v2}
\fmfv{decor.shape=circle,decor.filled=0,decor.size=75,fore=(0,,0,,0.9),background=(0.675,,0.675,,0.675)}{p3}
\fmfv{decor.shape=circle,decor.filled=0,decor.size=75,fore=(0,,0,,0.9),background=(0.675,,0.675,,0.675)}{p8}
\fmfv{decor.shape=circle,decor.filled=0,decor.size=60,l=${}^{\alpha\beta}\Pi_K^{\gamma\delta}$,l.a=0,l.d=-15,fore=(0.9,,0,,0),background=(0.675,,0.675,,0.675)}{v5}
\fmf{photon,rubout=0.5,tension=0.05,label=$\langle F_{\mu\nu}^A F_{\alpha\beta}^B\rangle_k$,label.side=left,fore=(0.59,,0.44,,0.84)}{v3,v1}
\fmf{photon,rubout=0.5,tension=0.3,label=$\langle F_{\gamma\delta}^B F_{\sigma\rho}^A\rangle_k$,label.side=left,fore=(0.59,,0.44,,0.84)}{v2,v4}
\fmfv{decor.shape=circle,decor.filled=full,decor.size=1.5thick,fore=(0.9,,0,,0)}{v1} 
\fmfv{decor.shape=circle,decor.filled=full,decor.size=1.5thick,fore=(0.9,,0,,0)}{v2} 
\fmfv{decor.shape=circle,decor.filled=full,decor.size=1.5thick,l=$e\Sigma_J^{\mu\nu}$,fore=(0,,0,,0.9)}{v3} 
\fmfv{decor.shape=circle,decor.filled=full,decor.size=1.5thick,l=$e\Sigma_J^{\sigma\rho}$,fore=(0,,0,,0.9)}{v4} 
\end{fmfgraph*}
\end{fmffile}
\caption{\label{f.KloopJJFAFB}One magnetic (red) loop correction to photon exchange between electric (blue) currents using $F^A_{\mu\nu}$$F^B_{\mu\nu}$ propagators. This propagator includes the Levi-Civita tensor, which is denoted by the purple color.}
\end{figure}

To see this, we now calculate the renormalization of the photon propagators using the field strength formalism. The electric-loop corrections to electric-electric scattering and the magnetic-loop corrections to magnetic-magnetic scattering have no mixed charge component and consequently produce the expected (and correct) results. Let us then consider the magnetic-loop correction to electric-electric scattering, as shown in Fig.~\ref{f.KloopJJFAFB}. We find
\begin{align}
&i\mathcal{M}_{1_K}^{JJ}\nonumber\\
=&ie\frac{J_e^\alpha n^\beta-J_e^\beta n^\alpha}{2n\cdot k}(-2i){}^{(k)}_{\alpha\beta}P_{(k)}^{\kappa\lambda}{}_{\ast\kappa\lambda}^{(n)}P^{(k)}_{\sigma\rho}{}^{\;\sigma\rho}_{(n)}P_{(k)}^{\xi\zeta}\frac{i\Pi_K}{2}{}_{\xi\zeta}I_{\phi\tau} {}^{\;\phi\tau}_{(k)}P_{(n)}^{\mu\nu}(-2i){}^{(k)}_{\mu\nu}P_{(k)}^{\theta\psi}{}_{\ast\theta\psi}^{(n)}P^{(k)}_{\gamma\delta}ie\frac{J_e^\gamma n^\delta-J_e^\delta n^\gamma}{2n\cdot k}\nonumber\\
=&-2ie^2\Pi_K\frac{J_e^\alpha n^\beta-J_e^\beta n^\alpha}{2n\cdot k}{}^{(k)}_{\alpha\beta}P_{(k)}^{\kappa\lambda}{}_{\ast\kappa\lambda}^{(n)}P^{(k)}_{\sigma\rho}{}^{\sigma\rho}I^{\mu\nu}{}^{(k)}_{\mu\nu}P_{(k)}^{\theta\psi}{}_{\ast\theta\psi}^{(n)}P^{(k)}_{\gamma\delta}\frac{J_e^\gamma n^\delta-J_e^\delta n^\gamma}{2n\cdot k}~.\label{e.JJMagLoop}
\end{align}
At this point we emphasize that, as we found in the single potential formalism, there are two ways to express the composition of the two-form maps on either side of the identity element:
\begin{align}
{}^{(k)}_{\mu\nu}P_{(k)}^{\alpha\beta}{}_{\ast\alpha\beta}^{(n)}P^{(k)}_{\sigma\rho}=&{}^{(k)}_{\mu\nu}P^{(k)}_{\ast\sigma\rho}-{}^{(k)}_{\mu\nu}P^{(n)}_{\ast\sigma\rho}~\label{e.PstarR}\\
=&-{}^{(k)}_{\ast\mu\nu}P^{(k)}_{\sigma\rho}+{}^{(n)}_{\ast\mu\nu}P^{(k)}_{\sigma\rho}~.\label{e.PstarL}
\end{align}
If we use Eq.~\eqref{e.PstarL} to couple to the left-most current of Eq.~\eqref{e.JJMagLoop} then the $n^\mu$ dependent term projects out. If we use Eq.~\eqref{e.PstarR} then we find a topological contribution to the amplitude that does \emph{not} project out! Therefore, in order to remove the topological terms that are confusing the perturbation series we must use Eq.~\eqref{e.PstarR} and separate out the ${}^{(k)}_{\mu\nu}P^{(n)}_{\ast\sigma\rho}$ term. This choice is \emph{required} because if we use Eq.~\eqref{e.PstarL} and remove the (projected out) topological term it is equivalent to using Eq.~\eqref{e.PstarR} and retaining $n^\mu$ dependent term. 

Similarly, if we use Eq.~\eqref{e.PstarR} to couple to the right-most current of Eq.~\eqref{e.JJMagLoop} then the topological term projects out. We must instead use Eq.~\eqref{e.PstarL} and remove by hand the $n^\mu$ dependent term. This ensures that we have truly eliminated all topological terms from the calculation. One then finds the one-loop result for a magnetic virtual particle
\begin{align}
\mathcal{M}_{1_K}^{JJ}=&2e^2\Pi_K\frac{J_e^\alpha n^\beta-J_e^\beta n^\alpha}{2n\cdot k}{}^{(k)}_{\alpha\beta}P^{(k)}_{\ast\sigma\rho}{}^{\sigma\rho}I^{\mu\nu}{}^{(k)}_{\ast\mu\nu}P^{(k)}_{\gamma\delta}\frac{J_e^\gamma n^\delta-J_e^\delta n^\gamma}{2n\cdot k}\nonumber\\
=&-e^2\Pi_K\frac{J_e^2}{k^2}~.
\end{align}
Thus we see that by fully removing the topological terms we find that the running due to a magnetic loop has the opposite sign to running due to an electric loop. The effect of an electric loop on magnetic-magnetic scattering follows exactly the same form
\beq
\mathcal{M}_{1_J}^{KK}=-b^2\Pi_J\frac{K_e^2}{k^2}~.
\eeq

These results also connect to the phenomenological calculations related to perturbatively charged magnetic particles in~\cite{Terning:2020dzg}. In that work it is shown that the spurious $1/(n\cdot k)$ poles that appear in tree-level electric-magnetic scattering are always cancelled when $n^\mu$ is taken to align with the tube of magnetic flux that joins the monopole-antimonopole pair. In the case of a virtual  monopole loop, however, it is not clear what value to assign to $n^\mu$ as there are no on-shell monopoles with fixed momenta. The above calculation shows that the correct monopole loop corrections to electric-electric scattering are independent of $n^\mu$, obviating the problem of deciding what direction to choose. 

\begin{figure}
\center
\begin{fmffile}{JloopFAFBJK}
\begin{fmfgraph*}(300,70)
\fmfpen{1.0}
\fmfstraight
\fmfset{arrow_len}{3mm}
\fmfleft{p1,p2,p3,p4,p5} \fmfright{p6,p7,p8,p9,p10}
\fmf{phantom,tension=1}{p2,p11,p7}
\fmf{phantom,tension=1}{p4,p12,p9}
\fmf{phantom,tension=1}{p3,v5,p8}
\fmffreeze
\fmf{phantom,tension=3.0}{p3,v3}
\fmf{phantom,tension=1.5}{v4,p8}
\fmf{phantom,tension=3.0}{p11,v1}
\fmf{phantom,tension=2.5}{p12,v2}
\fmfv{decor.shape=circle,decor.filled=0,decor.size=75,fore=(0,,0,,0.9),background=(0.675,,0.675,,0.675)}{p3}
\fmfv{decor.shape=circle,decor.filled=0,decor.size=75,fore=(0.9,,0,,0),background=(0.675,,0.675,,0.675)}{p8}
\fmfv{decor.shape=circle,decor.filled=0,decor.size=60,l=${}^{\alpha\beta}\Pi_J^{\gamma\delta}$,l.a=0,l.d=-15,fore=(0,,0,,0.9),background=(0.675,,0.675,,0.675)}{v5}
\fmf{photon,rubout=0.5,tension=0.05,label=$\langle F_{\mu\nu}^A F_{\alpha\beta}^A\rangle_k$,label.side=left,fore=(0,,0,,0.9)}{v3,v1}
\fmf{photon,rubout=0.5,tension=0.3,label=$\langle F_{\gamma\delta}^A F_{\sigma\rho}^B\rangle_k$,label.side=left,fore=(0.59,,0.44,,0.84)}{v2,v4}
\fmfv{decor.shape=circle,decor.filled=full,decor.size=1.5thick,fore=(0,,0,,0.9)}{v1} 
\fmfv{decor.shape=circle,decor.filled=full,decor.size=1.5thick,fore=(0,,0,,0.9)}{v2} 
\fmfv{decor.shape=circle,decor.filled=full,decor.size=1.5thick,l=$e\Sigma_J^{\mu\nu}$,fore=(0,,0,,0.9)}{v3} 
\fmfv{decor.shape=circle,decor.filled=full,decor.size=1.5thick,l=$b\Sigma_K^{\sigma\rho}$,fore=(0.9,,0,,0)}{v4} 
\end{fmfgraph*}
\end{fmffile}
\caption{\label{f.JloopJKFAFB}One electric (blue) loop correction to photon exchange between an electric current and a magnetic (red) current using $F^A_{\mu\nu}$ and $F^B_{\mu\nu}$ propagators. The electric-electric propagator is colored blue, while the mixed propagator is colored purple.}
\end{figure}
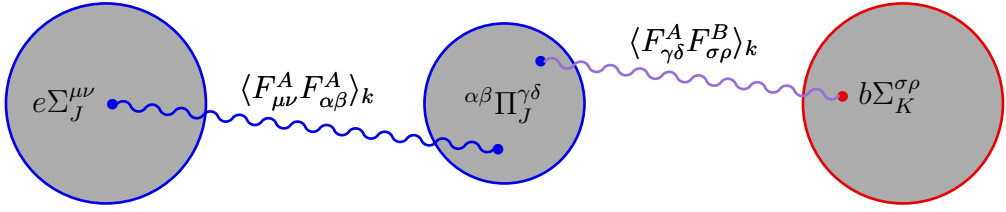

We now consider the one-loop corrections to electric-magnetic scattering. Beginning with a loop of electric particles, as in Fig.~\ref{f.JloopJKFAFB}, we find
\begin{align}
i\mathcal{M}^{JK}_{1_J}=&ie\frac{J_e^\alpha n^\beta-J_e^\beta n^\alpha}{2n\cdot k}(-2i){}^{(k)}_{\alpha\beta}P^{(k)}_{\sigma\rho}{}^{\;\sigma\rho}_{(n)}P_{(k)}^{\xi\zeta}\frac{i\Pi_J}{2}{}_{\xi\zeta}I_{\phi\tau} {}^{\;\phi\tau}_{(k)}P_{(n)}^{\mu\nu}(-2i){}^{(k)}_{\mu\nu}P_{(k)}^{\theta\psi}{}_{\ast\theta\psi}^{(n)}P^{(k)}_{\gamma\delta}ib\frac{K_e^\gamma n^\delta-K_e^\delta n^\gamma}{2n\cdot k}\nonumber\\
=&2ieb\Pi_J\frac{J_e^\alpha n^\beta-J_e^\beta n^\alpha}{2n\cdot k}{}^{(k)}_{\alpha\beta}P^{(k)}_{\sigma\rho}{}^{\sigma\rho}I^{\mu\nu}{}^{(k)}_{\mu\nu}P_{(k)}^{\theta\psi}{}_{\ast\theta\psi}^{(n)}P^{(k)}_{\gamma\delta}\frac{K_e^\gamma n^\delta-K_e^\delta n^\gamma}{2n\cdot k}~.\label{e.JKelLoop}
\end{align}
For the mixed propagator on the right side of the identity we must use Eq.~\eqref{e.PstarL} and remove the topological term. This leads to
\begin{align}
\mathcal{M}^{JK}_{1_J}=&-2eb\Pi_J\frac{J_e^\alpha n^\beta-J_e^\beta n^\alpha}{2n\cdot k}{}^{(k)}_{\alpha\beta}P^{(k)}_{\sigma\rho}{}^{\sigma\rho}I^{\mu\nu}{}^{(k)}_{\ast\mu\nu}P^{(k)}_{\gamma\delta}\frac{K_e^\gamma n^\delta-K_e^\delta n^\gamma}{2n\cdot k}\nonumber\\
=&0~.
\end{align}
The effect of a magnetic loop follows a similar pattern and also leads to no renormalization, $\mathcal{M}^{JK}_{1_K}=0$. 

In short, we find that after the topological terms have been removed
\begin{align}
\mathcal{M}^{JJ}_0+\mathcal{M}^{JJ}_{1_J}+\mathcal{M}^{JJ}_{1_K}=&e^2\frac{J_e^2}{k^2}\left[1+\Pi_J(k^2)-\Pi_K(k^2) \right]~,\\
\mathcal{M}^{KK}_0+\mathcal{M}^{KK}_{1_J}+\mathcal{M}^{KK}_{1_K}=&b^2\frac{K_e^2}{k^2}\left[1-\Pi_J(k^2)+\Pi_K(k^2) \right]~,\\
\mathcal{M}^{JK}_0+\mathcal{M}^{JK}_{1_J}+\mathcal{M}^{JK}_{1_K}=&eb\frac{\varepsilon_{\alpha\beta\gamma\delta}J_e^\alpha K_e^\beta n^\gamma k^\delta}{k^2(n\cdot k)}~.
\end{align}
Again, this shows that electric and magnetic couplings run inversely once the dynamical calculations are free from the topological phase terms. That is,
\beq
e=e_0\sqrt{Z}~, \ \ \ \ b=\frac{b_0}{\sqrt{Z}}~,
\eeq
with 
\beq
Z(k^2)\approx1/\left[1-\Pi_J(k^2)+\Pi_K(k^2) \right]~.
\eeq

This form of the running is also obtained in Seiberg-Witten theory \cite{SeibergWitten} in the region of moduli space where there are light magnetic charges \cite{Argyres:1995jj,Csaki:2010rv,Colwell:2015wna}.
In the case of only one type of light charge (either electric or magnetic) the theory is completely local and at weak coupling the perturbative running can be verified by comparing with the exact solution in the original duality frame or in the $S$-dual frame \cite{Colwell:2015wna}.

\subsection{Calculating with $F_{\mu\nu}$}
One can also derive these results using the actual field strength $F_{\mu\nu}$ as opposed to the $n^\mu$ dependent $F^A_{\mu\nu}$ and $F^B_{\mu\nu}$. We first use the fact that the source terms lie completely within the the two-form subspace spanned by ${}^{(n)}_{\;\mu\nu}P^{(n)}_{\sigma\rho}$. That is,
\beq
\frac{J^\mu n^\nu-J^\nu n^\mu}{2n\cdot k}{}^{(n)}_{\;\mu\nu}P^{(n)}_{\sigma\rho}=\frac{J_\sigma n_\rho-J_\rho n_\sigma}{2n\cdot k}~,
\eeq
and similarly for the magnetic currents. This allows us, by using the identities in Eq.~\eqref{e.FtoFaOrb}, to obtain
\begin{align}
 F^A_{\mu\nu}\frac{J^\mu n^\nu-J^\nu n^\mu}{2n\cdot k}=&F^A_{\mu\nu}{}_{(n)}^{\;\mu\nu}P_{(n)}^{\sigma\rho}\frac{J_\sigma n_\rho-J_\rho n_\sigma}{2n\cdot k}=F_{\mu\nu}{}_{(n)}^{\;\mu\nu}P_{(n)}^{\sigma\rho}\frac{J_\sigma n_\rho-J_\rho n_\sigma}{2n\cdot k}\nonumber\\
=& F_{\mu\nu}\frac{J^\mu n^\nu-J^\nu n^\mu}{2n\cdot k}~.
\end{align}
By a nearly identical calculation we also find
\beq
 F^B_{\mu\nu}\frac{K^\mu n^\nu-K^\nu n^\mu}{2n\cdot k}=  {}^\ast\!F_{\mu\nu}\frac{K^\mu n^\nu-K^\nu n^\mu}{2n\cdot k}~.
\eeq
That is, both $F^{A}_{\mu\nu}$ and $F_{\mu\nu}$ couple to the same electric source and both $F^{B}_{\mu\nu}$ and ${}^\ast\!F_{\mu\nu}$ couple to the same magnetic source. 

By using the definition of $F_{\mu\nu}$ in terms of $F^A_{\mu\nu}$ and $F^B_{\mu\nu}$ in Eq.~\eqref{e.FdefN} one can also determine the field strength propagator. This ends up being a quite simple result
\begin{align}
\langle F_{\mu\nu}F_{\sigma\rho}\rangle_k=&-2i{}_{(k)}^{\alpha\beta}P_{(k)}^{\gamma\delta}\left({}^{(n)}_{\mu\nu}P^{(n)}_{\alpha\beta}{}^{(n)}_{\gamma\delta}P^{(n)}_{\sigma\rho}+{}^{(n)}_{\ast\mu\nu}P^{(n)}_{\alpha\beta}{}^{(n)}_{\gamma\delta}P^{(n)}_{\ast\sigma\rho}\right)\nonumber\\
&+\langle F^{A\alpha\beta}F^{B\gamma\delta} \rangle_k\left({}^{(n)}_{\ast\mu\nu}P^{(n)}_{\alpha\beta}{}^{(n)}_{\gamma\delta}P^{(n)}_{\sigma\rho}-{}^{(n)}_{\mu\nu}P^{(n)}_{\alpha\beta}{}^{(n)}_{\gamma\delta}P^{(n)}_{\ast\sigma\rho} \right)\nonumber\\
=&-2i\left({}^{(n)}_{\mu\nu}P^{(n)}_{\alpha\beta}{}_{(k)}^{\alpha\beta}P_{(k)}^{\gamma\delta}{}^{(n)}_{\gamma\delta}P^{(n)}_{\sigma\rho} +{}^{(n)}_{\ast\mu\nu}P^{(n)}_{\alpha\beta}{}_{(k)}^{\alpha\beta}P_{(k)}^{\gamma\delta}{}^{(n)}_{\gamma\delta}P^{(n)}_{\ast\sigma\rho} +{}^{(n)}_{\ast\mu\nu}P^{(n)}_{\alpha\beta}{}_{(k)}^{\alpha\beta}P_{(k)}^{\ast\gamma\delta}{}^{(n)}_{\gamma\delta}P^{(n)}_{\sigma\rho}\right.\nonumber\\
&\left. -{}^{(n)}_{\mu\nu}P^{(n)}_{\alpha\beta}{}_{(k)}^{\alpha\beta}P_{(k)}^{\ast\gamma\delta}{}^{(n)}_{\gamma\delta}P^{(n)}_{\ast\sigma\rho} -{}^{(n)}_{\ast\mu\nu}P^{(n)}_{\alpha\beta}{}_{(k)}^{\alpha\beta}P_{(n)}^{\ast\gamma\delta}{}^{(n)}_{\gamma\delta}P^{(n)}_{\sigma\rho} +{}^{(n)}_{\mu\nu}P^{(n)}_{\alpha\beta}{}_{(k)}^{\alpha\beta}P_{(n)}^{\ast\gamma\delta}{}^{(n)}_{\gamma\delta}P^{(n)}_{\sigma\rho} \right) \nonumber\\
=&-2i{}^{(k)}_{\ast\mu\nu}P^{(k)}_{\ast\sigma\rho}-2i{}^{(n)}_{\mu\nu}P^{(n)}_{\sigma\rho}\\
=&-2i{}^{(k)}_{\mu\nu}P^{(k)}_{\sigma\rho}-2i{}^{(n)}_{\ast\mu\nu}P^{(n)}_{\ast\sigma\rho}~.\label{e.FFcorr}
\end{align}
In the second equality, the last two terms of the second line vanish because we are contracting indices that project into the $n^\mu$ subspace and those that project into the orthogonal space. Note that this shows that the topological part of the mixed propagator is projected out and plays no role in this computation. At the same time, the field strength propagator does depend on $n^\mu$. But, unlike the topological terms, there is no mixing of the $k^\mu$ terms and the $n^\mu$ terms. 

Correlators involving ${}^\ast\!F_{\mu\nu}$ can be obtained from Eq.~\eqref{e.FFcorr} by contracting with Levi-Civita tensors. Note that the final two equalities immediately imply that
\beq
\langle {^\ast\!}F_{\mu\nu}{}^\ast\!F_{\sigma\rho}\rangle_k=\langle F_{\mu\nu}F_{\sigma\rho}\rangle_k~.
\eeq
This underscores the fact that Zwanziger's formalism treats electric and magnetic currents in a duality covariant way, in stark contrast to the single potential formalism. One also finds that
\begin{align}
\langle {}^\ast\!F_{\mu\nu}F_{\sigma\rho}\rangle_k=-\langle F_{\mu\nu}{}^\ast\!F_{\sigma\rho}\rangle_k=&-2i{}^{(k)}_{\ast\mu\nu}P^{(k)}_{\sigma\rho}+2i{}^{(n)}_{\mu\nu}P^{(n)}_{\ast\sigma\rho}\nonumber\\
=&2i{}^{(k)}_{\mu\nu}P^{(k)}_{\ast\sigma\rho}-2i{}^{(n)}_{\ast\mu\nu}P^{(n)}_{\sigma\rho}~.
\end{align}

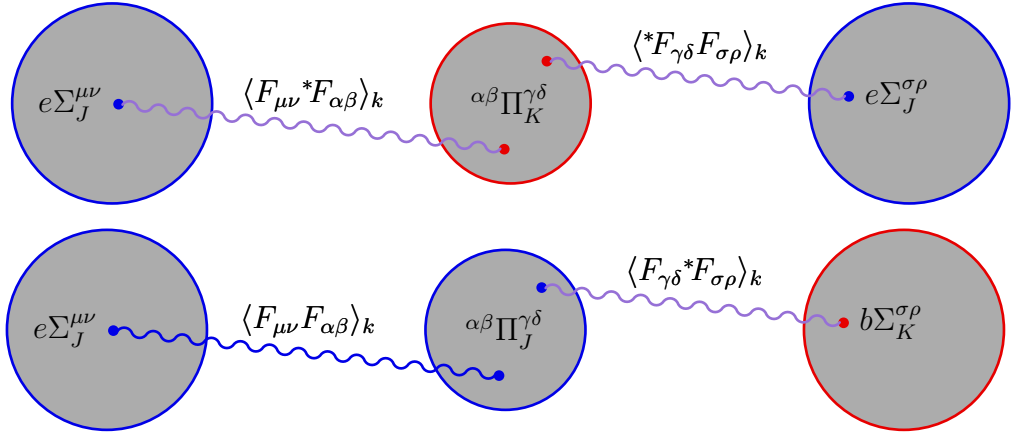
\begin{figure}
\center
\begin{fmffile}{KloopFFJJ}
\begin{fmfgraph*}(300,70)
\fmfpen{1.0}
\fmfstraight
\fmfset{arrow_len}{3mm}
\fmfleft{p1,p2,p3,p4,p5} \fmfright{p6,p7,p8,p9,p10}
\fmf{phantom,tension=1}{p2,p11,p7}
\fmf{phantom,tension=1}{p4,p12,p9}
\fmf{phantom,tension=1}{p3,v5,p8}
\fmffreeze
\fmf{phantom,tension=3.0}{p3,v3}
\fmf{phantom,tension=1.5}{v4,p8}
\fmf{phantom,tension=3.0}{p11,v1}
\fmf{phantom,tension=2.5}{p12,v2}
\fmfv{decor.shape=circle,decor.filled=0,decor.size=75,fore=(0,,0,,0.9),background=(0.675,,0.675,,0.675)}{p3}
\fmfv{decor.shape=circle,decor.filled=0,decor.size=75,fore=(0,,0,,0.9),background=(0.675,,0.675,,0.675)}{p8}
\fmfv{decor.shape=circle,decor.filled=0,decor.size=60,l=${}^{\alpha\beta}\Pi_K^{\gamma\delta}$,l.a=0,l.d=-15,fore=(0.9,,0,,0),background=(0.675,,0.675,,0.675)}{v5}
\fmf{photon,rubout=0.5,tension=0.05,label=$\langle F_{\mu\nu} {}^\ast\!F_{\alpha\beta}\rangle_k$,label.side=left,fore=(0.59,,0.44,,0.84)}{v3,v1}
\fmf{photon,rubout=0.5,tension=0.3,label=$\langle {}^\ast\!F_{\gamma\delta} F_{\sigma\rho}\rangle_k$,label.side=left,fore=(0.59,,0.44,,0.84)}{v2,v4}
\fmfv{decor.shape=circle,decor.filled=full,decor.size=1.5thick,fore=(0.9,,0,,0)}{v1} 
\fmfv{decor.shape=circle,decor.filled=full,decor.size=1.5thick,fore=(0.9,,0,,0)}{v2} 
\fmfv{decor.shape=circle,decor.filled=full,decor.size=1.5thick,l=$e\Sigma_J^{\mu\nu}$,fore=(0,,0,,0.9)}{v3} 
\fmfv{decor.shape=circle,decor.filled=full,decor.size=1.5thick,l=$e\Sigma_J^{\sigma\rho}$,fore=(0,,0,,0.9)}{v4} 
\end{fmfgraph*}
\end{fmffile}\\
\vspace{0.5cm}
\begin{fmffile}{JloopFFJK}
\begin{fmfgraph*}(300,70)
\fmfpen{1.0}
\fmfstraight
\fmfset{arrow_len}{3mm}
\fmfleft{p1,p2,p3,p4,p5} \fmfright{p6,p7,p8,p9,p10}
\fmf{phantom,tension=1}{p2,p11,p7}
\fmf{phantom,tension=1}{p4,p12,p9}
\fmf{phantom,tension=1}{p3,v5,p8}
\fmffreeze
\fmf{phantom,tension=3.0}{p3,v3}
\fmf{phantom,tension=1.5}{v4,p8}
\fmf{phantom,tension=3.0}{p11,v1}
\fmf{phantom,tension=2.5}{p12,v2}
\fmfv{decor.shape=circle,decor.filled=0,decor.size=75,fore=(0,,0,,0.9),background=(0.675,,0.675,,0.675)}{p3}
\fmfv{decor.shape=circle,decor.filled=0,decor.size=75,fore=(0.9,,0,,0),background=(0.675,,0.675,,0.675)}{p8}
\fmfv{decor.shape=circle,decor.filled=0,decor.size=60,l=${}^{\alpha\beta}\Pi_J^{\gamma\delta}$,l.a=0,l.d=-15,fore=(0,,0,,0.9),background=(0.675,,0.675,,0.675)}{v5}
\fmf{photon,rubout=0.5,tension=0.05,label=$\langle F_{\mu\nu} F_{\alpha\beta}\rangle_k$,label.side=left,fore=(0,,0,,0.9)}{v3,v1}
\fmf{photon,rubout=0.5,tension=0.3,label=$\langle F_{\gamma\delta} {}^\ast\!F_{\sigma\rho}\rangle_k$,label.side=left,fore=(0.59,,0.44,,0.84)}{v2,v4}
\fmfv{decor.shape=circle,decor.filled=full,decor.size=1.5thick,fore=(0,,0,,0.9)}{v1} 
\fmfv{decor.shape=circle,decor.filled=full,decor.size=1.5thick,fore=(0,,0,,0.9)}{v2} 
\fmfv{decor.shape=circle,decor.filled=full,decor.size=1.5thick,l=$e\Sigma_J^{\mu\nu}$,fore=(0,,0,,0.9)}{v3} 
\fmfv{decor.shape=circle,decor.filled=full,decor.size=1.5thick,l=$b\Sigma_K^{\sigma\rho}$,fore=(0.9,,0,,0)}{v4} 
\end{fmfgraph*}
\end{fmffile}
\caption{\label{f.loopFF}One loop corrections to photon exchange between currents using only $F_{\mu\nu}$ propagators within the Zwanziger formalism. Electric (magnetic) currents and interactions are colored blue (red). Mixed propagators are purple. }
\end{figure}

When contracted with external current Stokes surfaces of the form $J^\mu n^\nu-J^\nu n^\mu$ the terms in the propagators with a star on the $n$ indices project out. This leads to only the photon momentum, $k^\mu$, dependent parts of the propagators contributing and leading to the usual results for electric and magnetic interactions. The one-loop corrections also take the same form, as shown in Fig.~\ref{f.loopFF}. For electric-magnetic interactions the vacuum polarization function combines with the propagator, revealing the topological terms that do not project out. When these terms are removed we find the correct running coupling. In this way, one can see all of the tree-level and one-loop results as depending on only the propagator of the physical field strength, $F_{\mu\nu}$.

\section{Bounding Magnetically Charged Particles From Fine-Structure Running\label{s.fineBound}}
Now that the perturbative contribution of magnetic particles to electric running is understood, we can use the measured running of $e$ to probe the charge and mass of magnetic states. The most precise theory prediction of the running fine structure parameter is $\alpha(m_Z)^{-1}=128.951\pm0.045$~\cite{Burkhardt:2011ur}. The most precise measurement of $\alpha(m_Z)$ can be found from~\cite{L3:2005tsb}. From their figures we extract that at 68\% confidence
\beq
\alpha(m_Z)^{-1}\in \left[127.3,\,129.8 \right]~.
\eeq

Recall that the one-loop contribution of a Dirac fermion with electric charge $q$ to the running of $e$ comes from the coefficient of the logarithmically divergent part of the loop function
\beq
\frac{q^2e^2}{12\pi^2}~.
\eeq
For a scalar of charge $q$ this result is simply divided by 4.
This enters into the beta function for $e$ as
\beq
Q\frac{de}{dQ}=e\frac{q^2e^2}{12\pi^2}~,
\eeq
where $Q$ is the energy scale. Note that this can be written in terms of $\alpha$ as
\beq
Q\frac{d\alpha}{dQ}=\frac{2q^2\alpha^2}{3\pi} \ \ \ \ \text{ or } \ \ \ \ Q\frac{d\alpha^{-1}}{dQ}=-\frac{2q^2}{3\pi} ~.
\eeq

To obtain the effect of a particle with magnetic charge $g$ we use the same loop function formula with $+\to-$ and $eq\to4\pi g/e$. Thus, the coefficient of the logarithmically divergent part of the magnetic loop is
\beq
-\frac{4g^2}{3e^2}~,
\eeq
and the beta function is
\beq
Q\frac{de}{dQ}=-e\frac{4g^2}{3e^2}~.
\eeq
This can also be written in terms of $\alpha$. We find
\beq
Q\frac{d\alpha}{dQ}=-\frac{2g^2}{3\pi} \ \ \ \ \text{ and } \ \ \ \ Q\frac{d\alpha^{-1}}{dQ}=\frac{2g^2\alpha^{-2}}{3\pi} ~.
\eeq
Clearly, the magnetic charge contributes to the running of $\alpha^{-1}$ exactly the way an electric charge contributes to the running of $\alpha$.

\begin{figure}
\center
\includegraphics[width=0.7\textwidth]{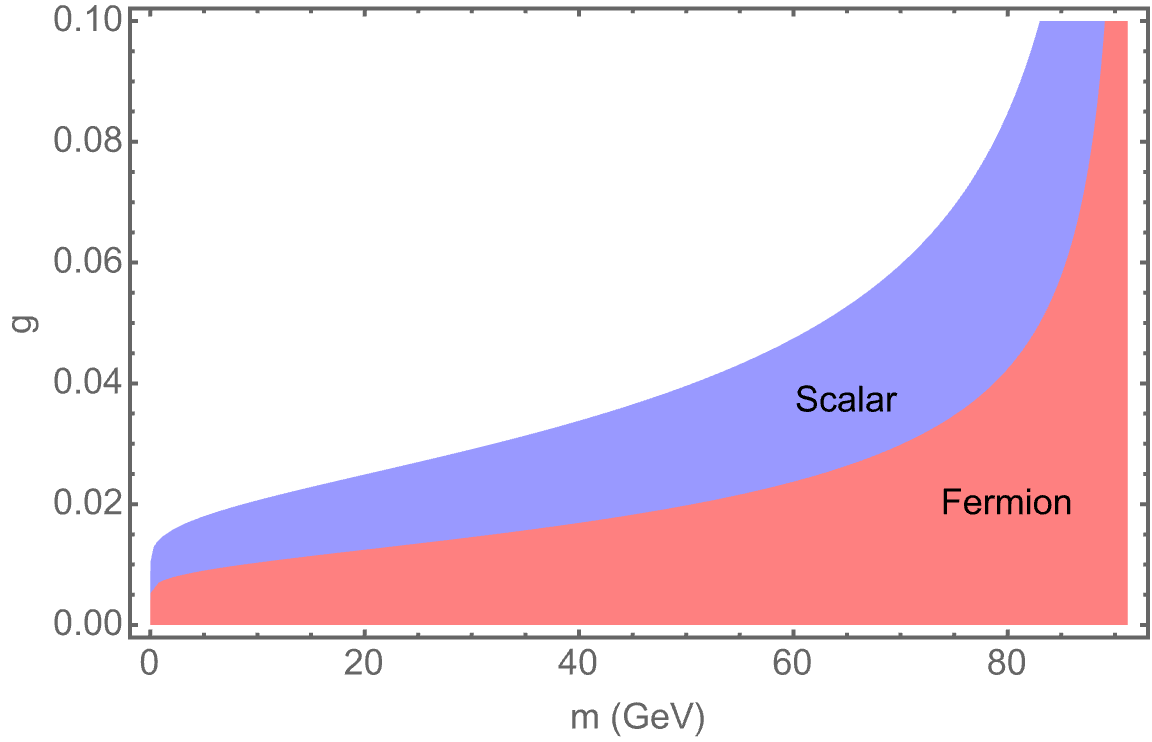}
\caption{\label{f.magRun} Region of magnetic charge and mass allowed by measurements of the running of $e$ for both scalar (blue) and fermionic (red) magnetic particles.}
\end{figure}

The one-loop contribution to the value of $\alpha(m_Z)$ from a magnetic fermion with charge $g$ and mass $m$ is simply
\beq
\alpha(m_Z)-\alpha(m)=-\frac{2g^2}{3\pi}\ln\frac{m_Z}{m}~.
\eeq
To agree with the experimental limits we require this modification of $\alpha(m_Z)$ to be less that $5\times10^{-5}$. This then bounds the possible charges and masses of the magnetic states. The allowed regions of parameter space are shown in Fig.~\ref{f.magRun} for both scalar (blue) and fermionic (red) magnetic monopoles. These bounds only apply to cases in which the magnetic charge is perturbative, such as what can result from kinetic mixing. The bounds we find are significantly weaker than those which have been already discussed in the literature~\cite{Hook:2017vyc,Terning:2019bhg,Graesser:2021vkr} but show how understanding the effects of virtual magnetic particles on various measured quantities allows for definite physical predictions.

\section{Conclusion\label{s.con}}
In this work we have outlined a momentum space formalism for tracking the photon's propagating degrees of freedom using field strengths, rather than vector potentials. This method is algebraically equivalent to the usual potential formalism, but provides a useful geometric picture where the field strength couples to a Stokes surface that has a worldline associated with the conserved current for its boundary. These surfaces can be chosen to have a simple implementation in momentum space, but then in position space correspond to an infinite surface along a direction $n^\mu$ that closes only at infinity.

This method reproduces all the standard results of QED when restricted to calculations with only electric charges. When perturbative magnetic charges are included (QEMD), both field strength and potential methods produce the same results. These results naively show that electric and magnetic couplings are renormalized in exactly the same way, as Schwinger thought. However, both methods also generate spurious ``contact'' terms between the currents, signaling a subtle error.

The field strength formalism allows us to carefully remove topological terms that appear in the perturbation series. Following this procedure we show that, when the topological terms are removed for resummation into an overall phase, the current-current ``contact'' terms are not generated. In addition, we find that the electric and magnetic couplings are renormalized exactly inversely, in agreement with Coleman, and the renormalization group invariance of the Dirac charge quantization.
 
 These results confirm that the standard method, using a vector potential to calculate the running of the coupling, is dangerously confused by these topological terms. Our field strength--momentum space formalism makes clear how these topological terms affect the results and how to remove them. It also provides a procedure for calculating the effects of magnetic monopole loops, when this is justified by perturbation theory. As an example we have obtained bounds on the mass and charge of ``dark'' magnetic particles from the measured running of the electric coupling.
 
It will be interesting to extend this analysis to the running of the CP violating $\theta$ parameter, which becomes physical in Abelian theories with both electric and magnetic charges. This formalism should also provide some insight into exactly how kinetic mixing with a ``dark'' magnetic sector can be induced by loops of heavy particles.

\section*{Acknowledgments}
We thank Katherine Fraser, Howie Haber, Mark Hughes, Seth Koren, Markus Luty, Michael Peskin, and Riccardo Rattazzi for enlightening discussions. J.T.  is supported in part by the DOE under grant DE-SC-0009999. C.B.V. is supported in part by the National Science Foundation under Grant No. PHY-2210067~.

\appendix


\section{The Field Strength Propagators\label{a.Fprop}}
In this section we derive the propagators for the field strength in single potential formulations of QED and for $F^A_{\mu\nu}$ and $F^B_{\mu\nu}$ in Zwanziger's formalism. We begin by deriving the propagator for $F_{\mu\nu}$ in single potential theories.  First, we employ the canonical formalism and then we obtain the same result from the path integral. Then we derive the propagators in the two potential formalism using the path integral.

\subsection{Single Potential: Canonical Formalism}
In single potential QED, we relate the time ordered product of field strengths to derivatives of time ordered products of potentials. From the Schwinger Dyson equations~\cite{Schwartz:2014sze} we find
\begin{align}
    \partial_0 \partial_0' \langle A_\mu (x) A_\nu (x') \rangle &=
    \langle \partial_0 A_\mu(x) \partial_0'A (x')_\nu \rangle
    + \delta(t-t') \langle 0 |[\partial_0 A_\mu(x), A_\nu(x')] |0\rangle~, \label{eq:2} \notag\\
    & - \delta'(t-t') \langle 0 | [A_\mu(t,\textbf{x}), A_\nu(t,\textbf{x}')] |0\rangle~, \notag \\
    \partial_0 \partial_i' \langle A_\mu(x) A_\nu(x') \rangle 
    &=
    \langle \partial_t A_\mu (x) \partial_i' A_\nu(x') \rangle
    + \delta(t-t') \partial_i' \langle 0 | [A_\mu(x), A_\nu(x')] |0\rangle~, \notag\\
    \partial_i \partial_0' \langle A_\mu(x) A_\nu(x') \rangle 
    &=
    \langle \partial_i A_\mu (x) \partial_{t'} A_\nu(x') \rangle
    - \delta(t-t') \partial_i \langle 0 | [A_\mu(x), A_\nu(x')] |0\rangle~, \notag\\
    \partial_i \partial_j' \langle A_\mu (x) A_\nu (x') \rangle &=
    \langle \partial_i A_\mu(x) \partial_j' A_\nu(x') \rangle~.
\end{align}
Here we are using the notation $\langle X Y\rangle \equiv \langle 0 | T\{ X Y \} | 0 \rangle$ to denote time-ordered, vacuum-to-vacuum amplitudes. In the standard $R_\xi$ gauges we have the following commutators for the vector potential and its time derivatives~\cite{Greiner:1996}:
\begin{align}
   \left [A_\mu(x), A_\nu(x')\right] &= 0~,\\
    [\partial_t A_\mu(t,\textbf{x}'), A_\nu(t,\textbf{x})]
    &= i \bigg(\eta_{\mu\nu}
    - \frac{\zeta - 1}{\zeta}
    \delta^0_\mu \delta^0_\nu
    \bigg) \delta^3(\textbf{x} - \textbf{x}')~,
\end{align}
where $\eta_{\mu\nu}$ is the Minkowski metric. From these commutators we compute the time-ordered, vacuum-to-vacuum correlation function of the field strength
\begin{align}
    &\langle F_{\mu\nu}(x) F_{\sigma\rho}(x') \rangle
    = \bigg( \partial_\mu \partial_\sigma'
    \langle A_{\nu}(x) A_{\rho}(x')
    \rangle\bigg)_{[\mu\nu][\sigma\rho]}
    -\bigg(\delta(t-t') 
    \delta^0_\mu \delta^0_\sigma
    \langle 0 | 
    [\partial_0 A_\nu(x),A_\rho(x') ]
    |0\rangle
    \bigg)_{[\mu\nu][\sigma\rho]} \notag\\
    &+\bigg(
    \delta^0_\mu \delta^0_\sigma
    \delta'(t-t')
    \langle 0 |
    [A_\nu(x),A_\rho(x')]
    |0\rangle
    \bigg)_{[\mu\nu][\sigma\rho]}
    + \bigg(
    \delta(t-t') 
    \delta^i_\mu \delta^0_\sigma
    \partial_i
    \langle 0 |
    [A_\nu(x),A_\rho(x')]
    |0\rangle
    \bigg)_{[\mu\nu][\sigma\rho]} \notag\\
    &-\bigg(
    \delta^0_\mu \delta^j_\sigma
    \delta(t-t')
    \partial_j'
    \langle 0 |
    [A_\nu(x),A_\rho(x')]
    |0\rangle
    \bigg)_{[\mu\nu][\sigma\rho]} \label{eq:4}
\end{align}
The notation $( \_ )_{[\_]...[\_]}$ means to antisymmetrize the parenthetical quantity across the bracketed indices.
In simplifying this result we find that all gauge dependent terms cancel
\begin{equation}
    \langle F_{\mu\nu}(x) F_{\sigma\rho}(x') \rangle
    = -\int \frac{d^4p}{(2\pi)^4} \frac{i}{p^2 +i\epsilon} p_{[\mu} \eta_{\nu][\rho} p_{\sigma]} e^{ip(x-x')}
    - i\delta_{[\mu}^0 \eta_{\nu][\rho} \delta^{0}_{\sigma]} \delta^4(x-x') ~,
    \label{eq:5}
\end{equation}
but also the appearance of a noncovariant contact term. 

How does this term affect the Feynman rules of this theory? In order to use the $\langle F_{\mu\nu} F_{\sigma\rho}\rangle$ correlator in the perturbation expansion we need to directly source the field strength. We take $\Sigma^{\mu\nu}$ to be whatever function of the fields couples to $F_{\mu\nu}$.
By the antisymmetry of $F_{\mu\nu}$ we see that only the antisymmetric part of $\Sigma^{\mu\nu}$ couples to $F$ thus take $\Sigma^{\mu\nu}$ to be antisymmetric in its indices.
Starting with the Lagrangian 
\begin{align}
    \mathcal{L}
    &= -\frac{1}{4} F^{\mu\nu}F_{\mu\nu}
    - F_{\mu\nu}\Sigma^{\mu\nu}~,
\end{align}
we find the canonical momenta are
\begin{align}
    \Pi^\mu
    &=
    \frac{\partial\mathcal{L}}
    {\partial(\partial_0 A_\mu)}
    = -F^{0\mu} - 2 \Sigma^{0\mu}~.
\end{align}
From the antisymmetry of $F_{\mu\nu}$ and $\Sigma_{\mu\nu}$ we find that $\Pi^0 = 0$; this is our first constraint.
The Hamiltonian is given by
\begin{align}
    \mathcal{H}
    &= 
    \Dot{A}_\mu    \Pi^\mu - \mathcal{L}
    \notag\\
    &=
    \frac{1}{2}
    \|\Pi\|^2
    + \Pi \cdot \nabla A^0
    + \frac{1}{4} F_{ij} F^{ij}
    + 2 \Pi^i \Sigma^{0i}
    + 2 \Sigma^{0i} \Sigma^{0i}
    + F_{ij}\Sigma^{ij}~.
\end{align}

The secondary constraint is found by demanding that the primary constraint does not evolve in time,
\begin{align}
    0 = \Dot{\Pi}^0
    = -\frac{\delta \mathcal{H}}{\delta A^0}
    = \nabla \cdot 
    \Pi
\end{align}
Including this in the Hamiltonian, and integrating by parts, we find
\begin{align}
    \mathcal{H}
    &=
    \frac{1}{2}
    \|\Pi\|^2
    + \frac{1}{4} F_{ij} F^{ij}
    + 2 \Pi^i \Sigma^{0i}
    + 2 \Sigma^{0i} \Sigma^{0i}
    +  F_{ij}\Sigma^{ij}~.
\end{align}
Thus, the Hamiltonian includes quadratic terms that are explicitly non-negative and independent of all current terms, this is the free Hamiltonian. The Hamiltonian's interaction terms are given by
\begin{align}
    \mathcal{H}_{\text{int}}
    &=
    2 \Pi^i \Sigma^{0i}
    + 2 \Sigma^{0i} \Sigma^{0i}
    +  F_{ij}\Sigma^{ij}~.
\end{align}
Having identified the free and interaction parts of the Hamiltonian we can move to the interaction picture.
Thus we substitute in our \emph{free field} definition for the canonical momentum ($\Pi^\mu =  -F^{0\mu}$) to find
\begin{align}
    \mathcal{H}_{\text{int}}
    =
    2\Sigma^{0i} \Sigma^{0i}
    +  \Sigma^{\mu\nu} F_{\mu\nu}
    =  - 2\Sigma_{0\nu} \Sigma^{0\nu}
    - \mathcal{L}_{\text{int}}~.
\end{align}

In the canonical picture, the Feynman rules are derived from the Dyson series which depend on the interaction \textit{Hamiltonian} not the interaction \textit{Lagrangian}. 
When one goes through the full derivation of the Feynman rules, one finds that the non-covariant terms in the interaction Hamiltonian cancel exactly with the non-covariant terms in the time ordered correlation function for $F_{\mu\nu}$. This same cancellation of noncovariant terms happens in other theories with derivative couplings \cite{Weinberg:1995mt}. 

Because the argument above was arbitrary with respect to what $\partial_{[\mu}A_{\nu]}$ couples to we see that it also holds for the single potential theory with magnetic monopoles, where the definition of $F_{\mu\nu}$ changes. The addition of the new couplings leads to new non-covariant terms in the interaction hamiltonian which exactly cancel with the terms introduced by new couplings of the non covariant terms in $\langle F_{\mu\nu}F_{\sigma\rho} \rangle $.
Thus, we find that in single potential theories that the field strength propagator, which enters into the Feynman rules, is given in momentum space as
\begin{align}
    \langle F_{\mu\nu} F_{\sigma\rho} \rangle_k
     = - 2i {}^{(k)}_{\mu\nu}P^{(k)}_{\sigma\rho}~.
\end{align}
This is precisely the form used in the body of the paper, motivated by much simpler arguments.

\subsection{Single Potential: Path Integral Formalism}
We now derive the $F_{\mu\nu}$ propagator in the path integral formalism and show that we can find it directly without having to worry about noncovariant terms.
Start with the generating functional for QED written as,
\begin{align}
    Z[J] \propto
    \int \mathcal{D}A\,
    \delta^\infty(\phi(A))
    \exp
    \left\{
    -i \int d^4x \frac{1}{4}
    F_{\mu\nu}F^{\mu\nu}
    - i\int d^4x
    A_\mu J^\mu
    \right\}
\end{align}
where $\phi = 0$ is our gauge fixing constraint and $\delta^\infty(f)$ a functional delta-function.
We can change variables from $A_\mu \to F_{\mu\nu}$ by taking the map
\begin{align}
    F_{\mu\nu} = 
    \partial_{[\mu} A_{\nu]}
    = \int d^4y \Delta_{\mu\nu,\alpha}(x-y) A^\alpha(y)~,
\end{align}
where
\begin{align}
    \Delta_{\mu\nu,\alpha}(x-y)
    = 
    \eta_{\alpha[\mu}
    \partial_{\nu]} \delta^4(x-y) ~.
\end{align}

We can consider making this change of variables in the path integral. 
The range of the map is not the space of all two forms, but only exact two forms. 
Thus, the measure of our transformed integral must include the Bianchi identity constraint through the delta-function
\begin{align}
    \delta^{\infty \times 4}(\partial_\mu {}^*F^{\mu\nu})
    \mathcal{D}F~.
\end{align}
Here $\mathcal{D}F$ is the measure on the space of all two form fields.
With our gauge fixed, this map is a bijection onto its image.
This means we can write the inverse map
\begin{align}
    A_\mu(x) =
    \int d^4y \Delta_{\mu,\alpha\beta}^{-1}(x,y;\phi) 
    F^{\alpha\beta}(y)~.
\end{align}
This is only properly defined when a particular gauge is chosen. Thus, the inclusion of the dependence on our gauge fixing constraint $\phi$.
Additionally, note that because our gauge fixing condition could be spacetime dependent $\Delta^{-1}$ may not be translation invariant therefore may not be a convolution operator;
this motivates the more general form $\Delta^{-1}(x,y;\phi)$ rather than $\Delta^{-1}(x-y;\phi)$.
Given this, the interaction term can be written
\begin{align}
    \int d^4x J^\mu(x)  A_\mu(x)
    &= 
    \int d^4 x J^\mu(x) 
    \int d^4y \Delta_{\mu,\alpha\beta}^{-1}(x,y;\phi) 
    F^{\alpha\beta}(y)\nonumber\\
     &= \int d^4y
     F^{\alpha\beta}(y)
     \Sigma_{\alpha\beta}(y)~,
     \label{a.invCoup}
\end{align}
where
\begin{align}
    \Sigma_{\alpha\beta}(y)
    \equiv \int d^4 x J^\mu(x)
     \Delta_{\mu,\alpha\beta}^{-1}(x,y;\phi)~.
\end{align}

In order to make this change of variables in the path integral we must include a factor of the functional Jacobian determinant in our path integral~\cite{Faddeev:1980be}.
The Jacobian is given by,
\begin{align}
	\det \left(\frac{\delta F_{\mu\nu}(x)}{\delta A_\gamma(y)} \right)
	= 
	\det \left(\Delta_{\gamma,\mu\nu}(x-y)\right)
	~.
\end{align}
Here we implicitly mean the determinant of the map from gauge fixed potentials to exact two forms, which cannot be zero as this map is invertible.
The determinant of the map from arbitrary gauge potentials to arbitrary two forms is certainly zero.
This determinant is field independent so it can be brought outside of the path integral and only leads to a change in normalization.
This simple result follows from the linear change of variables and would not hold for a nonlinear transformation. 
For example, the change of variables from potentials to field strengths would likely be significantly more complicated in a nonabelian gauge theory.

Given that $\Delta$ is a differential operator its inverse is necessarily nonlocal. We also see that there is a redundancy in the definition of the function which couples to $F_{\mu\nu}$ which comes from the gauge redundancy of $A$. If we suppose that the current comes from a closed particle trajectory and then apply Stokes' Theorem we find a coupling that looks like
\begin{align}
    \int d^4x A_\mu(x) J^\mu_{\text{Loop}}(x)
    =
    \int d^4 x
    F_{\mu\nu}(x) \Sigma^{\mu\nu}_{\text{Surface}}(x)~. \label{a.surfCoup}
\end{align}
As is discussed in Appendix~\ref{a.Fsource}, this coupling is invariant under a change of surface which leaves the boundary unchanged.
Comparing Eqs.~\eqref{a.invCoup} and~\eqref{a.surfCoup} we see that there is a correspondence between Stokes' surface invariance and gauge invariance. It is not clear, however, if this correspondence is one-to-one; for example the equations~\eqref{a.invCoup} and~\eqref{a.surfCoup} can differ by $2\pi$ times an integer and still correspond to the same physical coupling.

Still, given a gauge fixing constraint we can find the inverse defined above which leads to a particular coupling to $F_{\mu\nu}$. For example, choosing the familiar Lorenz gauge, $\phi(A) = \partial_\mu A^\mu$, produces
\beq
 A^\nu =\frac{1}{\Box} \partial_\mu F^{\mu\nu}~,
 \eeq
and so to
\begin{align}
    \Sigma^{\mu\nu}(x)
     = \frac{1}{2}\frac{1}{\Box}
     \partial^{[\mu}J^{\nu]}~.
\end{align}
Similarly, in axial gauge, $\phi(A) = n_\mu A^\mu$, we find
\beq
 A^\nu =\frac{1}{n\cdot\partial} n_\mu F^{\mu\nu}~,
 \eeq
and
\begin{align}
    \Sigma^{\mu\nu}(x)
     = \frac{1}{2}\frac{1}{n\cdot\partial}
     n^{[\mu}J^{\nu]}~.
\end{align}
We emphasize the agreement between this expression and the Stokes' surface form derived in Appendix~\ref{a.Fsource}.

Returning to the generating functional, we see that it can be written as a function of the source $\Sigma$
\begin{align}
    Z[\Sigma] \propto
    \int \mathcal{D}F
    \delta^{\infty\times 4}(\partial_\mu {}^*F^{\mu\nu})
    \exp
    \left\{
    -i \int d^4x \frac{1}{4}
    F_{\mu\nu}F^{\mu\nu}
    - i\int d^4x
    F_{\mu\nu} \Sigma^{\mu\nu}
    \right\}~.
\end{align}
It still remains to consider how to treat the delta function that enforces the Bianchi identity.
We write this constraint as
\begin{align}
    \delta^{\infty\times4}(\partial_\mu {}^*F^{\mu\nu})
    \propto \int \mathcal{D}\Lambda
    \exp \left\{
    - i\int d^4x
    \Lambda_\nu  
    \partial_\mu
    {}^*F^{\mu\nu}
    \right\}~,
    \label{a.delta}
\end{align}
where $\Lambda_\mu(x)$ is an auxiliary field which takes the form of a Lagrange multiplier. However, we choose to enforce the Bianchi constraint using a different method, similar to that used to derive the potential propagator in the $R_\xi$ gauges of QED.
We write the delta-function in Eq.~\eqref{a.delta} as the following limit
\begin{align}
     \delta^{\infty\times4}(\partial_\mu {}^*F^{\mu\nu})
     =
     \lim_{\xi \to \infty} \int \mathcal{D}\Lambda
    \exp \left\{
    - i\int d^4x
    \left[
    \Lambda_\nu  
    \partial_\mu
    {}^*F^{\mu\nu}
    -\frac{1}{2}
    \frac{1}{\xi}
    \partial_\nu \Lambda_\mu
     \partial^\nu \Lambda^\mu
    \right]
    \right\}~,
\end{align}
After integrating out the auxiliary field we find
\begin{align}
    \delta^{\infty\times4}(\partial_\mu {}^*F^{\mu\nu})
    &\propto \lim_{\xi \to \infty}
    \exp \left\{
    -i
    \frac{\xi}{8}
    \int d^4x
    F_{\mu\nu}
    \varepsilon^{\mu\nu\alpha\beta}
    \frac{1}{\Box}
    (\partial_\alpha
    \eta_{\beta\gamma}
    \partial_\delta)
    \varepsilon^{\gamma\delta\sigma\rho}
    F_{\sigma\rho}
    \right\}~.
\end{align}
which implies the generating functional is
\begin{align}
    &Z[\Sigma] \propto\\
    &\lim_{\xi \to \infty}
    \int \mathcal{D}F
    \exp
    \left\{
    -i \int d^4x
    \frac{F_{\mu\nu}}{2}
    \left(
    \frac{
    {}^{\mu\nu}I^{\sigma\rho}}{2}
    + 
    \frac{\xi}{4}
    \varepsilon^{\mu\nu\alpha\beta}
    (
    \frac{1}{\Box}
    \partial_\alpha
    \eta_{\beta\gamma}
    \partial_\delta)
    \varepsilon^{\gamma\delta\sigma\rho}
    \right)
    F_{\sigma\rho}
    - i\int d^4x
    F_{\mu\nu} \Sigma^{\mu\nu}
    \right\}~.\nonumber
\end{align}
This shows that to find the $F_{\mu\nu}$ propagator we must invert the kinetic operator given, in momentum space, as
\begin{align}
    \Tilde{\mathcal{K}}_{\mu\nu,\sigma\rho}
    = 
    \frac{i}{2}
    {}_{\mu\nu}I_{\sigma\rho}
    + 
    i
   \frac{\xi}{2}
    {}_{*\mu\nu}^{(k)}
    P{}_{*\sigma\rho}^{(k)}
    = 
    \frac{i}{2} \left(
    {}_{\mu\nu}^{(k)}
    P{}_{\sigma\rho}^{(k)}
    -\left(
     1 -\xi 
    \right){}_{*\mu\nu}^{(k)}
    P{}_{*\sigma\rho}^{(k)}
    \right).
\end{align}

Given the fact that $-{}_{*\mu\nu}^{(k)} P{}_{*\sigma\rho}^{(k)}$ is the orthogonal projector to ${}_{\mu\nu}^{(k)} P{}_{\sigma\rho}^{(k)}$ the inverse can be quickly determined.
Therefore, we find the $F_{\mu\nu}$ propagator is given by
\begin{align}
    \langle
    F_{\mu\nu}
    F_{\sigma\rho}
    \rangle_k
    &=
    \lim_{\xi \to \infty}
    -2i \left(
    {}_{\mu\nu}^{(k)}
    P{}_{\sigma\rho}^{(k)}
    -\left(
    1 -\xi 
    \right)^{-1}{}_{*\mu\nu}^{(k)}
    P{}_{*\sigma\rho}^{(k)}
    \right)
    =
    -2i
    {}_{\mu\nu}^{(k)}
    P{}_{\sigma\rho}^{(k)}~,
\end{align}
which is exactly what we found using canonical quantization methods in the previous subsection as well as through the heuristic methods in the body of the paper.
We note that this method is very similar to the derivation of the $A_\mu$ propagator in $R_\xi$ gauge. The only difference is that after inverting the kinetic operator we must take the limit $\xi \to \infty$ rather than leave $\xi$ as an arbitrary parameter.

\subsection{Two Potential Propagators}
While it may seem an interesting problem to derive the two potential field strength propagators in the canonical formalism, the process is much more tedious. The cancellation of noncovariant terms is not so simple due to the many redundant degrees of freedom that come from describing the two propagating modes by two vector potentials.
Rather than illustrate this process we show only the path integral derivation of the field strength propagators, where the extra non-covariant contact terms do not appear. 

This follows the same method as in the previous subsection, making the change of variables from $A_\mu \to F^A_{\mu\nu}$ and $B_\mu \to F^B_{\mu\nu}$. One finds that the generating functional can be written as
\begin{multline}
    Z[\Sigma_J,\Sigma_K] \propto
    \lim_{\xi_A \to \infty}
    \lim_{\xi_B \to \infty}
    \int \mathcal{D}F^A
    \mathcal{D}F^B
    \exp
    \bigg\{
    -\frac{1}{2} \int d^4x 
    F_{\mu\nu}^X
    \mathcal{K}_{XY}^{\mu\nu,\sigma\rho}
    F_{\sigma\rho}^Y
    \\\
    - i\int d^4x
    F_{\mu\nu}^A \Sigma^{\mu\nu}_J
    - i\int d^4x
    F_{\mu\nu}^B \Sigma^{\mu\nu}_K
    \bigg\}~,
\end{multline}
here indices $X,Y\in \{A,B\}$ denote the two types of potential. Here the kinetic term is expressed, in momentum space, as
\begin{align}
\Tilde{\mathcal{K}}^{XY}_{\mu\nu,\sigma\rho}
    = 
    \frac{i}{2}
    \begin{bmatrix}
        {}^{(n)}_{\mu\nu}
        P^{(n)}_{\sigma\rho}
        +
        \xi_A
        {}_{*\mu\nu}^{(k)}P_{*\sigma\rho}^{(k)}
        &
        -{}_{*\mu\nu}^{(n)}P_{\sigma\rho}^{(n)}
        \\
        {}_{*\mu\nu}^{(n)}P_{\sigma\rho}^{(n)}
        &
        {}^{(n)}_{\mu \nu}
        P^{(n)}_{\sigma\rho}
        +
        \xi_B
        {}_{*\mu\nu}^{(k)}P_{*\sigma\rho}^{(k)}
    \end{bmatrix}~.
\end{align}

While we have avoided the tedium of tracking the noncovariant terms that arise in canonical quantization, we do have to confront the inversion of this twelve-by-twelve matrix. However, it does take the form of six-by-six blocks so we can use the standard inversion of block matrices. That is, a block matrix of the form
\beq
\text{\bf M}=\left(\begin{array}{cc}
\text{\bf A} & \text{\bf B}\\
\text{\bf C} & \text{\bf D}
\end{array} \right)~,
\eeq
has an inverse which may be expressed in two equivalent ways
\beq
\text{\bf M}^{-1}=&\left(\begin{array}{cc}
\text{\bf A}^{-1}+\text{\bf A}^{-1}\text{\bf B}\left(\text{\bf D}-\text{\bf C}\text{\bf A}^{-1}\text{\bf B} \right)^{-1}\text{\bf C}\text{\bf A}^{-1} & -\text{\bf A}^{-1}\text{\bf B}\left(\text{\bf D}-\text{\bf C}\text{\bf A}^{-1}\text{\bf B} \right)^{-1}\\
-\left(\text{\bf D}-\text{\bf C}\text{\bf A}^{-1}\text{\bf B} \right)^{-1}\text{\bf C}\text{\bf A}^{-1} &   \left(\text{\bf D}-\text{\bf C}\text{\bf A}^{-1}\text{\bf B} \right)^{-1}
\end{array} \right)~,\nonumber\\
=&\left(\begin{array}{cc}
\left(\text{\bf A}-\text{\bf B}\text{\bf D}^{-1}\text{\bf C} \right)^{-1} & -\left(\text{\bf A}-\text{\bf B}\text{\bf D}^{-1}\text{\bf C} \right)^{-1}\text{\bf B}\text{\bf D}^{-1}\\
-\text{\bf D}^{-1}\text{\bf C} \left(\text{\bf A}-\text{\bf B}\text{\bf D}^{-1}\text{\bf C} \right)^{-1} & \text{\bf D}^{-1}+\text{\bf D}^{-1}\text{\bf C}\left(\text{\bf A}-\text{\bf B}\text{\bf D}^{-1}\text{\bf C} \right)^{-1}\text{\bf B}\text{\bf D}^{-1} 
\end{array} \right)~.\label{e.BlockInverse}
\eeq

To use this formula we make great use of the analysis of two-form mappings in Sec.~\ref{s.TwoForms}. In particular, we remind the reader of the identities
\beq
 {}^{(k)}_{\mu\nu} P^{(k)}_{\alpha\beta} {}_{(k)}^{\alpha\beta} P^{(n)}_{\sigma\rho}= {}^{(k)}_{\mu\nu} P^{(n)}_{\sigma\rho}~, \ \ \ \  {}^{(k)}_{\mu\nu} P^{(k)}_{\alpha\beta} {}_{(n)}^{\alpha\beta} P^{(k)}_{\sigma\rho}= {}^{(k)}_{\mu\nu} P^{(k)}_{\sigma\rho}~,\ \ \ \  {}^{(k)}_{\mu\nu} P^{(n)}_{\alpha\beta} {}_{(k)}^{\alpha\beta} P^{(n)}_{\sigma\rho}= {}^{(k)}_{\mu\nu} P^{(n)}_{\sigma\rho}~,
\eeq
along with those obtained by exchanging $n^\mu$ and $k^\mu$. It is then straightforward to check that the inverse of the {\bf A} block
\beq
 \text{\bf A}={}^{(n)}_{\mu\nu} P^{(n)}_{\sigma\rho}+ \xi_{A}{}_{*\mu\nu}^{(k)}P_{*\sigma\rho}^{(k)}~,
\eeq
is
\beq
  \text{\bf A}^{-1}={}^{(k)}_{\mu\nu} P^{(n)}_{\alpha\beta} {}_{(n)}^{\alpha\beta} P^{(k)}_{\sigma\rho}+\frac{1}{\xi_{A}} {}^{(n)}_{\ast\mu\nu} P^{(k)}_{\alpha\beta} {}_{(k)}^{\alpha\beta} P^{(n)}_{\ast\sigma\rho}~.
\eeq
We then find that
\beq
\text{\bf D}-\text{\bf C}\text{\bf A}^{-1}\text{\bf B} = {}^{(k)}_{\mu\nu} P^{(n)}_{\sigma\rho}+ \xi_{B}{}_{*\mu\nu}^{(k)}P_{*\sigma\rho}^{(k)}~,
\eeq
which has the inverse
\beq
\left(\text{\bf D}-\text{\bf C}\text{\bf A}^{-1}\text{\bf B}  \right)^{-1}={}^{(k)}_{\mu\nu} P^{(k)}_{\sigma\rho}+ \frac{1 }{\xi_{B}}{}_{*\mu\nu}^{(n)}P_{*\sigma\rho}^{(k)}~.
\eeq

Putting all the pieces together we find that the inverse of the entire matrix is given by
\begin{align}
    (\Tilde{\mathcal{K}}^{-1})^{XY}_{\mu\nu,\sigma\rho}
     = 
    -2i
    \begin{bmatrix}
        {}^{(k)}_{\mu\nu}
        P^{(k)}_{\sigma\rho}
        +
        \frac{1}{\xi_A}
        {}_{*\mu\nu}^{(n)}P_{*\sigma\rho}^{(k)}
        &
        {}_{\mu\nu}^{(k)}P_{*\sigma\rho}^{(k)}
        -
        {}_{\mu\nu}^{(k)}P_{*\sigma\rho}^{(n)}
        -
        \frac{1}{\xi_A}
        {}_{*\mu\nu}^{(n)}P_{\sigma\rho}^{(k)}
        \\
        -{}_{\mu\nu}^{(k)}P_{*\sigma\rho}^{(k)}
        +
        {}_{\mu\nu}^{(k)}P_{*\sigma\rho}^{(n)}
        +
        \frac{1}{\xi_B}
        {}_{*\mu\nu}^{(n)}P_{\sigma\rho}^{(k)}        
        &
        {}^{(k)}_{\mu\nu}
        P^{(k)}_{\sigma\rho}
        +
        \frac{1}{\xi_B}
        {}_{*\mu\nu}^{(n)}P_{*\sigma\rho}^{(k)}
    \end{bmatrix}.
\end{align}
We then take the limit $\xi_A \to \infty$ and $\xi_B \to \infty$ to find that the propagators are given by
\begin{align}
    \begin{bmatrix}
        \langle F^A F^A \rangle_k
        &
        \langle F^A F^B \rangle_k
        \\
        \langle F^B F^A \rangle_k
        &
        \langle F^B F^B \rangle_k
    \end{bmatrix}
    =
    -2i
    \begin{bmatrix}
        {}^{(k)}_{\mu\nu}
        P^{(k)}_{\sigma\rho}
        &
        {}_{\mu\nu}^{(k)}P_{*\sigma\rho}^{(k)}
        -
        {}_{\mu\nu}^{(k)}P_{*\sigma\rho}^{(n)}
        \\
        -{}_{\mu\nu}^{(k)}P_{*\sigma\rho}^{(k)}
        +
        {}_{\mu\nu}^{(k)}P_{*\sigma\rho}^{(n)}      
        &
        {}^{(k)}_{\mu\nu}
        P^{(k)}_{\sigma\rho}
    \end{bmatrix}.
\end{align}
We again find that the simple calculation in the main text is verified by this more rigorous derivation. 
This can be considered as a consequence of the fact that it is equivalent to invert an operator and then make a linear change of variables or make the change of variables and then invert the operator.
The simple calculation corresponds to the first method, while the more in depth calculation done in this appendix corresponds to the second.


\section{The Field Strength Source\label{a.Fsource}}
In this appendix we show that the object used to source the field strengths in this paper corresponds to a specific type of Stokes surface in the worldline formalism \cite{Strassler:1992zr,Schubert:1996jj}. In this formalism we consider conserved currents to be particle trajectories in spacetime, that is
\beq
 J^\mu(x) = q\int_0^T d\tau \delta^4(x-z(\tau)) \frac{dz^\mu}{d\tau}~,
 \eeq
where $z(\tau)$ is a parametrization of our path $\gamma = z([0,T])$.
The coupling of the current to a vector potential is
\begin{equation}
    \int d^4x J^\mu(x) A_\mu(x)= q\int_0^T d\tau A_\mu(z(\tau)) \frac{dz^\mu}{d\tau} = q\int_\gamma \bm{A}~,
\end{equation}
where we use the differential form notation $\bm{A} = A_\mu dz^{\mu}$ in the last equality.
We then assume that the trajectory forms a closed loop and apply Stokes' theorem to get
\begin{align}
    q \int_\gamma \bm{A}
    &=
    q \int_S d\bm{A}
    = q\int d\tau d\eta \partial_{[\mu}A_{\nu]}
    (s(\tau,\eta)) 
    \frac{1}{2}
    \frac{\partial s^{[\mu}}{\partial\tau} 
    \frac{\partial s^{\nu]}}{\partial\eta}
    \label{a.surfElem}~.
\end{align}
Here $s^\mu(\tau,\eta)$ is a parametrization of the surface $S$ bounded by the current loop. While it is not obvious, this result is independent of the choice of surface.

Consider integrals over two surfaces $S_1$ and $S_2$ which share the same boundary loop.
Together these surfaces bound some volume $V$. By applying Stokes' theorem to the difference of these two integrals we find
\begin{align}
    \int_{S_1} d\bm{A} - \int_{S_2} d\bm{A}
    = \int_V d^2 \bm{A}
     = \int d\tau d\eta d\zeta \partial_{[\mu}\partial_{\nu} A_{\sigma]} \frac{1}{6} \frac{\partial v^{[\mu}}{\partial\tau} 
    \frac{\partial v^{\nu}}{\partial\eta} 
    \frac{\partial v^{\sigma]}}{\partial\zeta} = 0~,
\end{align}
where $v(\tau,\eta,\zeta)$ is a parametrization of the volume $V$.

This result shows that a local coupling to a potential always has a manifestly surface invariant coupling to the field strength.
In particular, Zwanziger's formalism is useful not only because it has local couplings to potentials but also because it has manifest surface invariance in its couplings of the $F^A_{\mu\nu}$ and $F^B_{\mu\nu}$ fields strengths to sources. 
This is not true for all theories of electric and magnetic charges. In some cases the coupling between field strength and source is only Stokes surface invariant up to linking and intersection numbers~\cite{Blagojevic:1985sh}.

With this invariance in mind we return to surface defined in \eqref{a.surfElem}. The surface element in the worldline formalism is given by
\begin{align}
    \Sigma^{\mu\nu}(x) = 
    q\int d\tau d\eta  
    \delta^4(x-s(\tau,\eta))
    \frac{1}{2}
    \frac{\partial s^{[\mu}}{\partial\tau} 
    \frac{\partial s^{\nu]}}{\partial\eta}~.
\end{align}
In general this depends on our choice of spanning surface. We explore a particular choice in the next subsection that has relevance to the calculations made in this paper.

\subsection{Cylindrical Stoke's Surface}
Suppose that our surface parameterization is the extrusion of our current loop out to infinity along the direction $n^\mu$.
As a technical note, this is not a closed surface and so requires the usual assumption of vanishing boundary conditions in a neighborhood of infinity.
This would be parametrized by 
\begin{align}
    s^\mu(\eta,\tau) = z^\mu(\tau) + \eta \cdot n^\mu.
\end{align}
The surface element is then
\begin{align}
    \Sigma^{\mu\nu}(x) = 
    q\int_0^T d\tau \int_0^\infty d\eta  \,
    \delta^4(x-z(\tau) + \eta n)
    \frac{1}{2}
    \frac{dz^{[\mu}}{d\tau} 
    n^{\nu]}~.
\end{align}

For a semi-infinite representation of the Dirac string  we have that $(n\cdot \partial)^{-1}$ \cite{Zwanziger:1970hk} is given by
\begin{align}
    (n\cdot \partial)^{-1}(x)
    = -\int_0^\infty ds
    \delta^4(x + n s)~.
\end{align}
This allows us to write the surface as
\begin{align}
    \Sigma^{\mu\nu}(x) &= 
    -q\int_0^T d\tau
    \frac{dz^{[\mu}}{d\tau} 
    n^{\nu]}
    \frac{1}{2}
    \frac{1}{
    n\cdot \partial}( x - z(\tau))~.
\end{align}
In Fourier space we find 
\begin{align}
    \Sigma^{\mu\nu}(x) &= 
    q \int \frac{d^4k}{(2\pi)^4}   
    \int d\tau
    \frac{dz^{[\mu}}{d\tau}
    n^{\nu]}
    \frac{1}{2}
    \frac{i}{n \cdot k}
    e^{i k\cdot(x -z)}~.
\end{align}

We would like to express this in terms of the conserved current. To do this we must express the current in momentum space:
\begin{align}
    J(x) = q \int \frac{d^4k}
    {(2\pi)^4} 
    \int  d\tau
    \frac{dz^\mu}{d \tau} 
    e^{ik\cdot(x- z)}
    = \int \frac{d^4k}
    {(2\pi)^4} 
    e^{ik\cdot x}
    q\int dz^\mu 
    e^{-ik\cdot z}~,
\end{align}
which shows that the momentum space current is
\beq
\Tilde{J}(k) = q\int d\tau 
\frac{dz^\mu}{d\tau} 
e^{-ik\cdot z}~.
\eeq
Plugging this into the surface element produces
\begin{align}
    \Sigma^{\mu\nu}(x)
    = \int \frac{d^4k}
    {(2\pi)^4} 
    e^{ik\cdot x}
    \Tilde{J}^{[\mu} 
    n^{\nu]}
    \frac{1}{2}
    \frac{i}{n \cdot k}~.
\end{align}
Therefore, we find that the surface element can be expressed in momentum space in a way that depends only on the current 
\begin{align}
    \Tilde{\Sigma}^{\mu\nu}(k)
    = 
    \frac{1}{2}
    \frac{i}{n \cdot k}
    \Tilde{J}^{[\mu} 
    n^{\nu]}~.
\end{align}
Conversely, we see that the momentum space sources of this form correspond to surface integrals of cylinders created by extruding the current loop in the direction of $n^\mu$.

\section{Naive Potential Renormalization in Zwanziger Formalism\label{a.PotRenorm}}
In this appendix we calculate the renormalization of the Zwanziger propagators using the standard, potential-based methods. Our results agree the single potential calculations and with~\cite{Panagiotakopoulos:1982ne}, but contradict those of~\cite{Laperashvili:1999pu}.

To accurately calculate what follow it is useful to note that the matrix
\beq
\mathcal{M}_{\mu\nu}=\eta_{\mu\nu}+\frac{a}{n\cdot k}k_\mu n_\nu+\frac{b}{n\cdot k}k_\nu n_\mu+\frac{c}{n^2}n_\mu n_\nu+\frac{d}{k^2}k_\mu k_\nu+\frac{e}{n\cdot k}\varepsilon_{\mu\nu\alpha\beta}n^\alpha k^\beta ~,
\eeq
has the inverse
\beq
\mathcal{M}^{-1}_{\mu\nu}=x_0\left[\eta_{\mu\nu}+\frac{x_1}{D}\frac{k_\mu n_\nu}{n\cdot k}+\frac{x_2}{D}\frac{k_\nu n_\mu}{n\cdot k}+\frac{x_3}{D}\frac{n_\mu n_\nu}{n^2}+\frac{x_4}{D}\frac{k_\mu k_\nu}{k^2}-\frac{e}{n\cdot k}\varepsilon_{\mu\nu\alpha\beta}n^\alpha k^\beta \right]~,
\eeq
where
\begin{align}
x_0=&\left[1+e^2\left(1-\frac{n^2k^2}{(n\cdot k)^2} \right) \right]^{-1}~,\\
x_1=&(1+b+d)(a-e^2)-\left(d+e^2\frac{n^2k^2}{(n\cdot k)^2} \right)\left( a+c\frac{(n\cdot k)^2}{n^2k^2}\right)~,\\
x_2=&(1+a+c)(b-e^2)-\left(c+e^2\frac{n^2k^2}{(n\cdot k)^2} \right)\left( b+d\frac{(n\cdot k)^2}{n^2k^2}\right)~,\\
x_3=&(1+b+d)\left(c+e^2\frac{n^2k^2}{(n\cdot k)^2} \right)-(b-e^2)\left(c+a\frac{n^2k^2}{(n\cdot k)^2} \right)~,\\
x_4=&(1+a+c)\left(d+e^2\frac{n^2k^2}{(n\cdot k)^2} \right)-(a-e^2)\left(d+b\frac{n^2k^2}{(n\cdot k)^2} \right)~,\\
D=&\left(c+a\frac{n^2k^2}{(n\cdot k)^2} \right)\left(b+d\frac{(n\cdot k)^2}{n^2k^2} \right)-(1+a+c)(1+b+d)~.
\end{align}

First, we recall the standard one potential procedure. The propagator is
\beq
\Delta_{\mu\nu}=\frac{-i}{k^2}\left(\eta_{\mu\nu}-(1-\xi)\frac{k_\mu k_\nu}{k^2} \right)~,
\eeq
and the one-particle-irreducible corrections to the propagator take the form
\beq
i\Pi_{\mu\nu}=i\Pi(k)\left(k^2\eta_{\mu\nu}-k_\mu k_\nu \right)~.
\eeq
We then include all these effects as a geometric series
\begin{align}
\Delta^R_{\mu\nu}=&\Delta_{\mu\nu}+i\Delta^{\alpha}_\mu\Pi_{\alpha}^\beta\Delta_{\beta\nu}+\ldots=\Delta_{\mu}^\alpha\left[\eta_{\alpha_\nu}+i\Pi_{\alpha}^\beta\Delta_{\beta\nu}+\ldots \right]\nonumber\\
=&\Delta_{\mu}^\alpha\left[\eta^{\alpha\nu}-i\Pi^\alpha_\beta\Delta^{\beta\nu} \right]^{-1}=\frac{\Delta_{\mu}^\alpha}{1-\Pi}\left(\eta_{\alpha\nu}-\Pi\frac{k_\alpha k_\nu}{k^2} \right)\nonumber\\
=&\frac{\Delta_{\mu\nu}}{1-\Pi}+\frac{i\xi\Pi}{k^2(1-\Pi)}\frac{k_\mu k_\nu}{k^2}~.
\end{align}
Note that in general the renormalized propagator $\Delta^R$ is not simply a rescaling of the bare propagator, except for $\xi=0$. Or we could also define the renormalized $\xi_R=\xi(1-\Pi)$
\beq
\Delta^R_{\mu\nu}=\frac{-i}{k^2(1-\Pi)}\left(\eta_{\mu\nu}-(1-\xi_R)\frac{k_\mu k_\nu}{k^2} \right)~.
\eeq
Then, we connect this result to the running of the coupling $e$ by considering the interaction between two currents
\beq
eJ^\mu\Delta^R_{\mu\nu}eJ^\nu=-iJ^2\frac{e^2}{k^2(1-\Pi)}~.
\eeq
In general, we use the definition
\beq
\Delta^R_{\mu\nu}=Z^{-1}\Delta_{\mu\nu}~,
\eeq
where the $R$ label implies that the parameters inside are also renormalized. 

We now turn our attention to the two potential formalism. The Zwanziger Lagrangian in momentum space is
\begin{align}
\mathcal{L}=&-\frac{1}{2} A_{\mu}{\mathcal K}_{AA}^{\mu\nu}A_\nu-\frac{1}{2} B_{\mu}{\mathcal K}_{BB}^{\mu\nu}B_\nu+\frac{1}{2}A_{\mu}{\mathcal K}_{AB}^{\mu\nu}B_\nu-eJ^\mu A_\mu-bK^\mu B_\mu ,\label{e.PathInt}
\end{align}
where
\begin{align}
{\mathcal K}_{AA}^{\mu\nu}=&\frac{1}{n^2}\left[\eta^{\mu\nu}\left(n\cdot k \right)^2-\left(n\cdot k \right)\left(k^\mu n^\nu+k^\nu n^\mu \right)+\left(k^2-\xi_A^2\right)n^\mu n^\nu\right] , \label{e.MAA}\\
{\mathcal K}_{BB}^{\mu\nu}=&\frac{1}{n^2}\left[\eta^{\mu\nu}\left(n\cdot k \right)^2-\left(n\cdot k \right)\left(k^\mu n^\nu+k^\nu n^\mu \right)+\left(k^2-\xi_B^2\right)n^\mu n^\nu\right] ,\label{e.MBB}
\end{align}
are symmetric matrices including the gauge fixing terms~\cite{Gubarev:1998ss}
\beq
\mathcal{L}_\text{G.F.}=\frac{1}{2n^2}\left[\xi_A^2\left(n\cdot A \right)^2+\xi_B^2\left(n\cdot B \right)^2 \right]~,
\eeq
where $\xi_{A,B}$ have mass dimension one. We also have the antisymmetric tensor
\beq
\mathcal{K}_{AB}^{\mu\nu}=-\mathcal{K}_{BA}^{\mu\nu}=\frac{n\cdot k}{n^2}\varepsilon^{\mu\nu\alpha\beta}n_\alpha k_\beta~.
\eeq
The gauge diagonal matrices can be inverted to find
\begin{align}
{\mathcal K}_{AA}^{-1\mu\nu}=\frac{n^2}{(n\cdot k)^2}\left[\eta^{\mu\nu}-\frac{1}{n^2}n^\mu n^\nu-\frac{1}{\xi_A^2}k^\mu k^\nu \right]~,\\
{\mathcal K}_{BB}^{-1\mu\nu}=\frac{n^2}{(n\cdot k)^2}\left[\eta^{\mu\nu}-\frac{1}{n^2}n^\mu n^\nu-\frac{1}{\xi_B^2}k^\mu k^\nu \right]~.
\end{align}
But these are not the propagators, they do not take into account the mixing between the fields. We record the useful relation
 \beq
 \varepsilon_{\mu\sigma\alpha\beta}n^\alpha k^\beta\varepsilon_{\rho\nu\gamma\delta}n^\gamma k^\delta \eta^{\mu\nu}=\eta_{\sigma\rho}\left[k^2n^2-\left( n\cdot k\right)^2 \right]+\left(n\cdot k \right)\left(n_\sigma k_\rho+n_\rho k_\sigma \right)-k^2 n_\sigma n_\rho-n^2 k_\sigma k_\rho~.
 \eeq 

The propagators are obtained by writing the kinetic terms as
\beq
-\frac{1}{2}\left(A_\mu~,~B_\mu \right)\left(\begin{array}{cc}
\mathcal{K}_{AA}^{\mu\nu} & -\mathcal{K}_{AB}^{\mu\nu}\\[0.1cm]
-\mathcal{K}_{BA}^{\mu\nu} & \mathcal{K}_{BB}^{\mu\nu}
\end{array} \right)\left(\begin{array}{c}
A_\nu\\
B_\nu
\end{array} \right)\equiv -\frac{1}{2}V_\mu^T\,\bm{\mathcal{K}}^{\mu\nu}V_\nu~.\label{e.Vkin}
\eeq
This block matrix can be inverted using the formulae in Eq.~\eqref{e.BlockInverse}. We find
\beq
-i\bm{\mathcal{K}}^{-1}_{\mu\nu}=\bm{\Delta}_{\mu\nu}=\left(\begin{array}{cc}
\Delta^{AA}_{\mu\nu} & \Delta^{AB}_{\mu\nu}\\[0.1cm]
\Delta^{BA}_{\mu\nu} & \Delta^{BB}_{\mu\nu}
\end{array}\right)~,\label{e.VbareProp}
\eeq
where we have defined
\begin{align}
\Delta^{AA}_{\mu\nu}=&-i{\mathcal K}_{AA}^{-1\mu\alpha}\left[\delta^\alpha_\nu-\mathcal{K}_{AB}^{\alpha\beta}{\mathcal K}^{BB-1}_{\beta\gamma}\mathcal{K}_{BA}^{\gamma\delta}{\mathcal K}^{AA-1}_{\delta\nu} \right]^{-1}\nonumber\\
=&\frac{-i}{k^2}\left[\eta_{\mu\nu}-\frac{n_\mu k_\nu+n_\nu k_\mu}{n\cdot k} +\frac{n^2}{(n\cdot k)^2}\left(1-\frac{k^2}{\xi_A^2} \right)k_\mu k_\nu\right]~,\\
\Delta^{BB}_{\mu\nu}=&-i{\mathcal K}_{BB}^{-1\mu\alpha}\left[\delta^\alpha_\nu-\mathcal{K}_{BA}^{\alpha\beta}{\mathcal K}^{AA-1}_{\beta\gamma}\mathcal{K}_{AB}^{\gamma\delta}{\mathcal K}^{BB-1}_{\delta\nu} \right]^{-1}\nonumber\\
=&\frac{-i}{k^2}\left[\eta_{\mu\nu}-\frac{n_\mu k_\nu+n_\nu k_\mu}{n\cdot k} +\frac{n^2}{(n\cdot k)^2}\left(1-\frac{k^2}{\xi_B^2} \right)k_\mu k_\nu\right]~\\
\Delta^{AB}_{\mu\nu}=&\Delta^{AA}_{\mu\alpha}\mathcal{K}^{\alpha\beta}_{AB}\mathcal{K}^{BB-1}_{\beta\nu}=\mathcal{K}^{AA-1}_{\mu\alpha}\mathcal{K}^{\alpha\beta}_{AB}\Delta_{\beta\nu}^{BB}=-\frac{i}{k^2}\frac{\varepsilon_{\mu\nu\alpha\beta}n^\alpha k^\beta}{n\cdot k}~,\\
\Delta^{BA}_{\mu\nu}=&\Delta^{BB}_{\mu\alpha}\mathcal{K}^{\alpha\beta}_{BA}\mathcal{K}^{AA-1}_{\beta\nu}=\mathcal{K}^{BB-1}_{\mu\alpha}\mathcal{K}^{\alpha\beta}_{BA}\Delta_{\beta\nu}^{AA}=\frac{i}{k^2}\frac{\varepsilon_{\mu\nu\alpha\beta}n^\alpha k^\beta}{n\cdot k}~.
\end{align}

These propagators can also be obtained by treating $\mathcal{K}_{AB}$ as a mixing term. For instance, we consider the infinite series of mixings
\begin{align}
&{\mathcal K}_{AA}^{-1\mu\nu}+{\mathcal K}_{AA}^{-1\mu\alpha}\mathcal{K}^{AB}_{\alpha\beta}{\mathcal K}_{BB}^{-1\beta\gamma}\mathcal{K}^{BA}_{\gamma\delta}{\mathcal K}_{AA}^{-1\delta\nu}+\ldots\nonumber\\
=&{\mathcal K}_{AA}^{-1\mu\alpha}\left[\delta^\alpha_\nu-\mathcal{K}_{AB}^{\alpha\beta}{\mathcal K}^{-1BB}_{\beta\gamma}\mathcal{K}_{BA}^{\gamma\delta}{\mathcal K}^{-1AA}_{\delta\nu} \right]^{-1}=i\Delta_{AA}^{\mu\nu}
\end{align}
A nearly identical calculation leads to the $B_\mu$ to $B_\nu$ propagator. The mixed propagator is obtained from the series
\begin{align}
&\mathcal{K}_{AA}^{-1\mu\alpha}\mathcal{K}^{AB}_{\alpha\beta}\mathcal{K}_{BB}^{-1\beta\nu}+{\mathcal K}_{AA}^{-1\mu\alpha}\mathcal{K}^{AB}_{\alpha\beta}{\mathcal K}_{BB}^{-1\beta\gamma}\mathcal{K}^{BA}_{\gamma\delta}{\mathcal K}_{AA}^{-1\delta\rho}\mathcal{K}^{AB}_{\rho\sigma}\mathcal{K}_{BB}^{-1\sigma\nu}+\ldots\nonumber\\
=&\mathcal{K}_{AA}^{-1\mu\alpha}\mathcal{K}^{AB}_{\alpha\beta}\mathcal{K}_{BB}^{-1\beta\gamma}\left[\delta^\gamma_\nu-\mathcal{K}_{BA}^{\gamma\delta}{\mathcal K}^{AA-1}_{\delta\rho}\mathcal{K}_{AB}^{\rho\sigma}\mathcal{K}^{BB-1}_{\sigma\nu} \right]^{-1}=i\Delta_{AB}^{\mu\nu}~.
\end{align}

To track the effects of renormalization we take $A_\mu\to\sqrt{Z_A}A_\mu$ and $B_\mu\to\sqrt{Z_B}B_\mu$. By tracking these changes through the various matrices we find the expected results
\beq
\Delta_{AA}\to Z_A^{-1}\Delta_{AA}~, \ \ \ \ \Delta_{BB}\to Z_B^{-1}\Delta_{BB}~, \ \ \ \ \Delta_{AB}\to\frac{1}{\sqrt{Z_AZ_B}}\Delta_{AB}~.
\eeq
We assume the renormalized propagator $\bm{\Delta}^R_{\mu\nu}=-i\bm{\mathcal{K}}^{R-1}_{\mu\nu}$ is related to the bare propagator determined from Eq.~\eqref{e.VbareProp} by
\beq
\bm{\Delta}^{R-1}_{\mu\nu}(k)=\bm{\Delta}^{-1}_{\mu\nu}(k)-i\bm{\Pi}_{\mu\nu}~,
\eeq
where
\beq
i\bm{\Pi}_{\mu\nu}=i\left(\begin{array}{cc}
\Pi^{QQ}_{\mu\nu} &0 \\
0 & \Pi^{GG}_{\mu\nu} 
\end{array} \right)~.
\eeq
Here was have assumed there is no loop that directly connects the two potentials. We can then write the renormalized propagator in terms of the bare propagator
\beq
\bm{\Delta}^{R}_{\mu\nu}(k)=\bm{\Delta}_{\mu\alpha}(k)\left[\mathbb{I}_2\delta^\nu_{\alpha}-i\bm{\Pi}_{\alpha}^\beta\bm{\Delta}_{\beta}^\nu \right]^{-1}
\eeq
We explicitly find
\begin{align}
&\mathbb{I}_2 g_{\mu\nu}-i\bm{\Pi}_{\mu}^\alpha\bm{\Delta}_{\alpha\nu}=\left(\begin{array}{cc}
\mathcal{M}^{QQ}_{\mu\nu} & \mathcal{M}^{QG}_{\mu\nu} \\
\mathcal{M}^{GQ}_{\mu\nu} & \mathcal{M}^{GG}_{\mu\nu}
\end{array} \right)=\left(\begin{array}{cc}
\displaystyle \eta_{\mu\nu}-i\Pi_\mu^{QQ\alpha}\Delta^{AA}_{\alpha\nu} &\displaystyle -i\Pi_\mu^{QQ\alpha}\Delta^{AB}_{\alpha\nu} \\[0.1cm]
\displaystyle -i\Pi_\mu^{GG\alpha}\Delta^{BA}_{\alpha\nu} &\displaystyle \eta_{\mu\nu}-i\Pi_\mu^{GG\alpha}\Delta^{BB}_{\alpha\nu} 
\end{array} \right).
\end{align}
By using the block matrix inversion formulae in Eq.~\eqref{e.BlockInverse} we can write
\begin{align}
\Delta^{R}_{QQ}=&
\left(\Delta^{AA}-\Delta^{AB}\mathcal{M}_{GG}^{-1}\mathcal{M}^{GQ} \right)\left(\mathcal{M}_{QQ}-\mathcal{M}^{QG}\mathcal{M}_{GG}^{-1}\mathcal{M}^{GQ} \right)^{-1} ~,\\
\Delta^{R}_{QG}=&
\left(\Delta^{AB}-\Delta^{AA}\mathcal{M}_{QQ}^{-1}\mathcal{M}^{QG} \right)\left(\mathcal{M}_{GG}-\mathcal{M}^{GQ}\mathcal{M}_{QQ}^{-1}\mathcal{M}^{QG}\right)^{-1}~,\\
\Delta^{R}_{GQ}=&
\left(\Delta^{BA}-\Delta^{BB}\mathcal{M}_{GG}^{-1}\mathcal{M}^{GQ} \right)\left(\mathcal{M}_{QQ}-\mathcal{M}^{QG}\mathcal{M}_{GG}^{-1}\mathcal{M}^{GQ} \right)^{-1} ~,\\
\Delta^{R}_{GG}=&
\left(\Delta^{BB}-\Delta^{BA}\mathcal{M}_{QQ}^{-1}\mathcal{M}^{QG} \right)\left(\mathcal{M}_{GG}-\mathcal{M}^{GQ}\mathcal{M}_{QQ}^{-1}\mathcal{M}^{QG}\right)^{-1}~.
\end{align}

If we assume 
\beq
i\bm{\Pi}_{\mu\nu}=i\left(k^2\eta_{\mu\nu}-k_\mu k_\nu \right)\left( \begin{array}{cc}
\Pi^{QQ}(k^2) &0\\
0 & \Pi^{GG}(k^2)
\end{array}\right)~,
\eeq
is a two by two matrix that encodes the quantum corrections to propagators. We have here assumed a tensor structure that is independent of $n^\mu$. This is clearly justified if we confine ourselves to one-loop results, which we assume going forward. We then find
\beq
\left(\begin{array}{cc}
\displaystyle \eta_{\mu\nu}\left(1-\Pi^{QQ}\right)+\Pi^{QQ}\frac{n_\mu k_\nu}{n\cdot k} &\displaystyle  -\Pi^{QQ}\varepsilon_{\mu\nu\sigma\rho}\frac{n^\sigma k^\rho}{n\cdot k} \\[0.4cm]
\displaystyle  +\Pi^{GG}\varepsilon_{\mu\nu\sigma\rho}\frac{n^\sigma k^\rho}{n\cdot k} &\displaystyle  \eta_{\mu\nu}\left(1-\Pi^{GG}\right)+\Pi^{GG}\frac{n_\mu k_\nu}{n\cdot k}
\end{array} \right).
\eeq
These terms can be evaluated using the linear in the $\Pi$s results
\begin{align}
\mathcal{M}_{\mu\nu}^{QQ-1}=&\frac{1}{1-\Pi^{QQ}}\left[\eta_{\mu\nu}-\Pi^{QQ}\frac{n_\mu k_\nu }{n\cdot k}\right]~,\\
\mathcal{M}^{GG-1}_{\mu\nu}=&\frac{1}{1-\Pi^{GG}}\left[\eta_{\mu\nu}-\Pi^{GG}\frac{n_\mu k_\nu }{n\cdot k} \right]~.
\end{align}

This leads to
\begin{align}
\Delta^{R}_{QQ}=&\frac{-i}{k^2(1-\Pi^{QQ})(1-\Pi^{GG})}\left[\eta_{\mu\nu}-\frac{n_\mu k_\nu+n_\nu k_\mu}{n\cdot k} +\frac{n^2}{(n\cdot k)^2}k_\mu k_\nu\right]\nonumber\\
&+i\frac{n^2}{(n\cdot k)^2\xi_A^2}k_\mu k_\nu+\frac{i\Pi^{GG}}{(n\cdot k)^2}\left(n^2\eta_{\mu\nu}-n_\mu n_\nu \right)~,\\
\Delta^{R}_{QG}=&\frac{1}{(1-\Pi^{QQ})(1-\Pi^{GG})}\Delta^{AB}_{\mu\nu}~,\\
\Delta^{R}_{GQ}=&\frac{1}{(1-\Pi^{QQ})(1-\Pi^{GG})}\Delta^{BA}_{\mu\nu}~,\\
\Delta^{R}_{GG}=&\frac{-i}{k^2(1-\Pi^{QQ})(1-\Pi^{GG})}\left[\eta_{\mu\nu}-\frac{n_\mu k_\nu+n_\nu k_\mu}{n\cdot k} +\frac{n^2}{(n\cdot k)^2}k_\mu k_\nu \right]\nonumber\\
&+i\frac{n^2}{(n\cdot k)^2\xi_B^2}k_\mu k_\nu+\frac{i\Pi^{QQ}}{(n\cdot k)^2}\left(n^2\eta_{\mu\nu}-n_\mu n_\nu \right)~.
\end{align}
As emphasized in the text, it appears (if we focus only on terms with a $k^2$ pole) that $Z_A$ and $Z_B$ are equal. The extra terms in the charge-diagonal propagators are also unexpected. In the literature this confusion led to the result first obtained by Schwinger that the electric and magnetic charges are renormalized in the same way. It is has also led to many ad hoc reasons for dropping the terms proportional to the $n^\mu$ projector that do not appear in the tree level propagator. In fact, these results exactly match what we determined by using the field strength language when the topological portion of the electric-magnetic interactions are not fully removed from the calculation.


\begin{thebibliography}{99}

\bibitem{Thomson:1904}
J.J.~Thomson, {\it Elements of the Mathematical Theory of Electricity and Magnetism, 3rd Edition, Sec. 284}
\href{https://doi.org/10.1017/CBO9780511694141}{Cambridge University Press (1904)}

\bibitem{Dirac:1931kp}
P.~A.~M.~Dirac,
``Quantised singularities in the electromagnetic field,,''
\href{http://doi.org/10.1098/rspa.1931.0130}{Proc. Roy. Soc. Lond. A \textbf{133} (1931) no.821, 60-72}

\bibitem{Weinberg:1965rz}
S.~Weinberg,
``Photons and gravitons in perturbation theory: Derivation of Maxwell's and Einstein's equations,''
\href{http://doi.org/10.1103/PhysRev.138.B988}{Phys. Rev. \textbf{138} (1965), B988-B1002}


\bibitem{Terning:2018udc}
J.~Terning and C.~B.~Verhaaren,
``Resolving the Weinberg Paradox with Topology,''
\href{https://doi.org/10.1007/JHEP03(2019)177}{JHEP \textbf{03} (2019), 177}
\arXiv{1809.05102}{hep-th}.

\bibitem{Schwinger:1966zza}
J.~Schwinger,
``Electric- and Magnetic-Charge Renormalization. I,''
\href{http://doi.org/10.1103/PhysRev.151.1048}{Phys. Rev. \textbf{151} (1966), 1048-1054}

\bibitem{Schwinger:1966zzb}
J.~Schwinger,
``Electric- and Magnetic-Charge Renormalization. II,''
\href{http://doi.org/10.1103/PhysRev.151.1055}{Phys. Rev. \textbf{151} (1966), 1055-1057}

\bibitem{Coleman:1982cx}
S.~R.~Coleman,
``The Magnetic Monopole Fifty Years Later,''
\href{https://lib-extopc.kek.jp/preprints/PDF/1982/8211/8211084.pdf}{HUTP-82-A032.}


\bibitem{Terning:2018lsv} 
  J.~Terning and C.~B.~Verhaaren,
  ``Dark Monopoles and $SL(2,\mathbb{Z})$ Duality,''
 \href{https://doi.org/10.1007/JHEP12(2018)123}{JHEP \textbf{12} (2018), 123}
  \arXiv{1808.09459}{hep-th}.
  

\bibitem{Halpern:1978ik}
M.~B.~Halpern,
``Field Strength and Dual Variable Formulations of Gauge Theory,''
\href{http://doi.org/10.1103/PhysRevD.19.517}{Phys. Rev. D \textbf{19} (1979), 517}


\bibitem{Calucci:1981we}
G.~Calucci, R.~Jengo and M.~T.~Vallon,
``On the Quantum Field Theory of Charges and Monopoles,''
\href{https://doi.org/10.1016/0550-3213(82)90156-0}{Nucl. Phys. B \textbf{197} (1982), 93-112}.


\bibitem{Calucci:1982fm}
G.~Calucci, R.~Jengo and M.~T.~Vallon,
``A Quantum Field Theory of Dyons and Photons,''
\href{https://doi.org/10.1016/0550-3213(83)90186-4}{Nucl. Phys. B \textbf{211} (1983), 77-92}.


\bibitem{Calucci:1982wy}
G.~Calucci and R.~Jengo,
``On the Renormalization of the Quantum Field Theory of Point-like Monopoles and Charges,''
\href{https://doi.org/10.1016/0550-3213(83)90067-6}{Nucl. Phys. B \textbf{223} (1983), 501-524}.

\bibitem{Blagojevic:1985yd}
M.~Blagojevic and R.~Jengo,
``The Electron-Monopole Interaction as a Wess-Zumino Term,''
\href{https://doi.org/10.1016/0370-2693(85)91242-0}{Phys. Lett. B \textbf{165} (1985), 343-346}


\bibitem{GomezSanchez:2011orv}
C.~Gomez Sanchez and B.~Holdom,
``Monopoles, strings and dark matter,''
\href{http://doi.org/10.1013/PhysRevD.83.123524}{Phys. Rev. D \textbf{83} (2011), 123524}
\arXiv{1103.1632}{hep-ph}.

\bibitem{Hook:2017vyc}
  A.~Hook and J.~Huang,
 ``Bounding millimagnetically charged particles with magnetars,''
\href{http://doi.org/10.1103/PhysRevD.96.055010}{Phys.\ Rev.\ D {\bf 96} (2017)  055010}
\arXiv{1705.01107}{hep-ph}.

\bibitem{Terning:2019bhg}
J.~Terning and C.~B.~Verhaaren,
``Detecting Dark Matter with Aharonov-Bohm,''
\href{https://doi.org/10.1007/JHEP12(2019)152}{JHEP \textbf{12} (2019), 152}
\arXiv{1906.00014}{hep-ph}.

\bibitem{Terning:2020dzg}
J.~Terning and C.~B.~Verhaaren,
``Spurious Poles in the Scattering of Electric and Magnetic Charges,''
\href{https://doi.org/10.1007/JHEP12(2020)153}{JHEP \textbf{12} (2020), 153}
\arXiv{2010.02232}{hep-th}.

\bibitem{Graesser:2021vkr}
M.~L.~Graesser, I.~M.~Shoemaker and N.~T.~Arellano,
``Milli-magnetic monopole dark matter and the survival of galactic magnetic fields,''
\href{https://doi.org/10.1007/JHEP03(2022)105}{JHEP \textbf{03} (2022), 105}
\arXiv{2105.05769}{hep-ph}.


  \bibitem{SeibergWitten}
  N.~Seiberg and E.~Witten,
  ``Electric - magnetic duality, monopole condensation, and confinement in N=2 supersymmetric Yang-Mills theory,''
  \href{https://doi.org/10.1016/0550-3213(94)90124-4}{Nucl.\ Phys.\ B {\bf 426} (1994) 19}
   [Erratum-ibid.\ \href{https://doi.org/10.1016/0550-3213(94)00449-8 }{B {\bf 430} (1994) 485}]
  \arXivold{hep-th/9407087};
  ``Monopoles, duality and chiral symmetry breaking in N=2 supersymmetric QCD,''
  \href{https://doi.org/10.1016/0550-3213(94)90214-3}{Nucl.\ Phys.\ B {\bf 431} (1994) 484}
  \arXivold{hep-th/9408099}.

\bibitem{Colwell:2015wna}
K.~Colwell and J.~Terning,
``S-Duality and Helicity Amplitudes,''
\href{https://doi.org/10.1007/JHEP03(2016)068}{JHEP \textbf{03} (2016), 068}
\arXiv{1510.07627}{hep-th}.


  
  
\bibitem{Zwanziger:1970hk}
D.~Zwanziger,
``Local Lagrangian quantum field theory of electric and magnetic charges,''
  \href{http://doi.org/10.1103/PhysRevD.3.880}{Phys. Rev. D \textbf{3} (1971), 880}

\bibitem{Brandt:1977fa}
R.~A.~Brandt and F.~Neri,
``Remarks on Zwanziger's Local Quantum Field Theory of Electric and Magnetic Charge,''
\href{https://doi.org/10.1103/PhysRevD.18.2080}{Phys. Rev. D \textbf{18} (1978), 2080}

\bibitem{Deans:1981qs}
W.~Deans,
``Quantum Field Theory of Dirac Monopoles and the Charge Quantization Condition,''
\href{https://doi.org/10.1016/0550-3213(82)90294-2}{Nucl. Phys. B \textbf{197} (1982), 307-333}

\bibitem{Panagiotakopoulos:1982fp}
C.~Panagiotakopoulos,
``Infinity Subtraction in a Quantum Field Theory of Charges and Monopoles,''
\href{https://doi.org/10.1088/0305-4470/16/1/022}{J. Phys. A \textbf{16} (1983), 133}

\bibitem{Jengo:1982wx}
R.~Jengo and M.~T.~Vallon,
``Vacuum Effects on the Static Monopole-Anti-Monopole Interaction,''
\href{https://doi.org/10.1007/BF02816654}{Nuovo Cim. A \textbf{77} (1983), 249}

\bibitem{Goebel:1983we}
C.~J.~Goebel and M.~T.~Thomaz,
``Antishielding of Magnetic Charge,''
\href{https://doi.org/10.1103/PhysRevD.30.823}{Phys. Rev. D \textbf{30} (1984), 823}

\bibitem{Holdom:1985ag}
  B.~Holdom,
 ``Two U(1)'s and Epsilon Charge Shifts,''
\href{http://dx.doi.org/doi:10.1016/0370-2693(86)91377-8}{Phys.\ Lett.\  {\bf 166B} (1986) 196}.

\bibitem{Brummer:2009cs}
F.~Brummer and J.~Jaeckel,
``Minicharges and Magnetic Monopoles,''
\href{http://doi.org/10.1016/j.physletb.2009.04.041}{Phys. Lett. B \textbf{675} (2009), 360-364}
\arXiv{0902.3615}{hep-ph}.

  \bibitem{Bruemmer:2009ky}
  F.~Brummer, J.~Jaeckel and V.~V.~Khoze,
  ``Magnetic Mixing: Electric Minicharges from Magnetic Monopoles,''
  \href{https://doi.org/10.1088/1126-6708/2009/06/037}{JHEP {\bf 0906} (2009) 037}
  \arXiv{0905.0633}{hep-ph}.


\bibitem{DelZotto:2016fju}
M.~Del Zotto, J.~J.~Heckman, P.~Kumar, A.~Malekian and B.~Wecht,
 ``Kinetic Mixing at Strong Coupling,''
\href{https://doi.org/10.1103/PhysRevD.95.016007}{Phys.\ Rev.\ D {\bf 95} (2017) no.1,  016007}
 \arXiv{1608.06635}{hep-ph}.
  
\bibitem{Hook:2022pcf}
A.~Hook and J.~Huang,
``A Mass for the Dual Photon,''
\arXiv{2210.00015}{hep-ph}.



\bibitem{Blagojevic:1985sh}
M.~Blagojevic and P.~Senjanovic,
``The Quantum Field Theory of Electric and Magnetic Charge,''
\href{https://doi.org/10.1016/0370-1573(88)90098-1}{Phys. Rept. \textbf{157} (1988), 233}.

\bibitem{Strassler:1992zr}
M.~J.~Strassler,
``Field theory without Feynman diagrams: One loop effective actions,''
\href{https://doi.org/10.1016/0550-3213(92)90098-V}{Nucl. Phys. B \textbf{385} (1992), 145-184}
\arXivold{hep-ph/9205205}.

\bibitem{Schubert:1996jj}
C.~Schubert,
``An Introduction to the worldline technique for quantum field theory calculations,''
\href{https://www.actaphys.uj.edu.pl/R/27/12/3965/pdf}{Acta Phys. Polon. B \textbf{27} (1996), 3965-4001}
\arXivold{hep-th/9610108}.




\bibitem{Brandt:1977be}
R.~A.~Brandt, F.~Neri and D.~Zwanziger,
``Lorentz Invariance of the Quantum Field Theory of Electric and Magnetic Charge,''
\href{https://doi.org/10.1103/PhysRevLett.40.147}{Phys. Rev. Lett. \textbf{40} (1978), 147-150}


\bibitem{Brandt:1978wc}
R.~A.~Brandt, F.~Neri and D.~Zwanziger,
``Lorentz Invariance From Classical Particle Paths in Quantum Field Theory of Electric and Magnetic Charge,''
\href{https://doi.org/10.1103/PhysRevD.19.1153}{Phys. Rev. D \textbf{19} (1979), 1153}


\bibitem{Reece:2023czb}
M.~Reece,
``TASI Lectures: (No) Global Symmetries to Axion Physics,''
\href{https://doi.org/10.22323/1.439.0008}{PoS \textbf{TASI2022} (2024), 008}
\arXiv{2304.08512}{hep-ph}.


\bibitem{Dirac:1948um}
P.~A.~M.~Dirac,
``The Theory of magnetic poles,''
\href{https://doi.org/10.1103/PhysRev.74.817}{Phys. Rev. \textbf{74} (1948), 817-830}.

\bibitem{Blagojevic:1978zv}
M.~Blagojevic, D.~Nesic, P.~Senjanovic, D.~Sijacki and D.~Zivanovic,
``A New Approach to the Quantum Field Theory of Electric and Magnetic Charge,''
\href{http://doi.org/10.1016/0370-2693(78)90439-2}{Phys. Lett. B \textbf{79} (1978), 75-78}



\bibitem{Blagojevic:1979bm}
M.~Blagojevic and P.~Senjanovic,
``A One Potential Formulation of the Quantum Field Theory of Magnetic Poles,''
\href{http://doi.org/10.1016/0550-3213(79)90129-9}{Nucl. Phys. B \textbf{161} (1979), 112}


\bibitem{Argyres:1995jj}
P.~C.~Argyres and M.~R.~Douglas,
``New phenomena in SU(3) supersymmetric gauge theory,''
\href{http://doi.org/10.1016/0550-3213(95)00281-V}{Nucl. Phys. B \textbf{448} (1995), 93-126}
\arXivold{hep-th/9505062}.


\bibitem{Csaki:2010rv}
  C.~Csaki, Y.~Shirman and J.~Terning,
  ``Anomaly Constraints on Monopoles and Dyons,''
  \href{https://doi.org/10.1103/PhysRevD.81.125028}{Phys.\ Rev.\ D {\bf 81} (2010) 125028}
  \arXiv{1003.0448}{hep-th}.
  
\bibitem{Burkhardt:2011ur}
H.~Burkhardt and B.~Pietrzyk,
``Recent BES measurements and the hadronic contribution to the QED vacuum polarization,''
\href{https://doi.org/10.1103/PhysRevD.84.037502}{Phys. Rev. D \textbf{84} (2011), 037502}
\arXiv{1106.2991}{ hep-ex}.

\bibitem{L3:2005tsb}
P.~Achard \textit{et al.} [L3],
``Measurement of the running of the electromagnetic coupling at large momentum-transfer at LEP,''
\href{https://doi.org/10.1016/j.physletb.2005.07.052}{Phys. Lett. B \textbf{623} (2005), 26-36}
\arXivold{hep-ex/0507078}.

\bibitem{Schwartz:2014sze}
M.~D.~Schwartz,
``Quantum Field Theory and the Standard Model,''
Cambridge University Press, 2014.

\bibitem{Greiner:1996}
W.~Greiner and J. Reinhardt
``Field Quantization,''
Springer-Verlag Berlin Heidelberg, 1996.


\bibitem{Weinberg:1995mt}
S.~Weinberg,
``The Quantum theory of fields. Vol. 1: Foundations,''
Cambridge University Press, 2005.


\bibitem{Faddeev:1980be}
L.~D.~Faddeev and A.~A.~Slavnov,
``Gauge Fields. Introduction to Quantum Theory,''
Front. Phys. \textbf{50}, 1-232 (1980)



\bibitem{Panagiotakopoulos:1982ne}
C.~Panagiotakopoulos,
``Renormalization of the QEMD of a Dyon Field,''
\href{https://doi.org/10.1016/0550-3213(83)90600-4}{Nucl. Phys. B \textbf{212} (1983), 118-130}


\bibitem{Laperashvili:1999pu} 
  L.~V.~Laperashvili and H.~B.~Nielsen,
  ``Dirac relation and renormalization group equations for electric and magnetic fine structure constants,''
  \href{https://doi.org/10.1142/S0217732399002935}{Mod.\ Phys.\ Lett.\ A {\bf 14}, 2797 (1999)}
\arXivold{hep-th/9910101}

\bibitem{Gubarev:1998ss}
F.~V.~Gubarev, M.~I.~Polikarpov and V.~I.~Zakharov,
``Monopole - anti-monopole interaction in Abelian Higgs model,''
\href{https://doi.org/10.1016/S0370-2693(98)00957-5}{Phys. Lett. B \textbf{438} (1998), 147-153}
\arXivold{hep-th/9805175}.


\end{thebibliography}
\end{document}